\documentstyle[eqsecnum,epsf,aps,prb,multicol]{revtex}

\begin{document} 
\bibliographystyle{prsty}
\draft

\title{The N-chain Hubbard Model in Weak Coupling}
\author{Hsiu-Hau Lin}
\address{Department of Physics, University of
              California, Santa Barbara, CA 93106-9530}
\author{Leon Balents} 
\address{Institute for Theoretical Physics, University of California,
              Santa Barbara, CA 93106-4030}
\author{Matthew P. A. Fisher}
\address{Department of Physics, University of
               California, Santa Barbara, CA 93106\\
         and Institute for Theoretical Physics, University of California,
              Santa Barbara, CA 93106-4030}
\date{\today} 
\maketitle

\begin{abstract} 
  We present a systematic weak-coupling renormalization group (RG)
  technique for studying a collection of $N$ coupled one-dimensional
  interacting electron systems, focusing on the example of $N$-leg
  Hubbard ladders.  For $N=2,3$, we recover previous results, and find
  that also more generally broad regions of the phase space of these
  models are unstable to pairing, usually with approximate $d$-wave
  symmetry.  We show how these instabilities can be understood in
  terms of a fairly conventional ``gap'' function $\Delta$ at the
  discretized Fermi surface, and describe how this function is
  calculated.  The dimensional crossovers as $N \rightarrow \infty$
  and as many such ladders are weakly coupled together are also
  discussed.

\end{abstract}

\begin{multicols}{2}

\section{Introduction} 
Over the past few years, considerable interest has focused on systems 
of coupled chain conductors.  Early theoretical studies of Heisenberg 
ladders (appropriate for the strongly interacting, non-itinerant 
half-filled limit) revealed an interesting odd/even 
effect.\cite{Schulz86,Dagotto92,Rice93,Affleck94,White94}\  
If the number of legs of the ladder, 
$N$, is even, the system is expected to be a spin-liquid with a 
singlet ground state and a (spin) gap to the lowest lying excitations 
carrying angular momentum.  For odd $N$, the ground state has 
quasi-long-range antiferromagnetic order and a set of gapless 
spin-wave excitations, which puts it in the universality class of 
the single spin-$1/2$ Heisenberg chain.  Recent progress in the 
experimental preparation of relatively isolated spin ladders has begun 
to probe some of this rich physics and appears to have verified these 
expectations for $N=2,3$.\cite{Azuma94,Tennant95,Keren93,Satija80}\

The behavior of {\sl doped} ladders, i.e.  those with itinerant charge
carriers, is much richer.  Particular theoretical attention has been
paid to the case $N=2$, the two-leg
ladder.\cite{Dagotto92,Rice93,Finkel'stein93,Fabrizio93,Kuroki94,Balents96,Schulz96a,Nagaosa95,Orignac96}\
Early motivation stemmed from the possibility of realizing a concrete
example of resonating valence bond (RVB)
ideas.\cite{Kivelson87,Anderson87}\  According to this line of thought,
since the two-leg Heisenberg ladder is a spin liquid, the doped
carriers experience a short-range attractive interaction, leading to
pairing and the persistence of the spin gap.  Such behavior is indeed
observed in simulations of two-chain
Hubbard\cite{Noack94,Noack95,Dagotto96} and $t$-$J$
models,\cite{Dagotto92,Hayward95,Hayward96,Sano96,Troyer96}\ which push
the current computational limits of numerical methods working directly
at zero temperature.  Subsequent work by numerous authors has since
demonstrated the existence of such a spin gap phase for low dopings by
controlled analytical methods in weak
coupling.\cite{Balents96,Schulz96a,Fabrizio93}\  This weak-coupling
approach has the additional advantage that it provides a full picture
of the phase diagram, even away from half-filling.

In this paper, we extend this analysis to more general 
$N$-chain Hubbard 
models.\cite{Schulz96b,Kimura96a,Kimura96b,Penc96}\  
Such an extension is useful in two 
respects.  First, it allows a determination of the phase diagram for 
any small value of $N$, thereby elucidating the physics of doped spin 
liquids, the even/odd effect, and geometrical frustration.  
Furthermore, our equations allow a complete interpolation between one 
and two dimensions (along a particular path in parameter space -- see 
below).  An understanding of such a dimensional 
crossover\cite{Dagotto96}\ is a crucial first step in interpreting 
experiments in {\sl quasi}-one-dimensional conductors.

To determine the behavior in the weak interaction limit, we employ a
generalization of the renormalization group (RG) developed in
Ref.\onlinecite{Balents96}\ (the extension to the particular case $N=3$ has
already by studied by Arrigoni\cite{Arrigoni96a,Arrigoni96b}).  This
provides a systematic basis for treating the logarithmic divergences
arising in a naive perturbative analysis.  Coupled with the technique
of bosonization, the primary output of the RG is a ``gap'' function
$\Delta$, describing pairing and the relative phase among the various spin
and charge modes in the system.  In the limit of large $N$, $\Delta$
becomes identical to the gap function defined in the conventional BCS
theory of superconductivity, and one may thereby connect our results
directly with higher dimensional analogues.

The RG also determines the zero temperature behavior as the chain
length is taken to infinity.  Because such a system is, for any finite
$N$, still one-dimensional, it cannot sustain true off-diagonal
long-range order, but is instead a generalized Luttinger
liquid.  The particular Luttinger liquid phase, within a general
classification scheme developed in Ref.~\onlinecite{Balents96}\ also follows
from the gap function $\Delta$.  We will use this notation, in which a
phase with $m$ gapless charge and $n$ gapless spin modes is denoted
CmSn, in what follows.

The results of our calculations for positive $U$
Hubbard chains in the phase diagrams are summarized 
in Figs.[6-10].  We emphasize that the phase diagrams are valid for 
arbitrary filling $n$ and transverse hopping $t_{\perp}$ except at 
some specific lines (see below). 
First note the proliferation of phases as $N$
increases from $2$ to $4$.  We believe that this complexity persists
even in the $N\rightarrow\infty$ limit (but see below).  The crossover
to two dimensions is thus highly nontrivial.  Secondly, despite the
repulsive interactions, the majority of phases exhibit some degree of
reduction of gapless spin modes, i.e. pairing.  The symmetry of the
pair wavefunction is in most cases consistent with a d-wave form.
Unlike the two-chain case, however, as $N$ is increased, gapless spin
modes exist due to the presence of nodes in the pair wavefunction.
Under different circumstances, as can be seen from Figs.[6-10], both
$d_{x^2-y^2}$ and $d_{xy}$ states appear.

A number of special results are obtained for particular small values of $N$.
In the case $N=3$, the difference between open and periodic boundary
conditions is quite pronounced, due to the strong role of frustration
(and consequent absence of particle/hole symmetry) in the periodic
case.  As found previously by Arrigoni\cite{Arrigoni96a,Arrigoni96b}
and Schulz,\cite{Schulz96b}\ this Hubbard ``prism'' exhibits a spin gap
at half filling, which persists over a range of both particle and hole
doping.  An especially surprising effect occurs for $N=4$ with
periodic boundary conditions.  In certain regions of the phase
diagram, (singlet) Cooper pairs condense, not into the zero
center-of-mass momentum state, but rather into the $m=\pm 2$
center-of-mass (quasi-)angular momentum states around the four-chain
cylinder.  We therefore call this a Cylindrically EXtended (CEX)
d-wave phase.  Preliminary indications of the CEX d-wave phase have
been found in recent numerical calculations.\cite{ScalapinoUnpub}

Although detailed phase diagrams such as these have only been obtained
for $N=2,3,4$, our RG equations are valid for arbitrary $N$.  They can
be easily integrated numerically to any desired accuracy to determine
most features of the weak-coupling phase diagram for any $N$.  In the
limit $N\rightarrow\infty$, several connections can be made with other
work.\cite{Shankar94}
This limit may be taken in several ways.  For the simplest form
of our flow equations to remain valid, the interaction strength $U$ must be
scaled logarithmically to zero as $N\rightarrow\infty$.  Strictly
speaking, then, these RG equations do not describe truly
two-dimensional systems with finite, non-singular, interactions.  The
only logarithmic reduction of the domain of validity with increasing
$N$ suggests, however, that the two-dimensional limit may nevertheless
be well approximated in this scheme.  We present arguments that this is
indeed the case.  Firstly, in the large $N$ limit, our
RG equations reduce, up to an overall normalization of the
interaction strength, to those of Shankar,\cite{Shankar94}\ derived
directly in two dimensions.  Second, an extended set of RG equations
incorporating additional interactions, which we argue captures
completely the two-dimensional limit for small non-zero $U$, can be
shown to be equivalent to the previous ones for interactions with
non-singular momentum dependence at the Fermi surface.

Based on these analyses, we expect our RG equations to contain a
complete description of the dimensional crossover in the
weak-interaction limit.  In this limit, explicit analysis of the $1/N$
correction terms show that all the pairing instabilities occur only at
very low temperatures, $T_c(N) \sim \Delta_N \sim e^{-N}$.  Feedback of
the forward-scattering interactions into the Cooper channel,
responsible for the pairing instabilities in the smaller ladder
systems at weak coupling, is therefore insufficient to produce
superconductivity in the two-dimensional limit.  We conclude that {\sl
strong} and/or nearly-nested interactions are necessary
to explain superconductivity in 2d repulsive Hubbard systems.

The remainder of the paper is organized as follows: In Sec.~II, we
introduce the N-chain Hubbard model, its weak-coupling limit, and a
compact current-algebra notation for the allowed interactions.  In
Sec.~III, the RG equations are derived using the operator product
expansion to one loop order, and the numerical method used to study
these equations is explained.  In Sec.~IV, we show how the results of
these numerics can be understood using bosonization, introducing the
gap function $\Delta$ in this context.  These techniques are applied
in Sec.~V to determine the phase diagrams of 3- and 4-chain systems.
Our analysis of the 2d limit is given in Sec.~VI, and implications for
numerics and experiments are discussed in Sec.~VII.  Four appendixes
give further details of current algebra methods, RG equations for
umklapp couplings, initial values of coupling constants for the
Hubbard models, and Klein factors needed for the bosonization
calculations.

\section{ N-chain Hubbard Model} 
The N-chain Hubbard model is described
by the Hamiltonian $ H = H_0 + H_U$, 
\begin{eqnarray} 
  H_0 =
  \sum_{x,i,\alpha} \bigg\{ &&-t [d^{\dag}_{i \alpha} (x+1) 
  d^{\vphantom\dag}_{i\alpha}(x) + h.c. ] \nonumber\\ 
  &&-t_{\perp} [ d^{\dag}_{i+1 \alpha}(x) d^{\vphantom\dag}_{i \alpha}(x) +
  h.c.] \bigg\},\\ 
  H_U = \sum_{i,x}
  U &&: d^{\dag}_{i \uparrow}(x) d^{\vphantom\dag}_{i\uparrow}(x) d^{\dag}_{i
    \downarrow}(x) d^{\vphantom\dag}_{i \downarrow}(x):,
\end{eqnarray} 
where $ d^{\vphantom\dag}_{i} ( d_{i}^{\dag} ) $ is a fermion 
annihilation (creation) operator on chain $i$ ($i=1\ldots N$), and 
$\alpha =\uparrow, \downarrow $ is a spin index.  The parameters $t$ 
and $t_{\perp}$ are hopping amplitudes along and between the chains, 
and $U$ is an on-site Hubbard interaction.

We begin by diagonalizing the quadratic part of the Hamiltonian,
$H_0$, as appropriate in the weak-coupling limit, $U \ll
t,t_\perp$. This is accomplished by transforming to new fermion fields
$\psi_i$, where
\begin{equation}
  d_{j\alpha} = \sum_m S_{jm} \psi_{m\alpha}.
  \label{trans_matrix}
\end{equation}
The transformation matrix $S$ depends upon the boundary conditions in
the transverse ($y$) direction.  For periodic boundary conditions
(PBCs), the eigenfunctions are plane waves, and
\begin{equation}
  S_{jm}=\sqrt{\frac{1}{N}} \exp(\frac{2\pi i}{N} jm), \qquad
  \text{ (PBC)} 
  \label{SPBC}
\end{equation}
while for open boundary conditions (OBCs), the transverse eigenfunctions are
standing waves,
\begin{equation}
  S_{jm}=\sqrt{\frac{2}{N+1}} \sin(\frac{\pi}{N+1}jm), \qquad
  \text{(OBC)}
  \label{SOBC}
\end{equation}
This brings the Hamiltonian into diagonal form in momentum space:
\begin{equation} 
  H_0 = \sum_{i,\alpha} \int^{\pi}_{-\pi}
  \frac{dp}{2\pi} \epsilon_{i}(p) \psi^{\dag}_{i \alpha}(p) \psi_{i
    \alpha}(p),
\end{equation} 
The single-particle energy is
\begin{equation}
  \epsilon_{i}(p) = -2t \cos p -2t_{\perp} \cos(k_{yi}). 
  \label{dispersion}
\end{equation}  
A difference in the spectra between OBCs and PBCs arises due to a
difference in the set of allowed transverse momenta.  These are
\begin{eqnarray}
  k_{yi} =&& \frac{\pi}{N+1} i, \;\; i=1, 2,..., N, \:\:
  \text{for OBCs};
  \label{OBCmomenta}\\
  k_{yi} =&& \frac{2\pi}{N} i, \;\; i=0,\pm1,...,(\pm) [\frac{N}{2}], \:\:
  \text{ for PBCs}, \label{PBCmomenta}
\end{eqnarray}
where $[x]$ means the largest integer less than x.  In the case of
PBC, the momenta for $k_{yi} = \pm \pi$ are equivalent
(i.e. differ by $2\pi$) for $N$ even.  For this reason, we have
enclosed the final $\pm$ in Eq.~\ref{PBCmomenta} in parenthesis,
which indicates that for $N$ even, only one of these should be
included for the proper counting of modes.

Eq.~\ref{dispersion} defines $N$ bands, which, in weak-coupling, are
filled up to the chemical potential (Fermi energy) $\mu$.  For those
bands which are partially filled, this defines a set of Fermi points
$\{ k_{Fi} \}$ via
\begin{equation}
  \epsilon_i(k_{Fi}) = \mu.
  \label{chemical_potential}
\end{equation}
The chemical potential is fixed in terms of the physical density $n$
(measured as a particle number per site) by the implicit condition
\begin{equation}
  \sum_{i}k_{Fi} = \frac{\pi}{2}Nn \equiv N k_F.
  \label{filling_factor}
\end{equation}

We now turn to the treatment of interactions. 
It is useful to introduce a functional integral
formulation.  Correlation functions are calculated as averages
with respect to a ``Boltzmann weight'' $e^{-S}$, whose (Grassman)
integral is the partition function
(for example, see Ref.\onlinecite{Negele88})
\begin{equation}
  Z = {\rm Tr} e^{-\beta H} = \int [d\overline\psi][d\psi] e^{-S},
\end{equation}
where $\beta =(k_{\rm B}T)^{-1}$ is the inverse temperature.  Unless
explicitly stated otherwise, all calculations in this paper are
performed at zero temperature, i.e. $\beta=\infty$.  The (imaginary
time) action $S$ is
\begin{equation}
  S = \int_0^\beta d\tau \left[ \sum_{i\alpha x}
    \overline\psi_{i\alpha}(x)\partial_\tau 
    \psi^{\vphantom{i_{\alpha}}}_{i\alpha}(x) + H \right], 
\end{equation}
and $\overline\psi$ and $\psi$ are Grassman fields.  

In this formulation, it is straightforward to focus on the low-energy
properties of the system.  This is accomplished by integrating out
all Grassman variables corresponding to fermionic operators creating
excitations with substantial gaps.  In particular, we first integrate
out completely all $\psi_{i\alpha}$ and $\overline\psi_{i\alpha}$
corresponding to completely filled or empty bands, for which all
excitations are separated by a finite energy from the chemical potential.
When interactions are included, this gives rise to modifications of
the remaining effective action of O($U^2/t$, $U^2/t_\perp$), negligible
relative to the bare $O(U)$ couplings for $U \ll t$.  We denote the 
number of remaining partially filled bands by $N_f \leq N$.  
Secondly, we also integrate
most of the longitudinal momentum modes in the partially filled bands, leaving
only those in a width $2\Lambda$ around each Fermi point $k_{Fi}$ (we
will return to fix $\Lambda$ later). 
That is, we integrate out $\psi_{i\alpha}(p)$ and
$\overline\psi_{i\alpha}(p)$, provided $|p- k_{Fi}| > \Lambda$ {\sl
  and} $|p+k_{Fi}| > \Lambda$.  This again leads to a renormalization of
the ``bare'' couplings in the remaining effective action, this time
with the additional logarithmic factor
\begin{equation}
  U_R \simeq U\left[1 + {\rm const.} \times {U \over t} \ln
    (k_F/\Lambda) \right]. 
\end{equation}
This second step of integration is perturbatively controlled and makes
negligible modification to the bare couplings, provided
\begin{equation}
  U \ll \frac{t}{ \ln(k_F/\Lambda)}. \label{UcondA}
\end{equation}

Assuming Eq.~\ref{UcondA} is satisfied, the remaining fields have
longitudinal momentum in only a narrow shell near the Fermi points.
Within each shell, we can define chiral (right and left moving)
fermions as
\begin{eqnarray}
  \psi_{i \alpha} \sim &&\psi_{Ri \alpha}
  e^{ik_{Fi}x} + \psi_{Li \alpha} e^{-ik_{Fi}x}\:\:
  \text{for OBCs};\label{XfourierOBC}
  \\ 
  \psi_{i \alpha} \sim &&\psi_{Ri \alpha}
  e^{ik_{Fi}x} + \psi_{L\bar{i} \alpha} e^{-ik_{Fi}x} \:\:
  \text{ for PBCs}.\label{XfourierPBC}
\end{eqnarray}
Here we have introduced the notation $\bar{i} = -i$, which we will
continue to use throughout the remainder of the paper.  With
this definition, $\psi_{Ri}, \psi_{Li}$ have opposite momenta in the
PBC case (where transverse momentum is a good quantum number).  The
fields $\psi_{Ri}, \psi_{Li}$ may be thought of as ``slowly varying'',
due to their restricted range of momenta.

For small $\Lambda$, the dispersion may be
linearized within each momentum shell.  The effective Hamiltonian is then
\begin{equation} 
  H_0 = \sum_{i,\alpha} \int dx v_i \left[
    \psi^{\dag}_{Ri\alpha}i\partial_x \psi^{\vphantom\dag}_{Ri\alpha}
    -\psi^{\dag}_{Li\alpha}i\partial_x \psi^{\vphantom\dag}_{Li\alpha} \right], 
\label{H_0}
\end{equation}
where $v_i = 2t\sin k_{Fi}$. 

As it stands, the problem is formulated as an $N_f$-channel interacting 1d
Fermion system.  It will sometimes be useful, however, to view 
the system instead as a finite-width strip of a two-dimensional Hubbard 
model.  To translate between the two pictures, we recognize that in a 
finite-size system, only a discrete set of transverse momenta $k_{yi}$ 
are allowed.  One may think of these momenta as ``cutting'' the 2d Fermi 
surface, the intersections being the 1d Fermi points as shown
in Figs.[1,2]. This gives an 
intuitive connection to more familiar two-dimensional physics, and also 
helps in identifying the possible four-fermion interactions.  One caveat 
that should be kept in mind, however, is that for OBCs, the standing-wave 
transverse eigenfunctions are linear combinations of momenta $\pm 
k_{yi}$, so that a single pair of 1d Fermi points corresponds in this 
case loosely to {\sl four} points on the 2d Fermi surface.

\begin{figure}[hbt]
\epsfxsize=3.5in\epsfbox{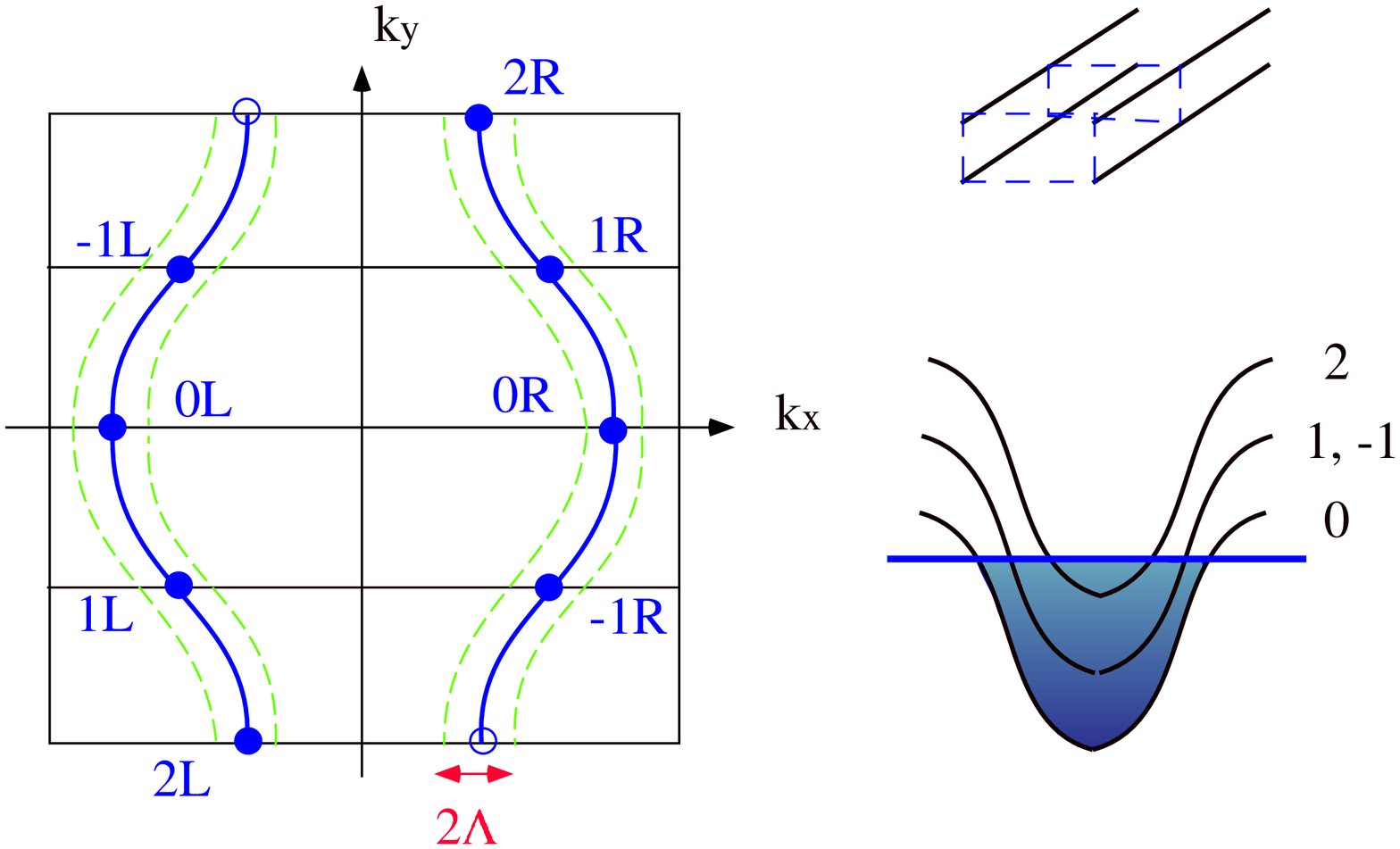}
{\noindent FIG. 1:  Band structure of the 4 chain Hubbard model with PBCs. Mapping onto
  the 2d BZ is shown at the left-hand side. The configuration of
  chains in real space is shown in the upper right.}
\end{figure}
\begin{figure}[hbt]
\epsfxsize=3.5in\epsfbox{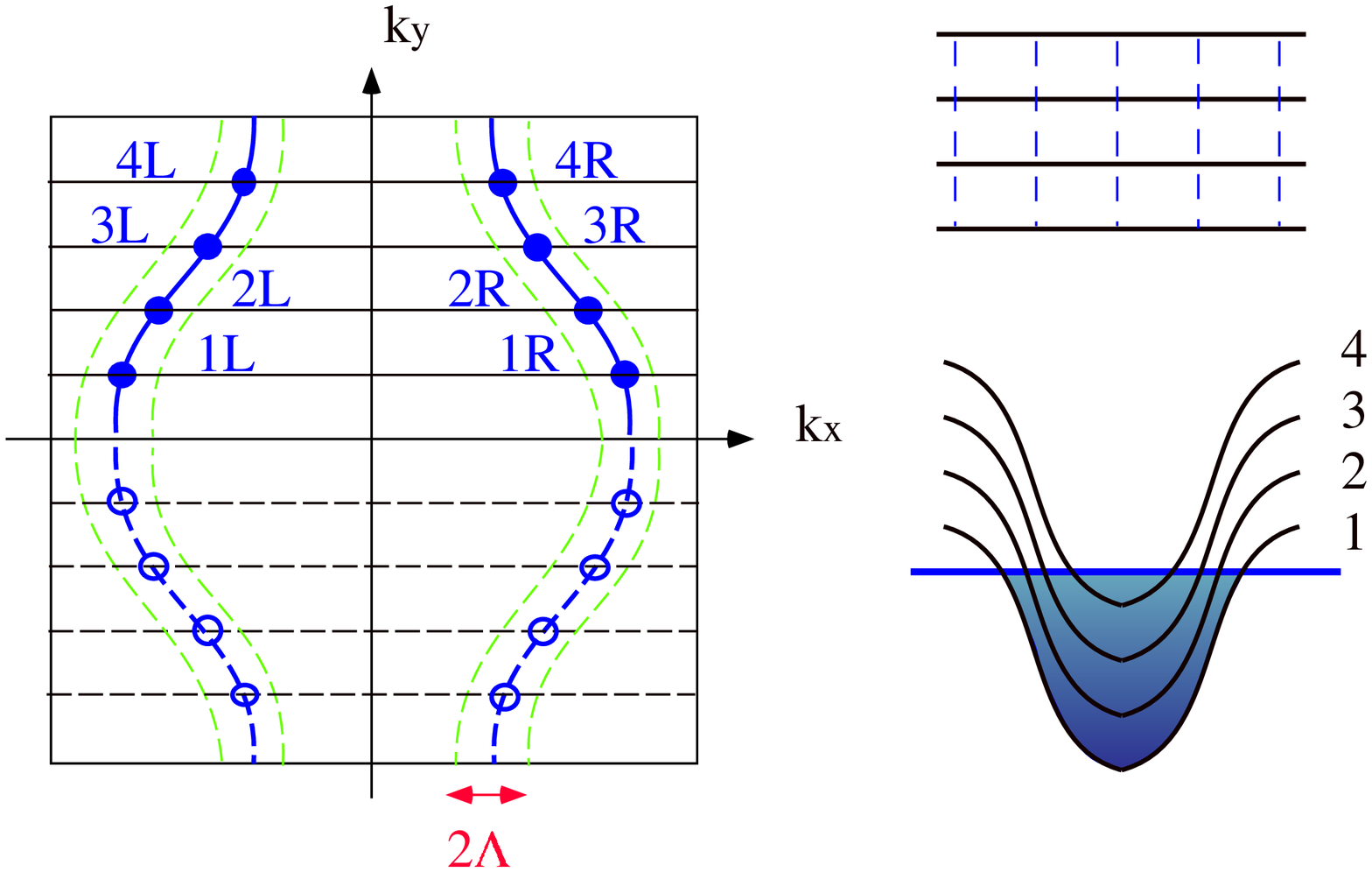}
{\noindent FIG. 2: Band structure of 4 chain Hubbard model with OBCs. Since the actual
  transverse eigenstates with OBCs are standing wave superpositions
  containing both $\pm k_y$, we have indicated each 1d Fermi point by
  two points (one closed and one open circle above) on the BZ.  The
  spatial configuration of chains is shown in the upper right.}
\end{figure}

We now try to write down all possible 4-point interaction terms
allowed by symmetry.  In addition to the U(1) $\times$ SU(2) symmetry
corresponding to charge and spin conservation, these terms must be
preserved by charge conjugation, time reversal, parity, and spatial
translation operations. The most general particle-conserving 4-point
vertex has the form
\begin{eqnarray}
  H_{int}=&&\int \prod_{a} {{dk_a} \over {2\pi}} \sum_{P_a, i_a} 
  \delta_{{\bf R}} (\bbox{Q}) V[\{P_a,i_a,k_a\}]
  \nonumber\\
  &&\psi^{\dag}_{P_1i_1}(k_1)
  \psi^{\dag}_{P_2 i_2}(k_2)\psi^{\vphantom\dag}_{P_3 i_3}(k_3)
  \psi^{\vphantom\dag}_{P_4 i_4}(k_4),
  \label{vertices}
\end{eqnarray}
where $P_i = \pm 1 \leftrightarrow (R/L)$, and spin indices are 
left implicit. The fermion fields $\psi_{P_a i_a}(k_a)$ are Fourier
transforms of the slowly varying chiral fields defined in 
Eqs.~\ref{XfourierOBC}-\ref{XfourierPBC}.
The strengths of the couplings are denoted by $V[\{P_a,i_a,k_a\}]$.
The total momentum transfer $\bbox{Q}$ is
\begin{eqnarray}
  Q_x = && -P_1 k_{F1}-P_2 k_{F2}+P_3 k_{F3}+P_4 k_{F4}
  \nonumber\\
  && -k_1-k_2+k_3+k_4, 
  \label{Qx_total}
  \\
  Q_y =&&\frac{2\pi}{N}(-P_1i_1-P_2i_2+P_3i_3+P_4i_4) \;\; 
  {\rm (PBC)}. 
  \label{Qy_total}
\end{eqnarray}
Note that $Q_y$ only appears for PBCs, since for OBCs,
transverse momentum is not a good quantum number. Momentum
conservation is implemented by the lattice delta function
\begin{equation}
  \delta_{\bf R}(\bbox{Q}) = \cases{\sum_{n_x} \delta(Q_x-2\pi n_x) &
    (OBCs) \cr
    \sum_{n_x,n_y} \delta^{(2)}(\bbox{Q}-2\pi(n_x,n_y)) & (PBCs)},
\end{equation}
where $n_x,n_y$ are integers.
Vertices with non-zero ${\bf n}$ are called umklapp interactions.

The dependence of the vertex function on $k_a$ is analyzed via the
Taylor expansion (we will assume it is differentiable)
\begin{equation}
  V[\{P_a,i_a,k_a\}] = V[\{P_a,i_a,0\}] + 
  \sum_a k_a \frac{\partial V}{\partial k_a} + O(k^2).
\end{equation}
We will see that, while the leading term is {\sl marginal} in the RG
sense, all the higher derivative corrections are in fact {\sl
  irrelevant}, and can be neglected to the (leading) order of accuracy
desired here.

\subsection{Interactions for OBCs}

Having obtained a general expression (Eq.~\ref{vertices}) embodying
the constraints on allowed vertices, we now turn to the classification
of the solutions of these constraints in the particular cases of
interest.  We will do this first for the case of OBCs, proceeding in
two steps.  First, we locate the possible combinations of the band
indices $\{P_a,i_a\}$, and second, we determine possible combinations
of spin indices, which are implicit in Eq.~\ref{vertices}, by SU(2)
symmetry.  Constraints from other symmetries are also discussed.

For OBCs, only momenta in the $k_x$ direction is conserved. 
We will assume $\Lambda$ is sufficiently small that the internal momenta 
$k_i$ may be neglected in Eq.~\ref{Qx_total}.  The condition for validity 
of this assumption will be derived at the end of the section.

Since  $k_{Fi} \leq \pi$, $n_x$ can take values, 0,$\pm1$,$\pm2$. 
For $n_x= \pm 2$, all the fermi momenta $k_{Fi} =\pi$.
This means that these bands must be completely filled, and, following the 
reasoning described earlier, do not survive in the low-energy theory.  
For $n_x=\pm1$, momentum balance is possible in partially filled bands.
However, at generic fillings, the
fermi momenta are incommensurate (in units of $2\pi$),
and cannot be made to sum up to $\pm 2\pi$.  More careful consideration 
shows that such interactions only exist 
on specific umklapp lines in  the $(n,t_\perp/t)$-plane. 
In this paper, we will restrict ourselves to the study of generic
fillings, for which these umklapp interactions in the $k_x$ direction
may be neglected.

\begin{figure}[hbt]
\epsfxsize=3.5in\epsfbox{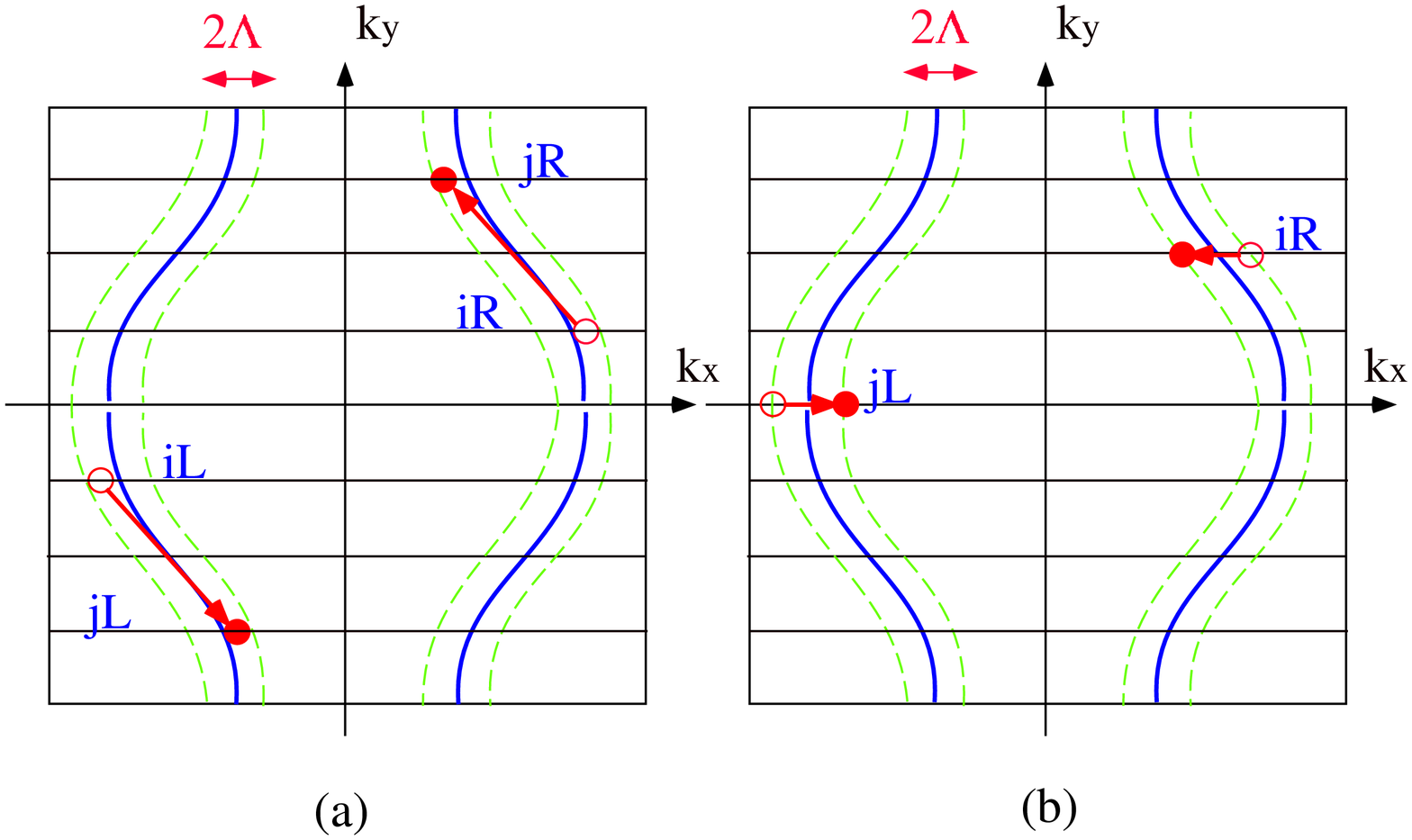}
{\noindent FIG. 3: Examples of Cooper scattering $c_{ij}$ (part
(a)) and forward scattering $f_{ij}$ (part (b)).}
\end{figure}

The last kind of vertices with $n_x=0$ conserve $x$-momenta
exactly. They may be found by plotting the 
interactions on the 2d Brillouin zone (BZ). These vertices satisfy
\begin{equation}
  Q_x(n,t_{\perp})=  -P_1 k_{F1}-P_2 k_{F2}+P_3 k_{F3}+P_4 k_{F4}=0.
  \label{Qx=0}
\end{equation} 
For a generic Fermi surface, 
two familiar classes of interactions are {\it always} allowed. 
The first comprises forward scattering interactions, which satisfy
\begin{eqnarray}
  (P_1, P_2)&&=(P_3, P_4)
  \nonumber\\
  (i_1,i_2) &&=(i_3,i_4),\;\;\text{forward scattering,}
  \label{forward_cond}
\end{eqnarray}
The second set is the Cooper (or backscattering) channel, defined by
\begin{eqnarray}
  P_1=\overline{P}_2, &&P_3=\overline{P}_4;
  \nonumber\\
  i_1=i_2, &&i_3=i_4, \;\;\text{Cooper scattering.}
  \label{Cooper_cond}
\end{eqnarray}
In Eqs.~\ref{forward_cond}-\ref{Cooper_cond}, and in the remainder of
the paper, $\overline{P} \equiv -P$ and $(x_1,x_2)=(x_3,x_4)$
indicates pairwise equality, i.e. either $x_1=x_3$, $x_2=x_4$, or
$x_1=x_4$, $x_2=x_3$.  The two possible solutions for forward
scattering actually describe the same vertices, up to a sign from the
Fermion ordering.  Referring to the 2d BZ (see Fig.[3]), one sees that
forward scattering interactions conserve the particle number
separately in each band, i.e. one electron is annihilated and created
in each of the bands $i_1$ and $i_2$.  In the Cooper scattering
channel, however, a pair of electrons is annihilated in band $i_1$ and
then scattered into band $i_3$.

There are, however, other vertices at specific fillings.  These
vertices correspond to special non-trivial solutions of
Eq.~\ref{Qx=0}. Such solutions exist only on specific lines in the
$(n, t_{\perp}/t)$ plane.  Because these lines form a set of measure
zero in the full phase space, the corresponding $V[\{P_a,i_a\}]$ will
be denoted ``minor'' vertices. Like the umklapp interactions, these
minor vertices can be excluded at generic fillings.

We have obtained the allowed vertices in momentum space.
However, since the couplings are not momentum dependent,
they can equally well be written in terms of a local Hamiltonian
density in coordinate space, i.e.
\begin{eqnarray}
  {\cal H}&&_{int}=\sum_{P_a,i_a} \bigg\{
  F[\{P_a,i_a\}] \psi^{\dag}_{P_1i_1}\psi^{\dag}_{P_2i_2}
  \psi^{\vphantom{P_2i_2}}_{P_1i_1}\psi^{\vphantom{P_2i_2}}_{P_2i_2} 
  \nonumber\\
  &&+C[\{P_a,i_a\}] \psi^{\dag}_{P_1i_1}\psi^{\dag}_{\overline{P}_1i_1}
  \psi^{\vphantom{P_2i_2}}_{P_2i_2}
  \psi^{\vphantom{P_2i_2}}_{\overline{P}_2i_2} \bigg\},   
  \label{realspace}
\end{eqnarray}
where all the $\psi$ and $\psi^\dagger$ are evaluated at the same space
point.  We have made the classification into forward and
Cooper-scattering channels explicit by the change of notation
$V[\{P_a, i_a, 0\}] \rightarrow F[\{P_a,i_a\}], C[\{P_a, i_a\}]$, as
appropriate.  In the case of forward scattering, as remarked earlier,
the two solutions of Eq.~\ref{Qx=0}\ lead to a single $F$ vertex.

Each vertex obtained so far has several possible generalizations once
the spin indices are included.  To make them explicit, we now
introduce charge and spin currents, in the scalar and vector
representations of SU(2), respectively.  These are
\begin{equation} 
  J_{ij} = \psi^{\dag}_{i\alpha} \psi^{\vphantom\dag}_{j\alpha}, \;\;
  \bbox{J}_{ij}=\frac12 \psi^{\dag}_{i\alpha} 
  \bbox{\sigma}^{\vphantom\dag}_{\alpha \beta}
  \psi^{\vphantom\dag}_{j\beta},
  \label{current_defs}
\end{equation}
where $\bbox{\sigma}$ denote Pauli matrices.  These currents satisfy
so-called Kac-Moody algebras, and this notation is therefore often
referred to as current algebra.  To regularize the composite operators
in Eq.~\ref{current_defs}, the currents are further defined to be
normal ordered (although we do not indicate the normal ordering
explicitly).  Four-point vertices can be written down as products of
two currents.  Such Bilinear current-current interactions are SU(2)
scalars (as appropriate for the Hamiltonian density) if and only if
they are formed by coupling two charge or two spin currents.  Each
vertex in Eq.~\ref{realspace}\ therefore has two counterparts once
spin is included.  The subset of these which couple right and left
movers is
\begin{eqnarray} 
  -{\cal H}^{(1)}_{int} = -&& \tilde{c}^{\rho}_{ij} J^{R}_{ij} J^{L}_{ij} +
  \tilde{c}^{\sigma}_{ij} \bbox{J}^{R}_{ij} \cdot \bbox{J}^{L}_{ij},
  \nonumber\\
  -&&\tilde{f}^{\rho}_{ij} J^{R}_{ii} J^{L}_{jj} +
  \tilde{f}^{\sigma}_{ij} \bbox{J}^{R}_{ii} \cdot \bbox{J}^{L}_{jj},
  \label{int1}
\end{eqnarray} 
where $\tilde{f}_{ij}$ and $\tilde{c}_{ij}$ denote the forward and
Cooper scattering amplitudes, respectively, between bands
$i$ and $j$.  Summation on $i, j$ is implied.

Since $f_{ii}, c_{ii}$ describe the same vertex, we choose the 
diagonal piece of the forward scattering amplitude to vanish,
i.e. $\tilde{f}_{ii}=0$, to avoid double countings.
Under charge conjugation  $J_{ij} \to
J_{ji}$, which implies $\tilde{c}_{ij} = \tilde{c}_{ji}$.  Similarly,
reflection symmetry (in $x$) implies $\tilde{f}_{ij} =
\tilde{f}_{ji}$.  While it is not obvious at this point,  the choice
of signs for the scalar and vector couplings in Eq.~\ref{int1}\ is such that
they are all positive for repulsive on-site interactions.

There are other interactions which are completely chiral,
e.g. $J^{R}_{ii}J^{R}_{jj}$.  As is well-known in conformal field
theory, such purely chiral interactions do not renormalize or generate
renormalization at leading order, and can be neglected in our
weak-coupling analysis.  Physically, they modify slightly the
``velocities'' of various charge and spin modes, which are already
order one in the bare theory.

\subsection{Interactions for PBCs}

When PBCs are imposed instead of OBCs, the
system retains a finite set of discrete transverse translational
symmetry operations.  Correspondingly, the transverse momentum $k_y$
(or more properly the exponential $e^{ik_y}$)
is a good quantum number, and the allowed interactions are further
constrained by the requirement $Q_y = 2\pi n_y$.
Since $k_{Fi},k_{yi} \leq
\pi$, $n_x,n_y$ can take the values, 0, $\pm1$, $\pm2$. As
explained in the last subsection, vertices with $n_x=\pm 2$ can be
ignored in the low-energy theory and those with $n_x =\pm 1$ only
live on specific umklapp lines and thus are not included.  It follows
that in the small $U$ limit at generic fillings, it is sufficient to
consider only vertices with ${\bf n}=(0, 0), (0, \pm 1), (0, \pm 2)$,
which conserve the $x$-momenta {\it exactly}.

The allowed vertices are found in two steps.  First, we find all
possible vertices which conserve $x$-momentum, and then we rule out
some of them by conservation of $y$-momentum.  To do the former, first
note that under a $y$-reflection $k_y \to -k_y$, which implies the
Fermi momenta satisfy $k_{Fi}=k_{F\bar {i}}$, where $\bar{i} =-i$.
This parity constraint, combined with conservation of $x$-momentum alone
allows vertices which satisfy
\begin{equation}
  (P_1, P_2)=(P_3, P_4),  \qquad (i_1,i_2) = (\pm i_3, \pm i_4)
  \label{nx=0PBCa}
\end{equation}
or
\begin{equation}
  P_1=\overline{P}_2,P_3=\overline{P}_4, 
  \qquad i_1=\pm i_2, i_3= \pm i_4.
  \label{nx=0PBCb}
\end{equation}
Note that Eqs.~\ref{nx=0PBCa}-\ref{nx=0PBCb}\ differ from their
counterparts for OBCs (Eqs.~\ref{forward_cond}-\ref{Cooper_cond}) by
the extra choice of $\pm$ sign for PBCs.  Physically, this arises
because with PBCs, plane waves with momenta $\pm k_y$ form two
independent allowed transverse eigenfunctions, while only the single
standing wave (superposition of the two) eigenfunctions satisfies OBCs.
As before, additional vertices exist on special lines in the phase
diagram, but will be ignored here.

Since the $y$-momentum is also conserved, not all of the vertices in
Eqs.~\ref{nx=0PBCa}-\ref{nx=0PBCb}\ are allowed.  Consider first the
corresponding constraint for odd $N$.  We must evaluate
Eq.~\ref{Qy_total}, which can be rewritten as
\begin{equation}
  -P_1 i_1 - P_2 i_2 + P_3 i_3 + P_4 i_4 = n_y N.
  \label{Qy_constraint_PBC}
\end{equation} 
Since all the band indices satisfy $|i| \leq (N-1)/2$, solutions with
$n_y = \pm 2$ do not exist.  Furthermore, after substitution of the
partial solutions in Eqs.~\ref{nx=0PBCa}-\ref{nx=0PBCb}, a little
algebra shows that the left-hand-side in Eq.~\ref{Qy_constraint_PBC}\
is always even, so that no solutions exists for $n_y = \pm 1$ either.
Therefore, for the odd-chain systems, we need only consider the
vertices which conserve momenta exactly, i.e. with ${\bf n} = (0,0)$.

Eq.~\ref{Qy_constraint_PBC}\ for $n_y=0$ is satisfied if and only if
the $+$ sign is chosen in Eqs.~\ref{nx=0PBCa}-\ref{nx=0PBCb}.  With
this restriction, the allowed interactions are precisely the same
forward and Cooper scattering ones that occur in the OBC case, and
may therefore be described as before by Eq.~\ref{int1}.

The situation is more complicated for $N$ even, because solutions of
Eq.~\ref{Qy_constraint_PBC}\ exist with $n_y=\pm 1$ and $n_y = \pm 2$.
The latter do not actually introduce any addition complications.  This
is because for $n_y=\pm 2$, all the band indices must satisfy $|i_a|
=N/2$. The bands $i_a  = \pm N/2$ are, however, equivalent, since their
$y$-momenta differ by $2\pi$. Therefore,  the band indices can instead be
chosen equal, and then satisfy $n_y=0$.  These
${\bf n}=(0,2)$ vertices are thus included in the ${\bf n}=(0,0)$ set
which will be discussed later.

The difficulty arises when ${\bf n}=(0, \pm 1)$, i.e. $n_y = \pm 1$.
\begin{equation}
  -P_1 i_1 -P_2 i_2 + P_3 i_3 +P_4 i_4 = \pm N.
\end{equation}
In this case, choosing the $-$ sign in
Eqs.~\ref{nx=0PBCa}-\ref{nx=0PBCb}\ leads to solutions of
Eq.~\ref{Qy_constraint_PBC}. The first set of these is similar to the
Cooper channel (see Fig.[4a]), with indices satisfying
\begin{eqnarray}
  P_1 = \overline{P}_2,&& P_3=\overline{P}_4;
  \nonumber\\
  i_1 = \bar{i_2}, &&i_3=\bar{i_4}, \qquad i_1-i_3 = \pm \frac{N}{2}.
  \label{cooper_umklapps}
\end{eqnarray}
The second set is similar to forward scattering (see Fig.[4b]), and has
\begin{eqnarray}
  (P_1, P_2)=&&(P_3, P_4); \nonumber
  \\
  (i_1, i_2)=&&(\bar{i_3}, \bar{i_4}), \qquad i_1-i_2= \pm \frac{N}{2}.
  \label{forward_umklapps}
\end{eqnarray}

\begin{figure}[hbt]
\epsfxsize=3.5in\epsfbox{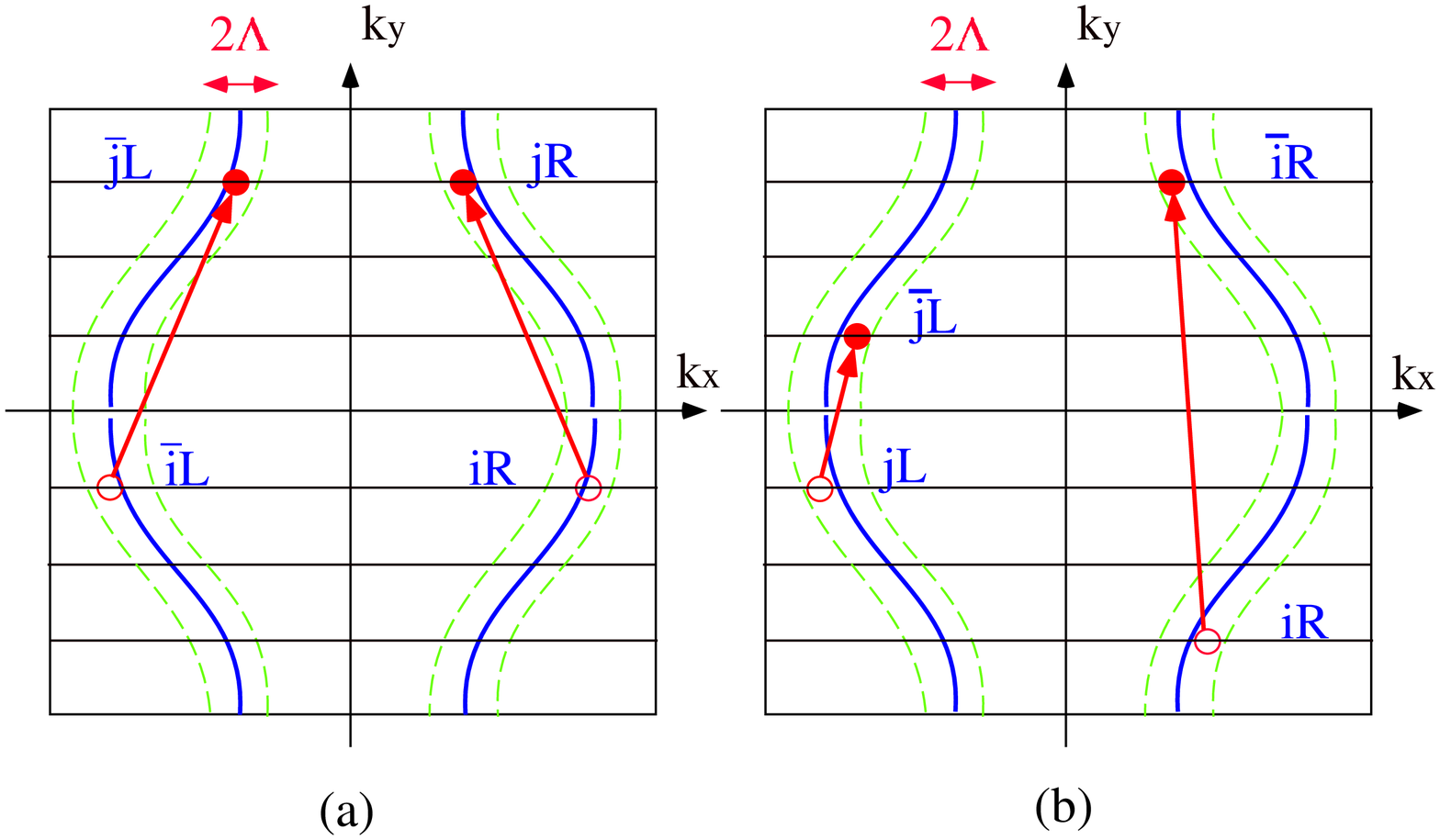}
{\noindent FIG. 4: Examples of transverse umklapp scattering 
  $u^{1}_{ij}$ (part (a)) and $u^{2}_{ij}$ (part (b)). As is
  clear from the figure, the momentum in the $k_{y}$ direction is not
  conserved.}
\end{figure}

The two kinds of umklapp interactions in the $k_y$ direction 
can also be written down as products of currents. 
Following the method we developed in last subsection,
those $y$-umklapp interactions which couple the right and left movers
can be described by,
\begin{eqnarray} 
  -{\cal H}^{(2)}_{int} =
  &&-\tilde{u}^{1\rho}_{ij} J^{R}_{ij} J^{L}_{\bar{i} \bar{j}} +
  \tilde{u}^{1\sigma}_{ij} \bbox{J}^{R}_{ij} \cdot \bbox{J}^{L}_{\bar{i} \bar{j}}
  \nonumber\\ 
  &&-\tilde{u}^{2\rho}_{ij} J^{R}_{i\bar{i}} J^{L}_{j\bar{j}}
  + \tilde{u}^{2\sigma}_{ij} \bbox{J}^{R}_{i\bar{i}} \cdot \bbox{J}_{j\bar{j}}.
  \label{int2}
\end{eqnarray} 
Implicit in this notation is the constraint (see
Eqs.~\ref{cooper_umklapps}-\ref{forward_umklapps}) that
the $\tilde{u}_{ij}$ are non-zero only for $ |i-j| =
\frac{N}2$.  Since $u^{1}_{i\bar{i}}, u^{2}_{i\bar{i}}$ describe the same
interaction, we choose ${\tilde u}^{2\rho}_{i\bar{i}}, {\tilde
  u}^{2\sigma}_{i\bar{i}}=0$ to avoid double counting.  Since under
charge conjugation, $J_{ij} \to J_{ji}$, ${\tilde u}_{ij} =
\tilde{u}_{ji}$.  Similarly, under a parity transformation, $(i,j) \to
(\bar{i}, \bar{j})$ and $R \to L$, so ${\tilde u}_{ij} = {\tilde
  u}_{\bar{i} \bar{j}}$.

Finally, of course, non-umklapp interactions also exist for even $N$.
Just as for $N$ odd, these ${\bf n}=(0, 0)$ vertices are simply the
forward and Cooper channel ones as for OBCs. Therefore, for even-chain
systems with PBCs, the full set of allowed vertices comprises forward,
Cooper, and ($Q_y = \pm 2\pi$) umklapp interactions, as given in
Eq.~\ref{int1} and Eq.~\ref{int2}.

\subsection{Constraints on momentum cutoff}
\label{sec:constraints_on_cutoff}

So far, we have determined all the interactions
allowed by symmetry, assuming that the momentum cutoff
$\Lambda$ is ``small enough'' to
neglect the $-k_1-k_2+k_3+k_4$ term in Eq.~\ref{Qx_total}.
In this section, we make this requirement precise, and
investigate how this begins to break down for larger $\Lambda$. 

In fact, the results in the last two subsections are strictly correct
only for $\Lambda=0$.  With a finite cutoff, the picture is modified
in two ways.  Firstly, the specific lines on which the ``minor''
couplings exist are widened and occupy a finite area in the $(n,
t_{\perp}/t)$ plane.  Secondly, for sufficiently large $\Lambda$, new
vertices (not included in the forward and Cooper scattering channels)
can arise for generic fillings (i.e. throughout the $(n,t_\perp/t)$
plane). 

Consider first the minor vertices, which exist in the region of the
phase diagram defined by
\begin{eqnarray}
  f&(n, t_{\perp}; ijkl)&  \equiv   
  -P_1k_{Fi}-P_2k_{Fj}+P_3k_{Fk}+P_4k_{Fl} , \nonumber \\
  \bigg|f&(n,t_\perp;ijkl)&\bigg|  \leq  4\Lambda.
  \label{minor_coupling_condition}
\end{eqnarray}
For $\Lambda=0$, as noted previously, solutions of this equation other
than the Cooper and forward scattering ones exist only on isolated
lines.  Since $f(n,t_\perp)$ is a smooth function of its arguments,
there will be finite neighborhoods around these lines which satisfy
Eq.~\ref{minor_coupling_condition}.  The widths $\delta n$ of these
neighborhoods can be estimated by Taylor expanding $f$ around  the
$f=0$ lines, i.e.
\begin{equation}
  \bigg|\frac{\partial f}{\partial n} \delta n\bigg| \sim  \Lambda.
\end{equation}
Since the derivatives of the fermi momenta, $\partial k_{F}/ \partial n$,
are order one, the width of the line is approximately as large as
the cutoff, 
\begin{equation}
  \delta n \sim \frac{\Lambda}{k_{F}}.
\end{equation}

To know the fraction of the phase diagram influenced by these minor
vertices, it is also necessary to determine the number (or
equivalently, the density) of these regions.  Consider first OBCs.
Since each region grows adiabatically from a $\Lambda=0$ line, we can
simply count the number of solutions to $f(n, t_\perp;ijkl)=0$ (at,
say, fixed $t_\perp$).  Roughly, the number of solutions may be
estimated as follows.  Picking a fixed $n$ and $t_\perp$, we choose three
of the band indices, say $i,j$ and $k$.  Then $f=0$ determines a Fermi
momentum $k_{Fl}$ for the last band.  Generally, however, this
momentum will not be one of the discrete set of Fermi momenta for this
$n$ and $t_\perp$.  Now begin varying say $t_\perp$, keeping $i,j,k$
and $n$ fixed.  As $t_\perp$ varies, so do $k_{Fi}$,$k_{Fj}$ and
$k_{Fk}$, and hence the required $k_{Fl}$.  As this happens, very soon
$k_{Fl}$ will pass through an allowed value, and we have found a
solution.  Given that, one may then vary {\sl both} $t_\perp$ and $n$
in order to keep $f=0$ for this particular $ijkl$, defining a curve in
the $(n,t_\perp/t)$ plane.  Since this can be repeated for each set of
$i,j,k$, the total number of such curves is $N_{\rm minor} \sim N^3$.
For PBCs, roughly the same argument holds, except that $k$ and $l$ are
related by conservation of $Q_y$.  The number of minor vertices for
PBCs thus reduces to $N_{\rm minor} \sim N^2$.

For large $N$, the fraction of the phase diagram in which minor
vertices contribute can be estimated simply by summing the widths of
these lines.  The resulting fraction $f_{\rm minor} \sim N_{\rm
  minor} \delta n$ is negligible ($f \ll 1$) provided
\begin{eqnarray}
  \frac{\Lambda}{k_{F}} \ll \frac{1}{N^3} \qquad && \text{(OBC)},
  \nonumber\\
  \frac{\Lambda}{k_{F}} \ll \frac{1}{N^2} \qquad && \text{(PBC)}.
  \label{lambda1}
\end{eqnarray}

For sufficiently large $\Lambda$, the one-dimensional bands associated
with adjacent points on the Fermi surface begin to overlap, and it
becomes possible to form new interactions by substituting one for the
other in the original forward and Cooper scattering channels.  If
$\Lambda$ is large enough to allow this, such interactions exist {\sl
  generically}, i.e. throughout the $(n,t_\perp/t)$ plane.  As an
example, consider the shifted terms (for PBCs) shown in 
Fig.[5]:
\begin{equation}
  -{\cal H}_{ij}^{s}(\delta) =  c^{\rho}_{ij}(\delta)
  J^{R}_{i+\delta,j+\delta} J^{L}_{ij} + f^{\rho}_{ij}(\delta)
  J^R_{i,i+\delta}J^L_{j,j+\delta} .
  \label{nearCooper}
\end{equation}
Here $\delta$ parameterizes the transverse momentum shift; for
$\delta=0$, the vertices reduce to the familiar Cooper and forward
scattering types.

\begin{figure}[hbt]
\epsfxsize=3.5in\epsfbox{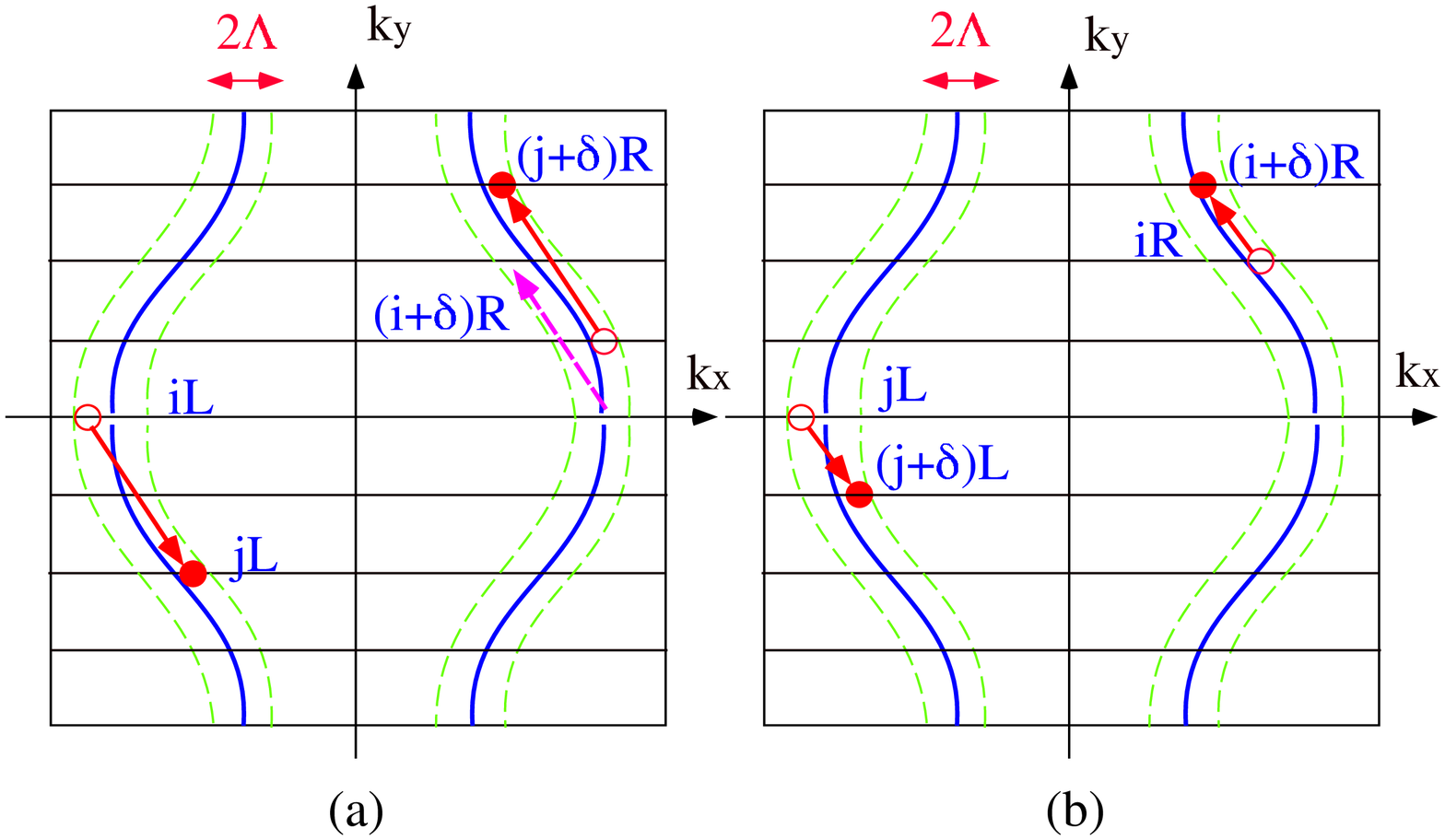}
{\noindent FIG. 5:  New vertices for finite cutoff $\Lambda$. It is possible to shift
  the Cooper vertices (shown by dashed lines) for a finite cutoff and
  still maintain momentum conservation. The maximum shift $\delta$ is
  estimated in Eq.~\ref{shift}.}
\end{figure}

Let us consider in detail the conditions under which ${\cal H}^s$ conserves
momentum in the cut-off theory (see Fig.[5]).  Because both $i$ and $j$
have been shifted by $\delta$ in 
Eq.~\ref{nearCooper}, $Q_y$ is already conserved by design. 
Taking into account $Q_x$ conservation by Taylor
expanding (for $\delta \ll N$) Eq.~\ref{minor_coupling_condition}\
gives the requirement
\begin{equation}
  \bigg| \frac{dk_x}{dk_y}(i)-\frac{dk_x}{dk_y}(j)\bigg| \frac{2\pi}{N} 
  \delta < \Lambda a.
\end{equation}
To obtain an order of magnitude estimate, we then approximate the mean
curvature between bands $i,j$ by its typical value $t_{\perp}/t$,
\begin{eqnarray}
  \bigg|\frac{dk_x}{dk_y}(i)-\frac{dk_x}{dk_y}(j)\bigg| &&\simeq 
  \frac{d^2k_x}{dk_y^2}\frac{2\pi}{N} |i-j|
  \nonumber\\
  && \simeq (\frac{t_{\perp}}{t}) \frac{2\pi}{N} |i-j|.
\end{eqnarray}
The maximum allowed shift $\delta_{\rm max}$ in band indices is therefore
\begin{equation}
  \delta_{\rm max} \sim (\frac{t}{t_{\perp}})\frac{N^2}{|i-j|}
  \frac{\Lambda}{k_{F}}.
  \label{shift}
\end{equation}
For $\delta_{\rm max} <1$, only the unshifted Cooper channel
interaction is allowed.  Demanding this leads to the constraint
\begin{equation}
  \frac{\Lambda}{k_{F}} < (\frac{t_{\perp}}{t}) (\frac{1}{N^2}).
  \label{lambda2}
\end{equation} 

Combining the constraint on the initial coupling strengths,
Eq.~\ref{UcondA}, and that on the momentum cutoff, Eqs.~\ref{lambda1},
\ref{lambda2}, we find the reduced set of interactions in
Eq.~\ref{int1}\ and \ref{int2}\ is sufficient provided the initial
coupling strengths satisfy
\begin{equation} 
  \frac{U}{t} \ll \frac{1}{\ln N}.
  \label{weakU}
\end{equation} 
Here we have dropped the order one factors in front of the logarithm
which are different for OBCs and PBCs.  Since the condition on $U$ is
only logarithmic in $N$, it is not a severe constraint on the initial
values of the couplings for finite chains.  Nevertheless, it is clear
that the true two-dimensional limit is rather subtle.  Indeed,
Eq.~\ref{weakU}\ indicates a possible non-commutativity of the order
of limits (at zero temperature) in taking $U \rightarrow 0$ and $N
\rightarrow \infty$.  We will return to the interesting and important
issues involved in the 2d limit in Sec. VI.  For the moment, we will
restrict ourselves to finite $N$ and discuss the corresponding
weak-coupling behavior of such systems, under the conditions of
Eq.~\ref{weakU}.  

\section{Renormalization Group flow equations} 

To analyze the behavior of the weakly interacting system, we employ
the RG approach.  In this section, we describe the scheme used and
present the resulting differential RG flow equations (analogous flow
equations for particular restricted cases were obtained in
Refs.~\cite{Varma85}\ and \cite{Penc90}).  Further details of the
calculations can be found in appendix A.  The general approach of the
RG is to progressively eliminate short-wavelength, high-energy degrees
of freedom.  To formulate this mode elimination, we first rewrite the
partition function in terms of an average,
\begin{eqnarray}
  Z = &&\int [d\overline\psi][d\psi] e^{-S_0-S_{int}}
  \nonumber\\
  =&& Z_0 \langle e^{-S_{int}} \rangle_0,
\end{eqnarray}
where $Z_0$ is the partition function without interactions, and
angular brackets with the subscript $0$ denotes an average with
respect to the quadratic action $S_0$ only.  This form may be
re-exponentiated using the cumulant
expansion, 
\begin{eqnarray}
  \langle e^{-S_{int}} \rangle_0 = &&\exp \bigg\{
  \langle -S_{int} \rangle_0 +\frac12 [ \langle S^{2}_{int} \rangle_0
  -\langle S_{int} \rangle_{0}^{2}] 
  \nonumber\\
  &&+ O(S_{int}^3)\bigg\}.
\end{eqnarray}

Up to this point, we have systematically derived the low-energy
Fermion model with a ``momentum-shell'' cut-off $\Lambda$.  While the
RG may be implemented directly with this model, it happens that the
one-loop RG equations needed here are in fact independent of the
cut-off scheme used.  This independence arises from the dominance of
logarithmically divergent terms at one-loop level, whose coefficients
are insensitive to the particular form of cut-off used.  We take
advantage of this property here by adopting instead a real-space cut-off
$a \approx 1/\Lambda$.  This distance then appears as an explicit
cut-off in all $x$-integrals, e.g.
\begin{equation}
  \langle S^{2}_{int} \rangle_0 = \int_a^\infty \prod_{i=1,2}
  d\tau_i dx_i \langle {\cal H}_{int}(\tau_1, x_1) 
  {\cal H}_{int}(\tau_2, x_2) \rangle_0,
\end{equation} 
using the compact notation $\int_A^B \equiv \int_{A<|x_1-x_2|<B}$.
Each integral is now separated into two parts: long-wavelength modes,
$|x_1-x_2|>ba$, and short-wavelength modes, $a<|x_1-x_2|<ba$, where
$b> 1$ is the rescaling parameter.  This separation is convenient
because in the latter integral, all fields are at nearby space points.
This allows the use of the operator product expansion (OPE) to replace
the products of such nearby operators by a series of local operators
(for review, see Ref.\onlinecite{Ludwig91,Cardy87}), i.e.
\begin{eqnarray}
  \int_a^{ba} \prod_{i=1,2}d\tau_i dx_i &&
  \langle {\cal H}_{int}(\tau_1, x_1) {\cal H}_{int}(\tau_2, x_2) \rangle_0
  \nonumber\\
  \simeq \int dx d\tau && \langle \delta{\cal H}_{int} \rangle_0.
\end{eqnarray}
The method to compute the effective interaction, $\delta{\cal
  H}_{int}$, can be found in Appendix A. 

As shown in appendix A, the effective interaction $\delta{\cal
  H}_{int}$ has the same form as the original Hamiltonian, and thus has
the effect of renormalizing the bare couplings.  The RG finishing with
a rescaling step, which attempts to bring the theory as much as
possible back to its original form.  To restore the original value of
the cut-off and maintain the original set of Fermi velocities
requires the change of variables
\begin{eqnarray}
  x' =\frac{x}{b}, & \:\: & \tau' = \frac{\tau}{b},
  \label{coord_rescaling} \\
  \overline\psi'(x,\tau) = b^{1/2}\overline\psi(x',\tau'), &&
  \psi'(x,\tau) = b^{1/2} \psi(x',\tau'). \label{field_rescaling}
\end{eqnarray} 
While this indeed preserves (at one-loop order) the form of $S_0$, the
interactions are of course changed.  The simplest way to keep track of
these changes is to perform the RG {\sl infinitesimally}, with the
rescaling factor $b= e^{dl}$.  Iterating both steps of the RG then
leads to differential RG flow equations for the coupling constants, as
a function of length scale $L(l)=e^l$.

For OBCs, the allowed couplings are forward and Cooper scattering. 
The RG equations governing them are
\begin{eqnarray} 
  {\dot f}^{\rho}_{ij} &&= (c^{\rho}_{ij})^2 +
  \frac{3}{16} (c^{\sigma}_{ij})^2,
  \label{RG1}\\ 
  {\dot f}^{\sigma}_{ij} &&=
  -(f^{\sigma}_{ij})^2 +2c^{\rho}_{ij}c^{\sigma}_{ij} -\frac12
  (c^{\sigma}_{ij})^2, \label{RG2}
\end{eqnarray} 
\begin{eqnarray} 
  {\dot c}^{\rho}_{ij} = -\sum _k &&\bigg\{ \alpha_{ij,k}
  (c^{\rho}_{ik}c^{\rho}_{kj} +\frac{3}{16}
  c^{\sigma}_{ik}c^{\sigma}_{kj}) \bigg\} 
  \nonumber\\
  +&&(c^{\rho}_{ij}h^{\rho}_{ij} +\frac{3}{16}
  c^{\sigma}_{ij}h^{\sigma}_{ij}),\label{RG3}
  \\
  {\dot c}^{\sigma}_{ij} = -\sum_k
  &&\bigg\{ \alpha_{ij,k} (
  c^{\rho}_{ik}c^{\sigma}_{kj}+c^{\sigma}_{ik}c^{\rho}_{kj} + \frac12
  c^{\sigma}_{ik}c^{\sigma}_{kj} ) \bigg\} 
  \nonumber\\
  +&&(c^{\rho}_{ij}h^{\sigma}_{ij}+c^{\sigma}_{ij}h^{\rho}_{ij} -\frac12
  c^{\sigma}_{ij}h^{\sigma}_{ij}),
  \label{RG4}
\end{eqnarray}
where $f_{ij}= \tilde{f}_{ij}/\pi(v_i+v_j)$ and the same for $c_{ij}$. 
Also, we define $h_{ij} \equiv 2f_{ij}+\delta_{ij}c_{ii}$ for convenience. 
The weight factor in the summation $\alpha_{ij,k} \equiv
\frac{(v_i+v_k)(v_j+v_k)}{2v_k(v_i+v_j)}$ is symmetrical in $i,j$.  The
dots indicate logarithmic derivatives with respect to the length scale,
i.e. $\dot{f} \equiv \partial f/ \partial l$.

For PBCs, if the number of chains is odd, the allowed vertices are the
same as in OBC, and Eqs.~\ref{RG1}-\ref{RG4}\ hold without
modification.  However, if the number of chains is even, the
additional umklapp interactions in the $k_y$ direction give rise to
additional flow equations and extra terms in the forward and Cooper
scattering equations above.  Because these augmented RG equations are
quite complicated and do not provide any obvious insights upon
inspection, we have relegated them to appendix B.

Eqs.~\ref{RG1}--\ref{RG4}\ must be supplemented by initial conditions
to completely specify the problem.  While it of course straightforward
to obtain these initial values from the bare couplings, it does
require a certain amount of algebra to work out the effect of the
unitary transformation in Eqs.~\ref{SPBC}-\ref{SOBC} (see appendix C).
Initially, then, all the couplings $f_{ij},c_{ij}$ are $O(U/t)$ (but
with specific ratios).

\section{Strong-Coupling Analysis} 

In this section, we describe the application of the RG equations to
the $N$-chain Hubbard models with both OBCs and PBCs for small $N$.
We will see that the RG flows almost always diverge, and will discuss
the interpretation of such divergences, using as input certain general
features of the numerical integrations.  The instabilities encountered
generally correspond to some degree of pairing, the notion of which we
make precise through the study of various pair fields.

\subsection{Classification of couplings}
\label{sec:relevant_couplings}

\subsubsection{General scheme}

With the initial values in hand, the RG flow equations can be
integrated to investigate the physics of the weak-coupling limit.  We
have done this using Mathematica on a Sun Sparc-4 workstation.  While
the specific solution found depends upon the details of the model
parameters (e.g. $N$, $t_\perp/t$, $n$), certain gross features of the
behavior are generic.  In particular, for almost all sets of initial
conditions, the solutions of the RG equations are singular, and
certain linear combinations of coupling constants diverge at some {\sl
  finite} $l_d$.  Since the RG equations were obtained perturbatively,
they are not valid arbitrarily close to such an apparent divergence.  To
obtain sensible results, we instead cut off the RG flow at some
specific length scale $l^{*} < l_d$, chosen so that $U/t \ll \max
\{f_{ij}, c_{ij}, u_{ij} \}\ll1$.  At this cut-off length scale, the
couplings may be classified into two groups.  The first set includes
those couplings which have become ``large" but still weak, i.e.  $U/t
\ll g_{i} \ll1$, which we call marginally {\sl relevant}.  The
remaining couplings do not grow under the RG, but remain $O(U/t)$ or
smaller, and will be called {\sl marginal} or {\sl marginally
  irrelevant} respectively.  At the length scale $l^*$, the system
thus exhibits a separation of energy scales, with the marginally
relevant interactions much larger than the marginal or marginally
irrelevant ones.  The phase diagram of the system may then be
determined simply by neglecting the latter interactions and studying
the states determined by the marginally relevant couplings alone.

\subsubsection{Strict $U \rightarrow 0^+$ limit}

In the truly asymptotic limit $U \rightarrow 0^+$, much of the
classification of couplings can be accomplished analytically.  To do
so, consider the formal solutions of the RG flow equations as
functions of $l$ and the Hubbard interaction $U$, which we will denote
$g_i(l;U)$, where $i$ is a composite index labelling all the
interactions.  The perturbative RG is valid provided all the {\sl
  renormalized} interactions are small, i.e. $|g_i(l;U)| \ll 1$, which
is certainly true initially.  To proceed, we need to assume something
about the form of $g_i(l;U)$ near the divergent scale $l_d$.  Assuming
a power-law first suggested in Ref. 30 (which can be verified analytically 
for simple cases and numerically quite generally), dimensional analysis 
essentially requires
\begin{equation}
  g_{i}(l;U) \simeq \frac{U G_{i}}{(l_0 - Ul)^{\gamma_i}},
\label{power_law}
\end{equation}
where $l_0 = U l_d$ and we have set $t=1$ here and in the remainder of
this section.  The order one coefficients
$G_i$ and exponents $\gamma_i$ remain to be determined.  Given the
form of Eq.~\ref{power_law}, it is clear that the formally
divergent couplings only become much greater than $U$ when $l$ is
very close to $l_d$.  We are thus actually interested in the {\sl
  asymptotic} behavior of Eqs.~\ref{RG1}--\ref{RG4}.  Unlike the full
integration of the RG flows, the less ambitious task of determining
these asymptotics can in fact be accomplished analytically, as we now
demonstrate.  

In order to fix the exponents $\gamma_i$, it is necessary to use some input
from numerics. In particular, in every case we have examined, the
vector part of the forward scattering interactions, $f^{\sigma}_{ij}$,
is always marginally {\sl irrelevant}.  From Eq.~\ref{RG2}, this
implies that the combination $2c^{\rho}_{ij}c^{\sigma}_{ij}
-\frac12(c^{\sigma}_{ij})^2$ is small, i.e. $\lesssim O(U^2)$.  Since
we are keeping only those interactions which scale to values of order
one, to this order of accuracy the renormalized couplings satisfy
\begin{eqnarray}
  f^\sigma_{ij}(l^*) & \approx & 0, \label{fs_equals_zero} \\
  c_{ij}^\sigma(l^*) & \approx & 4 c_{ij}^\rho(l^*). \qquad (i \neq j)
  \label{cooper_constraint}
\end{eqnarray}
Eqs.\ref{fs_equals_zero}-\ref{cooper_constraint}\ suffer corrections
of $O(U)$, but may be treated as equalities in the following
leading-order analysis.  

Eq.~\ref{fs_equals_zero}--\ref{cooper_constraint}\ can be understood
in a simple physical way.  Some simple algebra demonstrates that if
Eqs.~\ref{fs_equals_zero}--\ref{cooper_constraint}\ were replaced by
exact equalities, these conditions would be preserved by the RG flow.
Thus it is natural to suspect that under these conditions, the system
has acquired an additional symmetry.  Indeed, upon closer examination
one finds that the constraint implies {\sl independent} conservation
of spin within each band.  Although this is not an exact property of
the Hubbard model, it is approximately satisfied due to the on-site
nature of the interactions.  For on-site interactions, Fermi
statistics allow only a coupling of oppositely oriented spins, which
implies Eq.~\ref{cooper_constraint}.  Apparently the deviation from
this symmetry caused by the nonzero initial values of $f^\sigma_{ij}$
is sufficiently small to allow the symmetry to be asymptotically
restored at long distances.

Based on this observation, we will calculate the exponents $\gamma_i$
in the $U \to 0^{+}$ limit, taking as an example the Cooper scattering
vertices.  Other exponents can be obtained by similar means.
The RG equations for the Cooper couplings $c^{\sigma}_{ii}$ in
Eq.~\ref{RG3}\ can be re-written with the help of
Eq.~\ref{cooper_constraint} as
\begin{equation}
  {\dot c}^{\sigma}_{ii} \approx -( c^{\sigma}_{ii})^{2} - \sum_{k \neq i}
  \alpha_{ii,k} (c^{\sigma}_{ik})^{2}.
\end{equation}
Since all terms on the right--hand side are of the same sign
(negative), they cannot cancel one another, and balancing the two
sides of the equation then gives the constraint 
\begin{equation}
  \gamma^{c\sigma}_{ii}+1 = \max \{ 2 \gamma^{c\sigma}_{ii}, 2
  \gamma^{c\sigma}_{ik} \},
\label{exponent_constraint}
\end{equation}
where $\gamma^{c\sigma}_{ij} \equiv \gamma( c^{\sigma}_{ij})$.

Eq.~\ref{exponent_constraint}\ has two solutions.
The first is
\begin{equation}
  \gamma^{c\sigma}_{ik} \leq \gamma^{c\sigma}_{ii} = 1.
\end{equation} 
The second possibility is,
\begin{equation}
  \gamma^{c\sigma}_{ii}< \gamma^{c\sigma}_{ik} =
  (1+\gamma^{c\sigma}_{ii})/2 \leq 1.
\end{equation}
We thus conclude that all the exponents associated with Cooper
couplings are bounded above by one.  Similar considerations applied to
the other RG equations imply that {\sl all} the exponents are less
than or equal to one.   If one can probe arbitrarily near the divergent point
$l_d$ (i.e. if $U$ is put arbitrarily close to $0^+$), any couplings
with exponents $\gamma_i = 1$ will eventually outstrip any others with
smaller exponents (even those with larger prefactors).  In the strict
$U \rightarrow 0^+$ limit, therefore, the relevant
couplings are those with exponents $\gamma_{i}=1$.

$ $From Eq.~\ref{power_law}, the relevant couplings with initial
interaction $U$ thus satisfy
\begin{equation}
  g_{i}(l;U) \simeq \frac{U G_{i}}{l_0- Ul},
\label{power_one}
\end{equation}
for $l$ near the cutoff length scale $l_d$.  Substituting
Eq.~\ref{power_one} into Eqs.~\ref{RG1}-\ref{RG4}, the parameters $U$
and $l_{d}$ cancel out, leaving a set of {\sl algebraic} relations
between the {\sl constants} $G_{i}$.  These algebraic equations are
formally obtained from Eqs.~\ref{RG1}-\ref{RG4}\ simply by replacing
$g_i \rightarrow G_i$ and $\dot{g}_i \rightarrow G_i$.  The relative
strengths of various couplings in the asymptotic regime can then be
determined (relatively) easily using these algebraic relations.
Usually there is more than one solution for $G_{i}$.  The
identification of specific solutions with specific initial conditions
which depend on, e.g. filling factor $n$, can only be found using
numerical integration of the full RG equations.  We emphasize as well
that these results rely upon on the strict $U \rightarrow 0^+$ limit.
A priori, given the lack of other energy scales in the Hubbard model,
we would nevertheless expect such results to hold qualitatively
provided $U \lesssim 1$ (for fixed small finite $N$).  In
fact, given the large number of coupling constants involved,
apparently order one factors can conspire to render the limits of
validity of this strict weak-coupling limit considerably smaller (e.g.
$U/t \lesssim 10^{-5}$ in some regions of the phase diagram even for
$N=2$ -- see section V!).  For $U/t \lesssim 1$ but {\sl outside} the
strict $U \rightarrow 0^+$ limit, the algebraic relations do not hold,
but we nevertheless expect quantitatively correct results from the
{\sl numerical} integration of the full RG flows.

\subsection{Bosonization strategy}

Our strategy will be to bosonize relevant couplings in
the renormalized Hamiltonian at the
scale $l^*$.  This is done using the 
Bosonization formula\cite{Shankar95,Emery79}
\begin{equation}
  \psi_{R/Li\alpha}(l^*) =\sqrt{\frac{\Lambda}{2\pi}}\eta_{i\alpha}
  e^{(i \sqrt{4\pi}\phi_{R/Li\alpha})},
  \label{bosonization_formula}
\end{equation}
Here the chiral Boson fields
$\phi_{R/Li\alpha}$ obey the commutation relations
\begin{eqnarray}
  [\phi_{Ri\alpha}(x),\phi_{Rj\beta}(y)]  = && -
  [\phi_{Li\alpha}(x),\phi_{Lj\beta}(y)] \nonumber\\
  = \frac{i}4 {\rm sgn}(x-y) &&\delta_{ij} \delta_{\alpha\beta},\\
  \left[\phi_{Ri\alpha}(x),\phi_{Lj\beta}(y)\right] &&= \frac{i}4 
  \delta_{ij} \delta_{\alpha \beta}
\end{eqnarray}
The $\eta_{i\alpha}$ are Majorana (real) Fermions, known as Klein
factors, introduced to preserve the proper anticommutation relations
between Fermion fields with differing band and spin indices.  They
obey 
\begin{equation}
  \{ \eta_{i\alpha},\eta_{j\beta}\} = 2\delta_{ij}\delta_{\alpha\beta}.
\end{equation}
It is usually more convenient to trade the chiral Boson fields
pairwise for a conventional Bosonic phase field $\phi$ and its dual
(displacement-like) field $\theta$, defined by
\begin{eqnarray}
  \phi_{i\alpha} = &&\phi_{Ri\alpha}+\phi_{Li\alpha}
  \nonumber\\
  \theta_{i\alpha} =&&\phi_{Ri\alpha}-\phi_{Li\alpha}.
\end{eqnarray}
They satisfy $[\phi(x),\theta(y)] = -i \Theta (y-x)$. 
Physically, the $\theta(x)$ field describes the displacement of the
electrons, while the dual field $\phi(x)$ represents
their phase. We can make  
a further canonical transformation to 
\begin{eqnarray}
  (\phi,\theta)_{i\rho}=&&[(\phi,\theta)_{i\uparrow}+
  (\phi,\theta)_{i\downarrow}]/\sqrt{2},
  \nonumber\\
  (\phi,\theta)_{i\sigma}=&&[(\phi,\theta)_{i\uparrow}-
  (\phi,\theta)_{i\downarrow})]/\sqrt{2}.
  \label{spin_charge}
\end{eqnarray}
The $\rho$ fields then describe charged singlet excitations, while the
$\sigma$ fields describe neutral excitations carrying spin.  Carrying
through the change of variables in Eq.~\ref{bosonization_formula}\
carefully, one finds that the non-interacting Hamiltonian Eq.~\ref{H_0}
is equivalent to the Bosonic Euclidean action
\begin{equation} 
  S_0 =
  \sum_{i\nu}\int_{x,\tau} \frac{v_i}{2} [(\partial_x \phi_{i\nu})^2
  +(\partial_x \theta_{i\nu})^2] + i \partial_x \theta_{i\nu}
  \partial_{\tau} \phi_{i\nu},
  \label{bare_boson_action}
\end{equation} 
where $\nu = \rho, \sigma$.

Upon bosonizing with Eq.~\ref{bosonization_formula}, the four-Fermion
interactions are converted to linear combinations of gradient
couplings and sinusoidal functions of the phases.  The former give
rise to continuous shifts of the parameters of the low-energy
description: modifications of mode velocities, charge stiffnesses,
etc.  We say that these shifts leave the system in the same {\sl
  phase}.  The sinusoidal interactions, by contrast, can potentially
cause more drastic changes in the low-energy theory.  They tend to
``pin'' their arguments (linear combinations of the Bosonic phases) to
particular values, modulo $2\pi$.  We will treat the marginally
relevant couplings in this way, by {\sl expanding} the corresponding
harmonic functions around their minima, regarding the fluctuations
around them as massive.  This will effectively give gaps to some
of the modes of the original non-interacting Fermion system, resulting
in distinct new phases.

To perform this procedure and determine the nature of the gaps arising
from the interactions, then, it is sufficient to keep only those
marginally relevant couplings which become sinusoids upon
Bosonization.  There are still quite a number of potential such terms,
so it now pays to use some additional input from the numerics.

\subsection{Generic Instability}
\label{sec:generic_instability}

For simplicity, we will first focus on situations in which the
relevant couplings include only forward ($f$) and Cooper ($c$) scattering
vertices.  This is completely general for OBCs, but excludes certain
regions of phase space in the case of PBCs with even $N$, for which
the transverse Umklapp ($u$) couplings may become relevant.  This
alternate channel of instability will be returned to later in this
section.

\subsubsection{Semiclassical Hamiltonian and analysis}

As a preliminary step in the analysis, we first rewrite the
interactions in terms of the underlying Fermion fields.  Using the
SU(2) identity,
\begin{equation}
  \bbox{\sigma}_{\alpha \beta}\cdot
  \bbox{\sigma}_{\gamma \epsilon}
  = 2 \delta_{\alpha \epsilon} \delta_{\beta \gamma}
  -\delta_{\alpha \beta} \delta_{\gamma \epsilon},
\end{equation}
the scalar and vector parts from Eq.~\ref{int1}\ become
\begin{eqnarray}
  {\cal H}^{(1)}_{int} = &&  \sum_{ij} \bigg[ f^{\rho}_{ij}
  \psi^{\dag}_{Ri\alpha} \psi^{\vphantom\dag}_{Ri\alpha} 
  \psi^{\dag}_{Lj\beta} \psi^{\vphantom\dag}_{Lj\beta}
  \nonumber\\
  &&+(c^{\rho}_{ij}+ \frac14 c^{\sigma}_{ij})
  \psi^{\dag}_{Ri\alpha} \psi^{\vphantom\dag}_{Rj\alpha} 
  \psi^{\dag}_{Li\beta} \psi^{\vphantom\dag}_{Lj\beta}
  \nonumber\\
  &&-\frac12 c^{\sigma}_{ij}
  \psi^{\dag}_{Ri\alpha} \psi^{\vphantom\dag}_{Rj\beta} 
  \psi^{\dag}_{Li\beta} \psi^{\vphantom\dag}_{Lj\alpha} \bigg].
  \label{fermion_int1}
\end{eqnarray}
Here we have dropped the irrelevant forward scattering $f^{\sigma}_{ij}$.

Following the strategy above, we must now Bosonize the system.  In
doing so, the interactions in Eq.~\ref{fermion_int1}\ may be divided
into two sets.  The first consists of $f^{\rho}_{ij}$ and
$c^{\rho}_{ii}$, which only contribute gradient terms after
bosonization.  To determine the phase of the system, therefore, we
need include only the second set, which contains the Cooper couplings
$c^{\sigma}_{ii}$, $c^{\sigma}_{ij}$, and
$c^{\rho}_{ij}$.  Keeping only these terms, and imposing the
constraint in Eq.~\ref{cooper_constraint}, the interaction Hamiltonian
becomes
\begin{eqnarray}
  &&{\cal H}^{(1)}_{int}= \sum_i \frac12 c^{\sigma}_{ii}
  \psi^{\dag}_{Ri\alpha} \psi^{\dag}_{Li\bar{\alpha}}
  \psi^{\vphantom\dag}_{Ri\bar{\alpha}}
  \psi^{\vphantom\dag}_{Li\alpha} + \sum_{i\neq j} \frac12 
  c^{\sigma}_{ij} 
  \nonumber\\
  && \times\left[
    \psi^{\dag}_{Ri\alpha} \psi^{\dag}_{Li\bar{\alpha}}
    \psi^{\vphantom\dag}_{Rj\bar{\alpha}} \psi^{\vphantom\dag}_{Lj\alpha}
    +\psi^{\dag}_{Ri\alpha}\psi^{\dag}_{Li\bar{\alpha}} 
    \psi^{\vphantom\dag}_{Lj\bar{\alpha}}
    \psi^{\vphantom\dag}_{Rj\alpha}\right],  
  \label{relevant_int1}
\end{eqnarray}
where $\bar{\alpha} \equiv -\alpha$. The interactions with $\alpha =
\beta$ in the third and fourth terms in Eq.~\ref{fermion_int1} cancel
each other, giving rise to the aforementioned independent conservation
of spin in channels $i$ and $j$.  This ordering of fermion fields is
particularly convenient for Bosonization.

As shown in detail in Appendix D,  for this set of interactions the
Klein factors  can be represented by the identity $\eta_i = 1$. 
Inserting Eq.~\ref{bosonization_formula}\ in Eq.~\ref{relevant_int1},
we then obtain
\begin{eqnarray} 
  {\cal H}&&_{int}^{(1)} \sim
  \sum_{i} c^{\sigma}_{ii} \cos (\sqrt{8\pi} \theta_{i\sigma}) 
  +\sum_{i<j} 4 c^{\sigma}_{ij}
  \nonumber\\ 
  && \times
  \cos(\sqrt{4\pi} \phi^{\rho-}_{ij})
  \cos(\sqrt{2\pi} \theta_{i\sigma})
  \cos(\sqrt{2\pi}\theta_{j\sigma}),
  \label{boson_int1}
\end{eqnarray} 
where $\phi^{\rho \pm}_{ij} \equiv (\phi^{\rho}_{i} \pm
\phi^{\rho}_{j})/\sqrt{2}$.  As pointed out by Schulz,\cite{Schulz96b}\
this explicit form after bosonization depends on the representation
chosen for the Klein factors. In other words, we still have some
``gauge'' freedom left to shift the bosonic fields. However, the
physical correlation functions, which include these Klein factors, are
independent of the specific gauge choice.

We next locate the minima of Eq.~\ref{boson_int1}.  These
can a priori be nontrivial, but turn out to be very simple in
practice.  In fact, we find that in all cases, the numerically
determined values of the coefficients $c_{ii}^\sigma$ and
$c_{ij}^\sigma$ are such that each term in Eq.~\ref{boson_int1}\ can
be minimized {\sl separately}.  Most often this occurs because all the
relevant couplings occur in channels connecting only two specific
bands.  We will focus on this special case now, in order to present a
more detailed but containable exposition.  

To proceed, let us denote the indices of the two strongly interacting
bands by $a$ and $b$.  The numerical integration demonstrates that,
although they are initially positive (repulsive interactions), the
diagonal Cooper-scattering spin vertices are driven negative under the
RG.  That is,
\begin{equation}
  c^{\sigma}_{aa}, \: c^{\sigma}_{bb} <0.
  \label{num1}
\end{equation}
We also obtain the sign
\begin{equation}
  c^{\sigma}_{ab}=4c^{\rho}_{ab} >0.
  \label{num2}
\end{equation}

Given these signs, a global minimum of Eq.~\ref{boson_int1}\ is
\begin{eqnarray}
  \sqrt{2\pi}\langle\theta_{a\sigma}\rangle&=&l \pi,\\ 
  \sqrt{2\pi}\langle\theta_{b\sigma}\rangle &= &m \pi,\\
  \sqrt{4\pi} \langle\phi^{-\rho}_{ab}\rangle &=&(l+m+2n+1) \pi,
\end{eqnarray}
where $l, m, n$ are integers. Since all solutions give the same results
of correlation functions, we will pick the $l=m=n=0$ solution for 
convenience.
Fluctuations around this semiclassical solution are
massive, as can be seen by the change of variables
\begin{eqnarray}
  \theta_{a\sigma} & = & \langle\theta_{a\sigma}\rangle +
  \delta\theta_{a\sigma}, \\ 
  \theta_{b\sigma} & = & \langle\theta_{b\sigma}\rangle +
  \delta\theta_{b\sigma}, \\ 
  \phi_{ab}^{\rho -} & = & \langle\phi_{ab}^{\rho -}\rangle +
  \delta\phi_{ab}^{\rho -}. 
\end{eqnarray}
Expanding to quadratic order gives
\begin{eqnarray}
{\cal H}^{(1)}_{int} & \sim & \frac12 (m^{\sigma}_{a})^{2}
  (\delta\theta_{a\sigma})^2 \nonumber \\ && + \frac12 (m^{\sigma}_{b})^{2}
  (\delta\theta_{b\sigma})^2 + \frac12 (m^{\rho}_{ab})^{2}
  (\delta\phi_{ab}^{\rho-})^2, 
  \label{massive_hamiltonian}
\end{eqnarray}
up to a constant. The masses for spin and charge modes are
\begin{eqnarray}
m^{\sigma}_{a}&=& 2 \sqrt{2\pi} \big( |c^{\sigma}_{aa}| 
+ c^{\sigma}_{ab} \big)^{1/2},\\
m^{\sigma}_{b}&=& 2 \sqrt{2\pi} \big( |c^{\sigma}_{bb}| 
+ c^{\sigma}_{ab} \big)^{1/2},\\
m^{\rho}_{ab}&=& 4 \sqrt{\pi} \big( c^{\sigma}_{ab} \big)^{1/2}.
\end{eqnarray} 
Comparison with the corresponding quadratic frequency terms in the
non-interacting action (Eq.~\ref{bare_boson_action}) indicates that
Eq.~\ref{massive_hamiltonian}\ describes three {\sl gapful} modes,
$\theta_{a\sigma}$, $\theta_{b\sigma}$, and $\phi^{\rho-}_{ab}$.
The first two terms in Eq.~\ref{massive_hamiltonian}\ suppress
fluctuations in the spin-densities of the two channels, and correspond
to spin gaps with magnitudes proportional to $m_a^\sigma$ and
$m_b^\sigma$.  The third term ``locks'' together the relative phase of
the charge modes in the two channels.  The remaining linear
combination $\phi^{\rho +}_{ab}$ and its conjugate $\theta^{\rho
  +}_{ab}$ are not affected by ${\cal H}^{(1)}_{int}$, and continue to
describe a gapless total charge mode.

\subsubsection{Pair fields}

The existence of a spin gap naturally suggests pairing of electrons
with oppositely oriented spins.  To investigate this notion further, 
it is instructive to consider a pairing operator
\begin{equation}
  \hat\Psi(X,x) = \psi_{P i \alpha}(X+x/2) \psi_{P' j \beta}(X-x/2), 
  \label{basic_pair_ops}
\end{equation}
which annihilates two electrons with specified band indices and spin
at particular positions.  For compactness, we will omit explicit
labeling of the pairing operator $\hat\Psi$ unless necessary to avoid
ambiguity.  In a true superconductor, such pair fields {\sl condense},
so that $\langle \hat\Psi \rangle \neq 0$.  The Mermin-Wagner theorem
prohibits such a continuous symmetry breaking in $1+1$ dimensions, but
$\hat\Psi$ can have power-law correlations (quasi-long-range order).  Of
physical interest are correlation functions of two or more of the
various $\hat\Psi$ operators at different well-separated points (in $X$).
For instance,
\begin{equation}
  C_{AB}(X,x) = \langle \hat\Psi_A^\dagger (X,x) \hat\Psi_B^{\vphantom\dag}(0,x)
  \rangle,
  \label{example_pair_correlator}
\end{equation}
where the composite index $A=(P_A i_A \alpha_A,P'_A j_A \beta_A)$, and
$B$ is defined in the same way.  Following the RG strategy,
correlation functions such as $C_{AB}(x)$ are evaluated in several
steps.  First, we employ the perturbative RG, integrating out Fermion
modes until the scale $l^*$.  Ignoring perturbative corrections from
the mode integration, each pairing operator then picks up just the
rescaling factor 
\begin{equation}
  \hat\Psi(X,x) \approx {1 \over {\Lambda\xi}}
  \hat\Psi(X/\Lambda\xi,x/\Lambda\xi;l^*), 
\end{equation}
where we have defined the {\sl coherence length} $\xi = \Lambda^{-1}e^{l^*}$.
At this point, the relevant couplings have become of order one, and
may be safely bosonized.  Carrying this out gives
\begin{eqnarray}
  \hat\Psi(X,x) &\simeq&
 {1 \over {\Lambda\xi}} \eta_{i\alpha}\eta_{j\beta} \exp
  \bigg[i\sqrt{4\pi} \big(
  \phi_{Pi\alpha}[(X + x/2)/\Lambda\xi]
 \nonumber \\
  && +\phi_{P'j\beta}[(X-x/2)/\Lambda\xi]\big)\bigg]. 
  \label{bosonize_it}
\end{eqnarray}
The next step is to insert this in the desired correlation function,
e.g. Eq.~\ref{example_pair_correlator}, and integrate out the massive
modes using Eq.~\ref{massive_hamiltonian}.  Since the masses in
Eq.~\ref{massive_hamiltonian}\ are order one, the correlation
functions of the gapped phase variables will decay exponentially over
(rescaled) distances of order one.  This allows us to make the
requirement of ``well-separated'' points more precise.  Since the
internal coordinates of the fields have themselves been rescaled by
$\xi$, it follows that for $|X| \gg \xi,$ the massive phase variables
in two pair fields separated by a distance $|X|$ are exponentially
decorrelated (provided the ``internal'' coordinates satisfy $|x| \ll
|X|$), and may be integrated out independently for each $\hat\Psi$.
For instance,
\begin{equation}
  \left\langle \hat\Psi_A^\dagger (X,x) \hat\Psi_B^{\vphantom\dag}(0,x') \right\rangle_g
  \approx \left\langle \hat\Psi_A^\dagger (X,x)\right\rangle_g \left\langle
    \hat\Psi_B^{\vphantom\dag}(0,x') \right\rangle_g,
\end{equation}
for $|X| \gg \xi$, where the subscript $g$ indicates an average over
the gapped phase fields.  

It is thus sufficient to study the partially averaged pairing
operators $\langle \hat\Psi(X,x) \rangle_g$.
We will do this carefully for the case of the two bands $a$ and $b$.
The average is carried out with respect to the action $S=\tilde{S}_0 +
S_1$, where
\begin{equation} 
  \tilde{S}_0 =
  \sum_{s}\int_{x,\tau} \frac{v_i}{2} [(\partial_x \phi_s)^2
  +(\partial_x \theta_s)^2] + i \partial_x \theta_s
  \partial_{\tau} \phi_s,
  \label{rediag_boson_action}
\end{equation} 
where $s=a\sigma,b\sigma,ab\rho-$, as obtained from
Eq.~\ref{bare_boson_action}, and
\begin{equation}
  S_1 = \int_{x,\tau} {\cal H}_{int}^{(1)}.
\end{equation}
$ $From Eqs.~\ref{massive_hamiltonian}\ and \ref{rediag_boson_action},
the small deviations $\delta\theta_{a\sigma}$,
$\delta\theta_{b\sigma}$, and $\delta\phi^{\rho-}_{ab}$ are decoupled
(from each other, but not from their conjugate fields) at the
quadratic level, and may therefore be averaged independently.

In carrying out this average, it is important to note the appearance
of the conjugate fields $\delta\phi_{a\sigma}, \delta\phi_{b\sigma},$
and $\delta\theta^{\rho -}_{ab}$.  By the uncertainty principle, since
$[\phi,\theta] = O(1)$, these variables are wildly fluctuating.  This
implies that any complex exponential containing one of these fields will
average to zero, unless it appears in the form of a ``neutral''
difference at nearby points, in which its fluctuation mean value is
automatically subtracted.  If such a subtraction does occur, the
average will decay exponentially with the separation of the subtracted
fields.

\end{multicols}

In fact, only four pairing operators satisfy this strong neutrality
constraint, and are therefore non-vanishing.  These are
\begin{eqnarray}
  \langle\hat\Psi_{d+}(X,x)\rangle_{g} =&&
  (\Lambda\xi)^{-1}\langle\psi_{Rd\uparrow}[(X+x/2)/\Lambda\xi] 
  \psi_{Ld\downarrow}[(X-x/2)/\Lambda\xi]\rangle_g,
  \nonumber\\
  \langle\hat\Psi_{d-}(X,x)\rangle_{g} =&&
  (\Lambda\xi)^{-1}\langle\psi_{Rd\downarrow}[(X+x/2)/\Lambda\xi] 
  \psi_{Ld\uparrow}[(X-x/2)/\Lambda\xi]\rangle_g,
  \label{neutral_pairs}
\end{eqnarray}
where again $d =a, b$.  As expected, only electrons
with opposite spin tend to form pairs.  

Inserting Eq.~\ref{bosonize_it}\ into Eq.~\ref{neutral_pairs}, we
obtain three averages.  In the relative charge sector,
\begin{eqnarray}
  &&\bigg\langle \exp\bigg(\pm i {\sqrt{\pi} \over 2}
  \left[\phi_{ab}^{\rho-}(\chi/2) + \phi_{ab}^{\rho-}(-\chi/2) +
    \theta_{ab}^{\rho-}(\chi/2) - \theta_{ab}^{\rho-}(\chi/2) \right]\bigg)
  \bigg\rangle_g \nonumber \\
  && = \pm i \bigg\langle \exp\bigg(\pm i {\sqrt{\pi} \over 2}
  \left[\delta\phi_{ab}^{\rho-}(\chi/2) + \delta\phi_{ab}^{\rho-}(-\chi/2) +
    \delta\theta_{ab}^{\rho-}(\chi/2) - \delta\theta_{ab}^{\rho-}(\chi/2)
  \right]\bigg) \bigg\rangle_g \nonumber \\
  && = \pm i g_{ab}^{\rho -}(\Lambda\chi), \label{mean1}
\end{eqnarray}
where $\chi = x/\Lambda\xi$, and we have without loss of generality chosen
$X=0$, since the average is independent of $X$ by translational
invariance.  The function $g_{ab}^{\rho-}(\chi)$ satisfies
\begin{equation}
  0<g_{ab}^{\rho-}(0)\lesssim 1, \qquad g_{ab}^{\rho-} \sim
  C_1 e^{-C_{2} \chi} \hspace{0.7cm} \chi \gg 1,
  \label{nice_function}
\end{equation}
with $C_1$ and $C_2$ order one constants.  The exponential decay
arises from the separation of the two $\theta_{ab}^{\rho-}$ fields,
whose rapid fluctuations exponentially suppress the average.  More
formally, the correlator involves a ``string'' connecting $x= \pm
\chi/2$, which carries an action per unit length (string tension) of
$1/\xi$.  

Similar reasoning leads to the results in the spin sector:
\begin{equation}
  \bigg\langle \exp\bigg(\pm i \sqrt{\pi \over 2}
  \left[\phi_{d\sigma}(\chi/2) - \phi_{d\sigma}(-\chi/2) +
    \theta_{d\sigma}(\chi/2) + \theta_{d\sigma}(\chi/2) \right]\bigg)
  \bigg\rangle_g = g_{d\sigma}(\Lambda\chi), \label{mean2}
\end{equation}
where $g_{d\sigma}(\chi)$ are functions with the same properties as
$g_{ab}^{-\rho}$, Eq.~\ref{nice_function}.

\begin{multicols}{2}
  
The fourth factor emerging from the averages in
Eq.~\ref{neutral_pairs}\ is the exponential of the $\phi_{ab}^{\rho+}$
field, which is not averaged over.  At this point, therefore, the
$\langle\hat\Psi_{d\pm}\rangle_{g}$ fields are still operators.  Using
Eqs.~\ref{mean1}-\ref{mean2}, they may be cast into the form
\begin{equation}
  \langle\hat\Psi_{d\pm}(X,x)\rangle_{g} = \pm \Delta_{d}(x) \;\; e^{i\phi(X)}
  (i\eta_{d\uparrow}\eta_{d\downarrow}),
  \label{pair_op1}
\end{equation}
where $\phi(X) = \sqrt{\pi} \phi^{\rho+}_{ab}(X/\Lambda\xi)$, and we have
used the fact $x \lesssim \xi$ (enforced by $g_{ab}^{-\rho}$ and
$g_{d\sigma}$) to neglect the $x$ dependence of $\phi$.  The overall
sign arises from re-ordering the Klein factors.  Physically, we may
now interpret $\phi$ as the usual $U(1)$ phase of the superconducting
order parameter.  The prefactor $\Delta_d$ is what is conventionally
interpreted as the pair wavefuncion in a superconductor, and has the
form
\begin{eqnarray}
  \Delta_{a}(x) &&\equiv \frac{1}{2\pi \xi}
  g_{ab}^{\rho-}(x/\xi)g_{a\sigma}(x/\xi), 
  \\
  \Delta_{b}(x) &&\equiv -\frac{1}{2\pi \xi}
  g_{ab}^{\rho-}(x/\xi)g_{b\sigma}(x/\xi).
\end{eqnarray}
The relative minus sign between $\langle\hat\Psi_{d+}\rangle_{g}$,
$\langle\hat\Psi_{d-}\rangle_{g}$ implies that the pair
wavefunction is a spin singlet, as is demonstrated by rewriting this
result as
\begin{equation}
  \langle \hat\Psi_{Ri\alpha Lj\beta}(X,x) \rangle_{g} =
  \delta_{ij} \Delta_i (x) 
  (\delta_{\alpha \uparrow} \delta_{\beta \downarrow}
  -\delta_{\alpha \downarrow} \delta_{\beta \uparrow})e^{i\phi(X)}.
  \label{d_pair}
\end{equation}
As in conventional superconductors, 
the ground state of the system is a singlet and,
therefore, SU(2) invariant. 

The relative sign between $\langle\hat\Psi_{a+}\rangle_{g}$,
$\langle\hat\Psi_{b+}\rangle_{g}$ indicates that the pair
wavefunction has $d$-wave symmetry in momentum space.  This is
illustrated in Figs.[6-10].  The precise nature of the wavefunction, i.e
the distinction between $d_{x^{2}-y^{2}}$ and $d_{xy}$ pairing,
depends on the positions of $(a, b)$ on the Fermi surface.  To
emphasize this point, we now calculate the pair wavefunction in real
space.

The most general pairing operator
in coordinate space is, 
\begin{equation}
  \Psi({\bf R}, {\bf r},\alpha,\beta) = 
  \psi_{\alpha}({\bf R}+\frac{{\bf r}}{2}) 
  \psi_{\beta}({\bf R}-\frac{{\bf r}}{2})
  \label{rs_pair_ops}
\end{equation}
where ${\bf R}=(X,Y)$ is the coordinate of the center of mass, 
and ${\bf r}=(x,y)$ is the relative distance between the pairing electrons.
Here $\alpha,\beta$ are the spin indices of the electron pair.

The pair fields $\hat\Psi$ (Eq.~\ref{basic_pair_ops}) and $\Psi$
(Eq.~\ref{rs_pair_ops}) are essentially related by a Fourier transform
in the transverse ($P,i$) indices.
To make this explicit, we must keep track of boundary conditions.
For OBCs, the right and left movers are standing waves in the 
transverse direction, and 
\begin{eqnarray}
  \psi_{\alpha}({\bf r}) \sim
  \big\{\psi_{Ri\alpha}(x) &&e^{ ik_{Fi} x}
   +\psi_{Li\alpha}(x) e^{ -ik_{Fi} x}\big\}\nonumber\\
 \times && \sin(k_{yi} y),
  \label{OBCtrans}
\end{eqnarray}
where $k_{yi}$ are the transverse
momenta defined in Eq.~\ref{OBCmomenta}. 
For PBCs, because the system is translational invariant,
the decomposition is the usual Fourier one,
\begin{equation}
  \psi_{\alpha}({\bf r}) \sim
  \psi_{Ri\alpha}(x) e^{ i{\bf k}_{Fi}{\bf r} }
  +\psi_{L{\bar i}\alpha}(x) e^{ -i{\bf k}_{Fi}{\bf r} },
  \label{PBCtrans}
\end{equation}
where the fermi vector ${\bf k}_{Fi} \equiv (k_{Fi}, 2\pi i/N)$.
  
\end{multicols}

Consider first OBCs.
Using Eqs.~\ref{rs_pair_ops}--\ref{OBCtrans},
\begin{eqnarray}
  \left\langle\Psi({\bf R,r},\alpha,\beta)\right\rangle_{g} 
  & \simeq & \sum_{i,j}
  \bigg \{ \sin(k_{yi}y_1) \sin(k_{yj}y_2)
  \psi_{i\alpha}(X+x/2)\psi_{j\beta}(X-x/2) 
  \bigg\}\nonumber \\
  & \approx &  \sum_{ij} \sin (k_{yi} y_1) \sin (k_{yj} y_2) \bigg[
  \left\langle \hat\Psi_{Ri\alpha Lj\beta}(X,x)\right\rangle_g
  e^{ik_{Fi}(X+x/2) -ik_{Fj}(X-x/2)} \nonumber \\
  && + \left\langle \hat\Psi_{Li\alpha Rj\beta}(X,x)\right\rangle_g
  e^{-ik_{Fi}(X+x/2) +ik_{Fj}(X-x/2)} \bigg]
  \label{obc_intermediate}
\end{eqnarray}
In the second line, we have used the fact that the non-vanishing
$\hat\Psi$ operators pair right and left moving fermions, and have thereby
dropped $\hat\Psi_{RR}$ and $\hat\Psi_{LL}$ contributions.
Using $\hat\Psi_{Li\alpha Rj\beta}(X,x) = -\hat\Psi_{Rj\beta L
  i\alpha}(X,-x)$, and $\Delta_i(-x) = \Delta_i(x)$ and
Eq.~\ref{d_pair}, Eq.~\ref{obc_intermediate}\ leads to the
familiar form
\begin{equation}
  \left\langle\Psi( {\bf R,r},\alpha,\beta)\right\rangle_{g}= 
  \Phi_{d}(Y,{\bf r}) \chi_{\alpha\beta} e^{i\phi}.
  \label{cooper_wf}
\end{equation}
The spatial and spin parts of the Cooper pair wave function are,
\begin{eqnarray}
  \Phi_{d}^{\rm OBC}(Y,{\bf r}) =&&\sum_{i=a,b}
  2\Delta_i(x)\cos(k_{Fi}x) \sin(k_{yi}y_1) \sin(k_{yi}y_2),
  \nonumber\\
  \chi_{\alpha\beta} =&& \delta_{\alpha\uparrow} \delta_{\beta\downarrow}
  -\delta_{\alpha\downarrow} \delta_{\beta\uparrow}.
  \label{dwaveOBC}
\end{eqnarray}
The positions of the electrons are denoted as $y_{1,2} = (Y \pm
\frac{y}{2})$.  Because of the hard-wall boundary conditions, the
spatial part of wave function depends on the transverse center-of-mass
coordinate $Y$.  Because $\Delta_a \Delta_b <0$, $\Phi_d^{\rm
  OBC}({\bf r})$ has d-wave symmetry in real space.

\begin{multicols}{2}
  
For PBCs, the results are quite similar.  The pairing operator
retains the same form of Eq.~\ref{cooper_wf}, with instead
\begin{equation}
  \Phi_{d}^{\rm PBC}({\bf r}) = \sum_{i=a,b} 2\Delta_{i}(x) \cos({\bf
    k}_{Fi}{\bf r}). 
  \label{dwavePBC}
\end{equation} 
In this case, the wave function only depends on the relative coordinate
${\bf r}$ because the system is translational invariant. Once again, the
symmetry is d-wave-like.

\subsection{Even chain PBCs}
\label{sec:even_pbc}

Turning to the case of even $N$ with PBCs, the generic presence of
transverse umklapp ($u$) interactions allows for a new RG instability.
This possibility is realized for $N=4$, as we have found by
numerically integrating the extended equations of appendix B.  In
certain regions of the phase diagram, the Cooper ($c$) couplings
become asymptotically irrelevant, and instead the transverse umklapp
and forward scattering interactions become dominant.  As before, this
occurs in only two bands, which we will again denote $a$ and $b$.  In
order for these two bands to be connected by two-particle umklapp
processes, they must satisfy $|a-b|=N/2$.
More careful attention to the numerics shows that, in addition, these 
relevant couplings satisfy
\begin{eqnarray}
  u^{1\rho}_{ab}=\frac14 u^{1\sigma}_{ab} &&>0, \label{tnum1}\\
  f^{\sigma}_{a\bar{a}}<0, f^{\sigma}_{b\bar{b}} && <0. \label{tnum2}
\end{eqnarray}

Eqs.~\ref{tnum1}--\ref{tnum2}\ appear quite similar to 
Eqs.~\ref{num1}--\ref{num2}, already encountered in the generic case.
In fact, careful study shows that the instability encountered here is
mathematically equivalent, after a relabeling of the bands, to the
earlier case.  Instead of repeating the analysis of the previous
subsection {\sl ad nauseum}, we will therefore instead only sketch the
essential points of the parallel treatment needed here.

To account for the change in paired bands, we combine the chiral boson
modes into the modified canonically conjugate fields
\begin{eqnarray}
  \bar{\theta}_{i\alpha} =&& \phi_{Ri\alpha} - \phi_{L\bar{i}\alpha}
  \nonumber\\
  \bar{\phi}_{i\alpha} =&& \phi_{Ri\alpha} + \phi_{L\bar{i}\alpha}.
\end{eqnarray}
Defining spin and charge bosons as in Eq.~\ref{spin_charge},
the interaction terms become
\begin{eqnarray} 
  &&{\cal H}_{int}^{(1)} +{\cal H}_{int}^{(2)}\sim
  \sum_{i=a,b} f^{\sigma}_{i\bar{i}} \cos (\sqrt{8\pi}
  \bar{\theta}_{i\sigma})  
  \nonumber\\ 
  &&+4 u^{1\sigma}_{ab}
  \cos(\sqrt{4\pi} \bar{\phi}^{\rho-}_{ab})
  \sin(\sqrt{2\pi} \bar{\theta}_{a\sigma})
  \sin(\sqrt{2\pi}\bar{\theta}_{b\sigma}),
\end{eqnarray} 
where $\bar{\phi}^{\rho \pm}_{ab} \equiv (\bar{\phi}^{\rho}_{a} \pm
\bar{\phi}^{\rho}_{b})/\sqrt{2}$.  This is of the same form as
Eq.~\ref{boson_int1}, and the semiclassical analysis is identical,
with $\theta \leftrightarrow \overline\theta$ and $\phi
\leftrightarrow \overline\phi$.  All the subsequent steps of the
analysis carry through with small modifications.  The non-vanishing
partially-averaged pair-fields expressed in terms of band indices are
\begin{equation}
  \langle \hat\Psi_{Ri\alpha Lj\beta}\rangle_{g}= 
  \delta_{i\bar{j}} \overline{\Delta}_i(x)
  \chi_{\alpha\beta}e^{i\phi(X)},
\end{equation}
where $\phi(X) = \sqrt{\pi} \overline\phi_{ab}^{\rho +}(X/\Lambda\xi)$, 
and the gap functions in momentum space are
\begin{eqnarray}
  \overline{\Delta}_{a} &= & {\Lambda \over {2\pi\xi}}
  \overline{g}_{ab}^{\rho -}(x/\xi)\overline{g}_{a\sigma}(x/\xi),
  \nonumber\\
  \overline{\Delta}_{b} &= & - {\Lambda \over {2\pi\xi}}
  \overline{g}_{ab}^{\rho -}(x/\xi)\overline{g}_{b\sigma}(x/\xi).
  \label{Delta_bars}
\end{eqnarray}
Notice that electrons with opposite $k_{x}$ but the ${\it same}$
$k_{y}$ are paired.  This implies that the Cooper pair carries
nonzero transverse {\sl quasi}-momentum.  To clarify the situation
further, we now specialize to the case of primary interest, $N=4$.
Following the previous subsection, we can again put the pairing
operator into real space.  Eq.~\ref{cooper_wf}\ continues to hold, but
with
\begin{eqnarray}
  \overline{\Phi}_{N=4}(Y,x) & = & \sum_{i=a,b} 2 \overline{\Delta}_{i}(x)
  \cos(k_{Fi}x)   e^{i2k_{yi} Y} \nonumber \\
  & = & -4|\overline{\Delta}_a|(x)\cos(k_{Fa}x)\sin(\pi Y),
  \label{TE}
\end{eqnarray}
where we have used reflection symmetry which implies
$\overline{g}_{a\sigma} = \overline{g}_{b\sigma}$ in
Eq.~\ref{Delta_bars}, and hence $\overline{\Delta}_a = -
\overline{\Delta}_b$, as well as $k_{Fa} = k_{Fb} = \pi/2$ in this
case.  Note that $Y$ takes on integer and half-integer values, so that
$\overline{\Phi}$ is real but can vary in sign.  If one imagines
wrapping the four chains around into a cylinder, the $Y$ dependence is
simply a superposition of the $m=\pm 2$ angular momentum states, i.e.
\begin{equation}
  \sin(\pi Y) = \sin (2\Theta).
\end{equation}
where $\Theta = \pi Y/2$ is the angle around the cylinder.  For this
reason, we call this a Cylindrically EXtended (CEX) d-wave 
state.  Note that since the superposition here is purely real, the
state does not carry any spontaneous current.

\section{Phase Diagrams} 
\label{sec:phase_diagrams} 

In the previous sections, we have described the RG and bosonization
technology necessary to analyze a weakly-interacting one-dimensional
Fermi system for any generic set of parameter values.  We have, of
course, applied these methods to study the particularly interesting
case of the $N$-chain Hubbard models.  The detailed calculations involve
lengthy but straightforward numerical integrations of the RG equations
and  mapping out the ensuing pairing instabilities as a function of
$N$, $n$, and $t_\perp/t$.  The primary {\sl results} of this work are
the phase diagrams shown in Figs.[6-10].  For the most part these
stand on their own, but we will comment on a few points. 

\begin{figure}[hbt]
\epsfxsize=3.5in\epsfbox{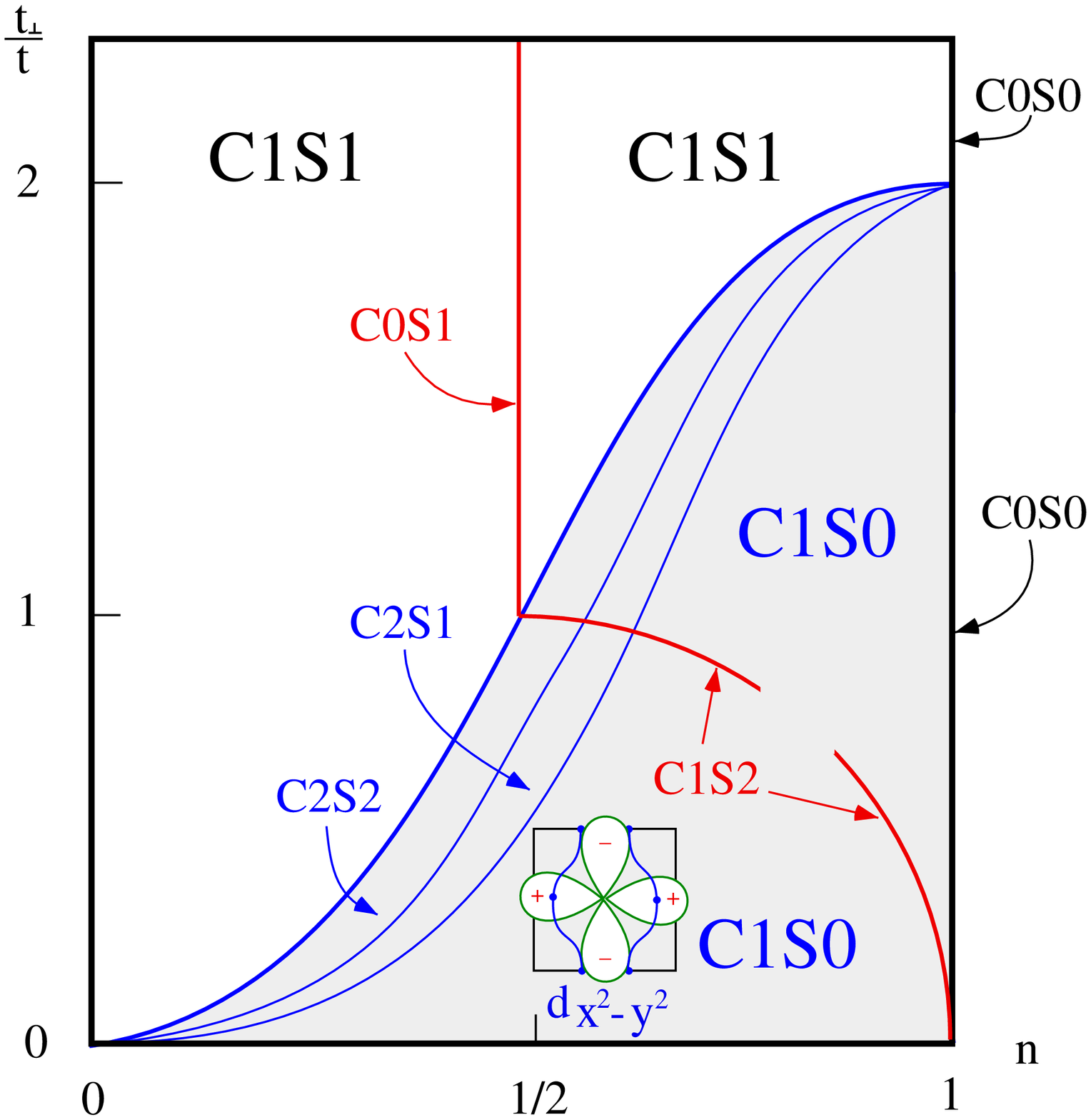}
{\noindent FIG. 6: Phase diagram of the two-chain Hubbard model with
  weak repulsive interactions. }
\end{figure}

\begin{figure}[hbt]
\epsfxsize=3.5in\epsfbox{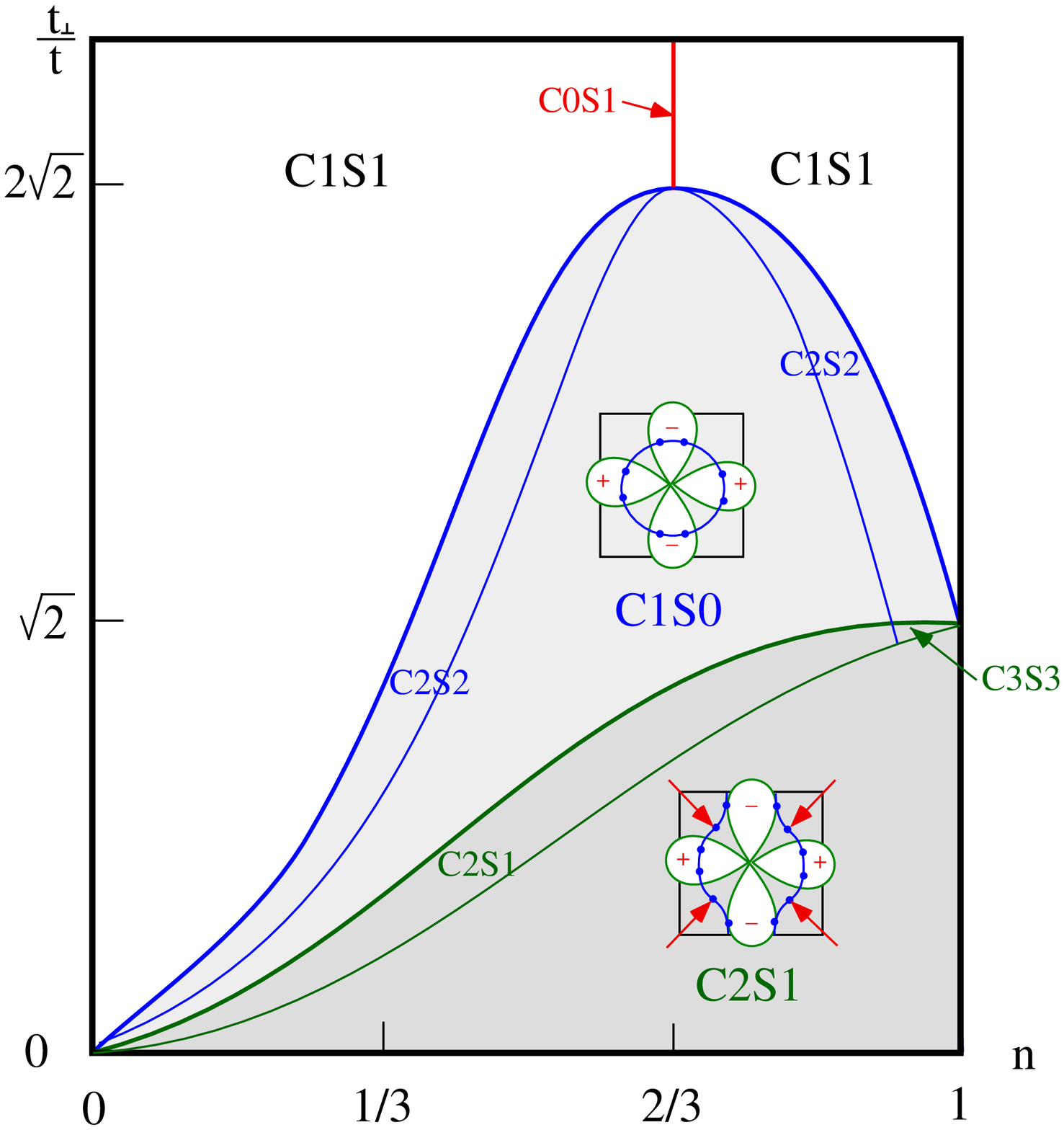}
{\noindent FIG. 7: Phase diagram of the 3-chain Hubbard model for OBCs.}
\end{figure}

\begin{figure}[hbt]
\epsfxsize=3.5in\epsfbox{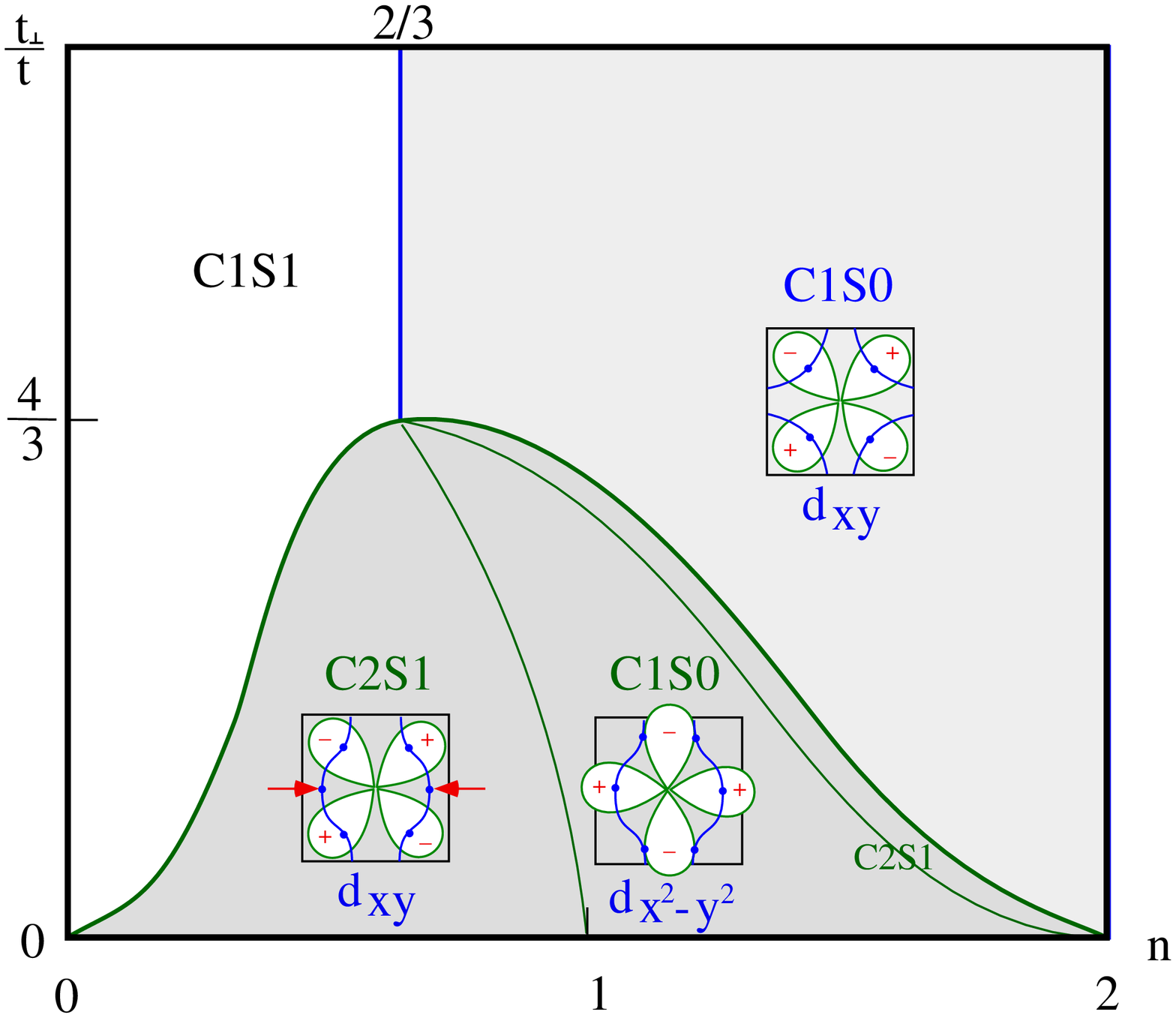}
{\noindent FIG. 8:  Phase diagram of the 3-chain Hubbard model for PBCs.}
\end{figure}

\begin{figure}[hbt]
\epsfxsize=3.5in\epsfbox{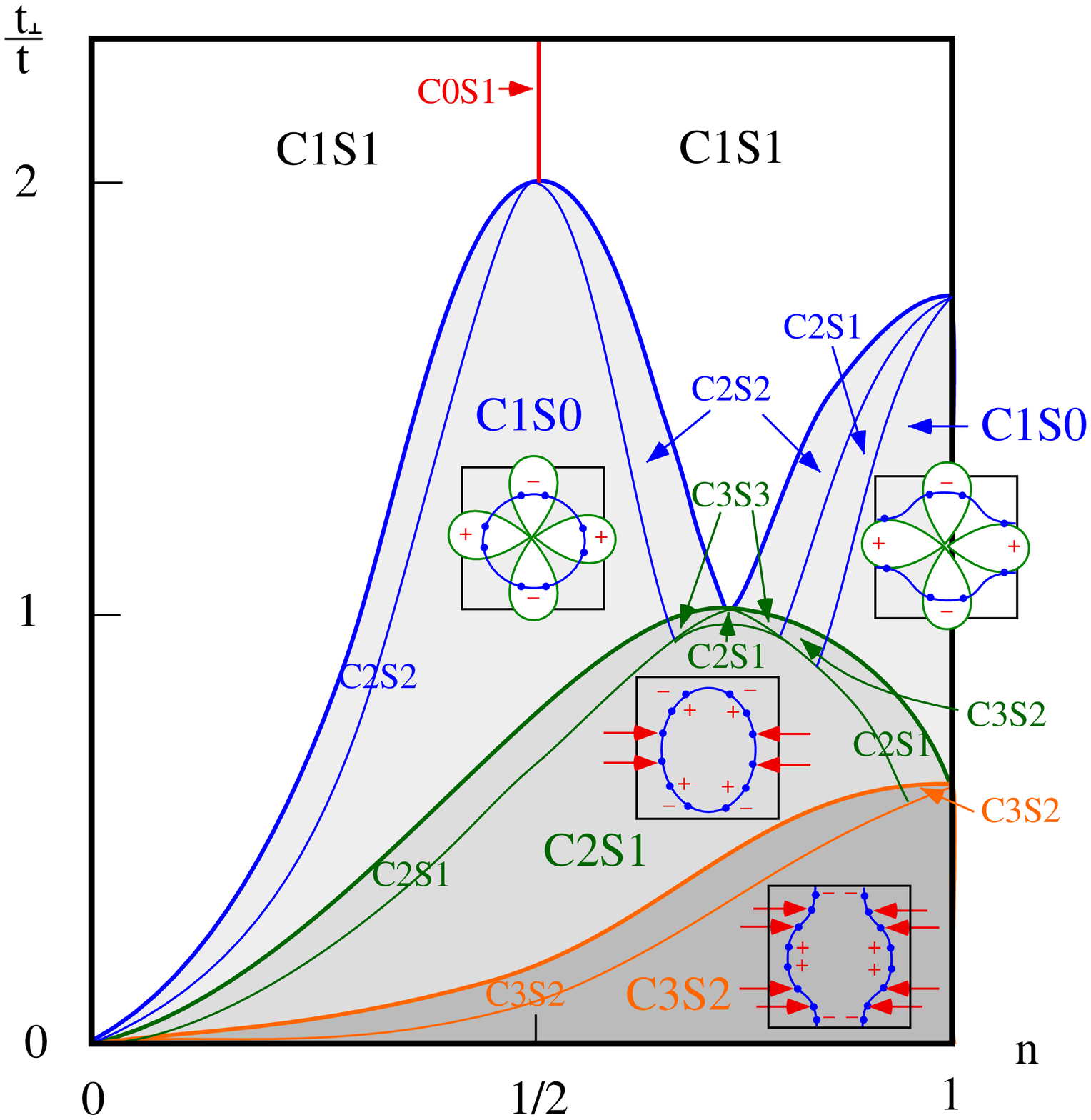}
{\noindent FIG. 9: Phase diagram of the 4-chain Hubbard model for OBCs.}
\end{figure}

\begin{figure}[hbt]
\epsfxsize=3.5in\epsfbox{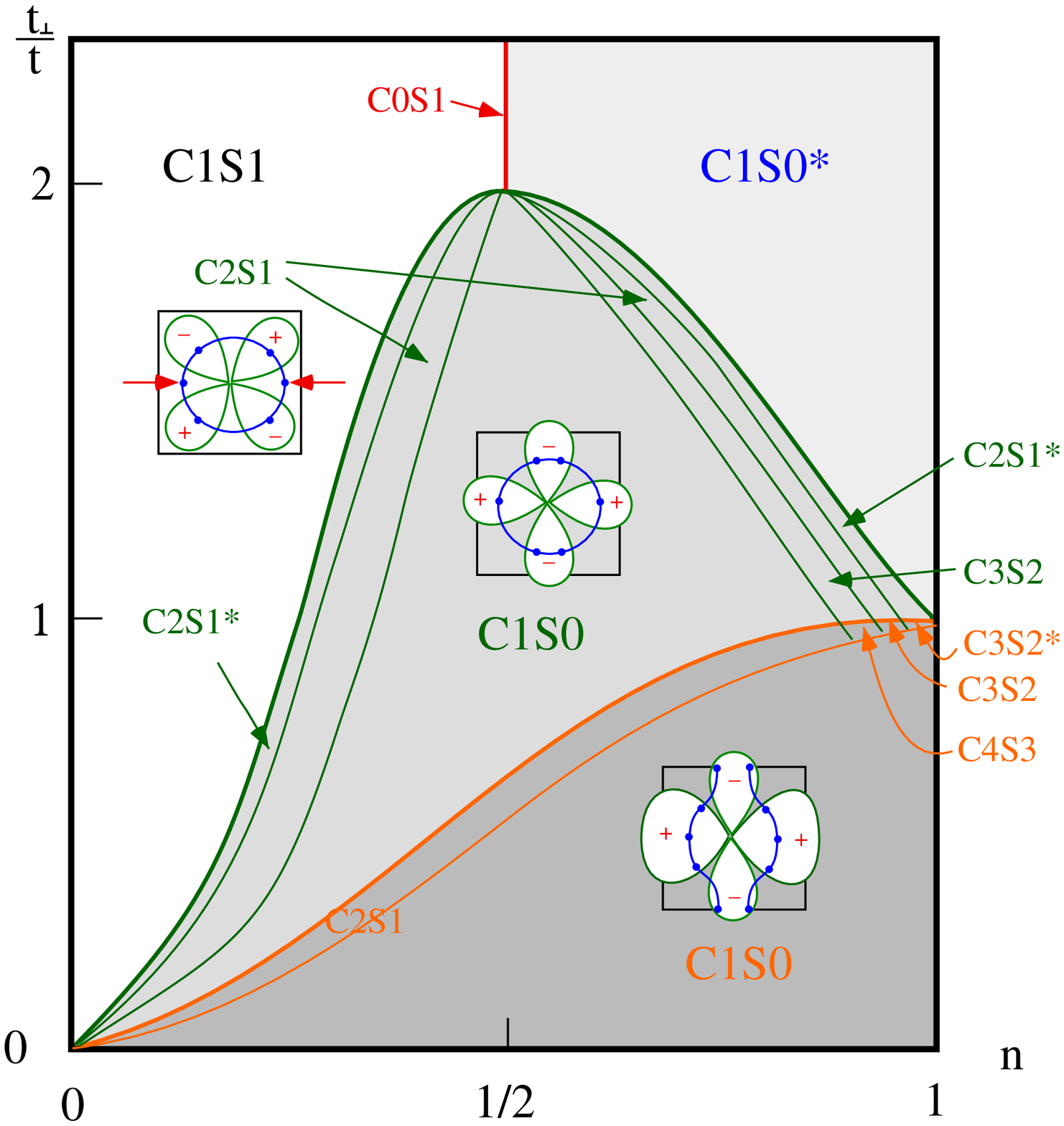}
{\noindent FIG. 10: Phase diagram of the 4-chain Hubbard model for
  PBCs.  The states marked with an asterisk exhibit CEX pairing.}
\end{figure}

\subsection{Commonalities}

\subsubsection{Band transitions}

In the weak-coupling limit, it is natural that the gross features of
the phase diagrams are dictated by the non-interacting band
structure.  In particular, the $n$--$t_\perp/t$ plane is divided into
distinct regions, in each of which a particular number $N_f$ of 1d
bands are {\sl partially} filled (and hence not inert).  The
boundaries between these regions constitute band transitions, which
generally survive as phase boundaries in the weakly interacting
system.  

While the band transitions need not be the only phase boundaries in
the interacting system, they usually form the most noticable divisions
of the phase space.  To locate them, one must solve
Eqs.~(\ref{dispersion}--\ref{OBCmomenta},
\ref{chemical_potential}--\ref{filling_factor}) for the lines along
which $\epsilon_a(0) = \mu$ (band $a$ is just empty) or
$\epsilon_a(\pi) = \mu$ (band $a$ is just full), for each $a$.  These
curves are indicated in the figures by heavy lines.  The shading of
the regions separated by the heavy lines indicates the number $N_f$ of
partially filled bands, with white corresponding to $N_f=1$ and the
darkest shade corresponding to $N_f=N$.

\subsubsection{Band edge phases}

On one side of such a band transition, the ``critical'' band (let us
denote its index by $a=E$) is almost empty or almost filled.  This
gives rise to a very small Fermi velocity, $v_E \ll t$, in the
weak-coupling RG.  For such a small velocity, the dimensionless
couplings acting purely within the band $E$ are greatly enhanced:
$g_{EE} = \tilde{g}_{EE}/(2\pi v_E) \gg g_{aE}$, for $a \neq E$.
Physically, this increased scattering is due simply to the large
density of states near the 1d van Hove singularity at the edge of the
band.  

Strictly speaking, the RG equations as we have derived them are not
valid directly at the band transition.  This is because the spectrum
of an empty/full band is not relativistic but quadratic ($\omega \sim
k^2/2m$), demanding a different (anisotropic) scaling.  We can,
however, approach very close to the band transition in the
weak-coupling limit.  That is, provided that we keep $U/v_E \ll 1$
(a much more stringent requirement than $U/t \ll 1$),
Eqs.~\ref{RG1}--\ref{RG4}\ remain valid.

In this region, the nearly vanishing Fermi velocity provides a useful
small parameter.  Indeed, those couplings within band $E$ are (at
least initially) much larger than those involving any of the
non-critical bands.  For the initial stages of integration of the RG
flows, then, these ``edge-band'' interactions dominate the evolution
of the couplings.  In particular, they lead to a decoupling of the
critical band from the remaining degrees of freedom.  To see this,
consider the evolution of the interactions between band $E$ and
another band $i$.  In the Cooper channel,
\begin{equation}
\left( \begin{array}{c}
{\dot c}^{\rho}_{iE}\\ {\dot c}^{\sigma}_{iE}
\end{array} \right) \approx
-\left( \begin{array}{cc}
c^{\rho}_{EE} & \frac{3}{16} c^{\sigma}_{EE}\\
c^{\sigma}_{EE} & c^{\rho}_{EE}+\frac12 c^{\sigma}_{EE}
\end{array} \right)
\left( \begin{array}{c}
c^{\rho}_{iE}\\ c^{\sigma}_{iE}
\end{array} \right).
\label{edge_other}
\end{equation}
Because the initial value for the Hubbard model,
\begin{equation}
\left( \begin{array}{c}
c^{\rho}_{iE}(0)\\ c^{\sigma}_{iE}(0)
\end{array} \right) = 
\left( \begin{array}{c}
1 \\ 4 \end{array} \right) c^{\rho}_{iE}(0),
\end{equation}
happens to be one of the eigenvectors of the matrix in
Eq.~\ref{edge_other}, the solution is particularly simple:
\begin{equation}
c^{\sigma}_{iE}(l) = 4 c^{\sigma}_{iE}(l) \simeq C_{1} 
e^{-\frac{C_2}{v_{E}} l}, \label{exponential_decay}
\end{equation}
where $C_{1}, C_{2}$ are constants of $O(U)$. 

This exponential decay holds provided the {\sl running} couplings
within band $E$ ($c_{EE}^\rho$, $c_{EE}^\sigma$) remain large compared
to the inter-band couplings.  Examination of
Eqs.~\ref{RG3}--\ref{RG4}\ shows that the intra-edge-band couplings
relax logarithmically to zero in the initial flow regime.  The above
equations are thus valid for
\begin{equation}
  c_{EE}^\rho, c_{EE}^\sigma \sim {1 \over l} \gg c_{kE}^\rho,
  c_{kE}^\sigma \sim {U \over v_k}.
\end{equation}
Following Eq.~\ref{exponential_decay}\ out to this, we see that the
couplings to the critical band are indeed exponentially suppressed 
\begin{equation}
  c^\sigma_{iE}(l = v_k/U) = 4 c^\sigma_{iE}(l=v_k/U) \simeq
  \tilde{C}_1 \exp[- \tilde{C}_2 v_k/v_E], \label{suppression}
\end{equation}
where $\tilde{C}_1$ and $\tilde{C}_2$ are order one constants.  

To complete the argument that band $E$ becomes decoupled from the
others, we must now show that this suppression persists into the
regime of divergence of other couplings, i.e. to the cut-off scale
$l^*$.  To do so, we employ the algebraic relations valid in the
asymptotic regime (Eq.~\ref{power_one}).  Consider, for example, the
relation derived from Eq.~\ref{RG3}.  The contributions on the right
hand side can be separated into singular and non-singular parts,
\begin{equation}
  C^{\sigma}_{kl} = -\alpha_{kl,E} C^{\sigma}_{kE}C^{\sigma}_{lE}
  +(\text{non-singular terms}),
\end{equation}
where $k,l \neq E$.  Here we have used the relaton $C^{\sigma}_{ij}=
4C^{\rho}_{ij}$ as usual. Since the factor $\alpha_{kl,E} \sim
1/v_{E}$ is singular when $v_{E} \to 0$, there are only two possible
options.  Either the singularity in $\alpha_{kl,E}$ is cancelled, and
$C_{kE} \sim \sqrt{v_E}$, or the singular and non-singular parts equal zero
separately and
\begin{equation}
  C^{\sigma}_{kE} =0. \label{vanishing}
\end{equation}
The former possibility is inconsistent with the exponential
suppression in Eq.~\ref{suppression}, so we expect that instead
Eq.~\ref{vanishing}\ holds and the couplings $c_{kE}$ will flow to zero in
the asymptotic regime.  This is indeed observed in all numerical
integrations of the full RG equations near a band transition.

This decoupling implies that the low energy structure of the system is
obtained by adding the single gapless charge and spin modes of the
critical band to the low energy structure of the remaining bands that
would have occurred were band $E$ inert.  To determine the phase of
the Hubbard model in the band-edge regime, therefore, we may simply
add $C1S1$ to the gapless mode content on the other side of the band
transition, in which the critical band is indeed inert.  If the phase on
this side is $C_N S_M$, then the band-edge result on the other side of
the transition line is
\begin{equation}
  C_{N}S_{M} + C1S1 = C_{N+1}S_{M+1}.
\label{band_edge_phase}
\end{equation} 

\subsection{Specific features}

\subsubsection{``d-wave'' pairing}

Probably the most striking aspect of the phase diagrams is the
ubiquity of paired states -- i.e. gapping out of the spin modes in at
least some of the bands.  Following the methods of section IV, these
pairing instabilities can be associated with a gap function defined at
the allowed discrete points on the 2d Fermi surface.  Except in
certain regions of phase space in the four-chain model, this gap
function has an approximate ``d-wave'' form.  For $N \geq 3$, a d-wave
gap has an interesting consequence in this context: the discrete
transverse wavevectors $k_{ya}$ can coincide with the nodes in the
pair wavefunction.  This indeed occurs, e.g. near half-filling for
$N=3$ with OBCs for $t_\perp < \sqrt{2} t$, giving rise to
simultaneous dominant superconducting correlations and power-law
antiferromagnetism.  The d-wave interpretation begins to break down,
however, for $N=4$ with OBCs, where several gapless spin modes are
present for small $t_\perp/t$.  The unusual distribution of gapped and
ungapped modes on the Fermi surface in this case is, we expect, a
consequence of the one-dimensional weak-coupling limit taken here,
which becomes rather restrictive for larger $N$.  A more complete
discussion of the approach to a two-dimensional weak-coupling limit is
described in section VI.

\subsubsection{Peculiarities of PBCs}

The systems with PBCs exhibit a number of (theoretically!) interesting
peculiarities.  For $N=3$, some of these have been pointed out by
Arrigoni,\cite{Arrigoni96a,Arrigoni96b}\ who performed a similar weak-coupling
analysis.  For odd $N$, the effects of PBCs can be expected to be
rather severe.  In the Hubbard model, they break particle-hole
symmetry, which has the effect of eliminating reflection symmetry of
the phase diagram around half-filling.  Furthermore, half-filling no
longer coincides with the conditions needed for umklapp processes at
the Fermi level.  This has the happy consequence that our generic
treatment (which ignores these umklapp interactions) remains valid at
$n=1$.  One then finds the rather surprising result that the
half-filled system has gapless charge excitations and a spin gap,
precisely the opposite of what is expected in the strong-coupling
limit with OBCs, where there is a charge gap ($\sim U$), and the
effective Heisenberg model (with odd $N$) is expected to have a gapless
spin mode.  Some partial understanding can be gained from the fact
that an odd-chain Heisenberg model with PBCs is {\sl frustrated}, and
can be shown (at least for $N=3$) to indeed have a spin gap.
However, the absence of a charge gap is a weak-coupling result, and
indicates the existence of a {\sl metal-insulator transition} at
half-filling as $U$ is increased.  Interestingly, the weak-coupling
paired state can be either of $d_{xy}$ or $d_{x^2-y^2}$ type, as
indicated in Fig.[8].  

A different sort of feature arises for PBCs with $N=4$.  Although this
situation retains particle-hole symmetry, there nevertheless exist
regions (the largest occurs for weak doping with $t_\perp > t$) in
which the finite transverse size has a severe effect.  This
CEX d-wave phase has a pair wavefunction whose phase
depends upon the {\sl transverse center-of-mass coordinate} of the
pair.  This phenomena is described in section IVD, and is certainly
special to the one-dimensional cylindrical geometry considered here.

\subsubsection{Extreme asymptotic instability of the C1S0 phase}

A number of authors have predicted the existence of a C1S0 paired
state for a weakly interacting two-chain Hubbard ladder.  As noted (in
proof) in Ref.~\onlinecite{Balents96}\ in the asymptotic limit $U/t
\rightarrow 0^+$, the C1S0 phase in fact occurs only for infinitesimal
doping, being replaced everywhere else by the $C2S1$ phase.  For
reasonable (but still small) values in the range $10^{-6} < U/t \lesssim
1$, however, the C1S0 phase still appears as dominant.

To understand this result requires a more detailed examination of the
asymptotic regime near the RG divergence.  To do so, we again consider
the algebraic relations described in 
Sec.~\ref{sec:relevant_couplings}.

$ $Using Eq.~\ref{power_one}, we look for a solution of the resulting
algebraic equations for which $C^{\sigma}_{12} \neq 0$, corresponding
to a C1S0 phase.  Some straightforward calculations give the unique answer
\begin{eqnarray}
&C&^{\sigma}_{11}= C^{\sigma}_{22} = -\frac12 
(1+ \sqrt{1-4 \alpha_{11,2} C^{\sigma}_{12}}), \label{equal_cooper}\\
&1&+\frac32 C^{\sigma}_{11} -\frac12 
(1+\alpha_{11,2})C^{\sigma}_{12}=0.
\end{eqnarray}

Now consider the stability of this solution.  From Eq.~\ref{RG4}, the
difference between $c^{\sigma}_{11}$ and $c^{\sigma}_{22}$ obeys
\begin{equation}
\frac{d}{dl} (c^{\sigma}_{11}-c^{\sigma}_{22}) =
-(c^{\sigma}_{11}-c^{\sigma}_{22}) (c^{\sigma}_{11}+c^{\sigma}_{22}).
\end{equation}
Since $c^{\sigma}_{ii}$ are negative near the divergent point
$l^{*}$, the difference of Cooper couplings
$c^{\sigma}_{11}-c^{\sigma}_{22}$ is relevant, and the above solution
is in fact {\sl unstable}.  For the special initial value
$c^{\sigma}_{11}-c^{\sigma}_{22} = 0$, which is attained in the limit
$n \rightarrow 1$, the system is specially tuned to an unstable
equilibrium, and the C1S0 phase enjoys a small region of existence.
Moving away from half-filling, however, the inequality of Fermi velocities
in bands $1$ and $2$ destroys this fine-tuning,
driving the system away from the C1S0 state.  In the asymptotic limit,
the stable solution is in fact the much simpler $C2S1$ flow.  This
point has been missed in other calculations, owing to the assumption
of equal Fermi velocities\cite{Schulz96a,Schulz96b}\ and the lack of a careful
stability analysis of the asymptotic
regimes.\cite{Fabrizio93,Arrigoni96a,Arrigoni96b}\ 
We emphasize, however, that numerically this instability is extremely
weak.  For even relatively weak couplings with $10^{-6} < U/t \lesssim
1$, we find that the C1S0 phase remains quasi-stable, occupying in
fact the majority of the two-chain phase diagram.

\section{Dimensional Crossover}

In this section, we discuss how the system approaches 2d behavior as
$N \rightarrow \infty$.  The limit is actually quite subtle, and we
will consider two distinct ways of performing it.  The simplest
procedure is simply to attempt to preserve the validity of the RG as
presented here (Eqs.~\ref{RG1}--\ref{RG4}).  This {\sl 1d weak
  coupling limit} can be realized in principle for any fixed (but
large) $N$ for sufficiently small $U$, but taking $N \rightarrow
\infty$ actually requires that the interactions vanish as well.  A
more physically appealing approach is the true {\sl 2d weak coupling
  limit}, in which $N\rightarrow \infty$ for a fixed (but small) $U$.
In this case, the RG as constructed so far must in principle be
supplemented by additional interactions.

\subsection{One-dimensional weak coupling limit}

We first consider the naive limit of the RG flows
(Eqs.~\ref{RG1}--\ref{RG4}) as $N \rightarrow \infty$.  Recalling the
results of section II (Eq.~\ref{weakU}), to retain the validity of
these equations, the interactions must be simultaneously taken to
zero, with $U \lesssim t/\ln N$.  Since this constraint is only
logarithmic in $N$, this is actually not a strong restriction even for
reasonably large values of $N$, and might indeed be physically
relevant in some systems.

In the large $N$ limit, the RG flows are dominated by those terms
involving sums over intermediate band indices; which effectively
increase these terms by a factor of $N$.  To make the largest terms of
order one for large $N$, we introduce the rescaled coupling constants
\begin{equation}
  c^{\rho}_{ij} =
  \frac{1}{N} \frac{2\sqrt{v_i
      v_j}}{v_{i}+v_{j}}\hat{c}^{\rho}_{ij},
  \label{2d_rescale}
\end{equation}
and similarly for the other interaction channels.
Note that, with Hubbard initial conditions, the values of
$\hat{c}^\rho_{ij}$ are of an order of one (in $N$).  
Inserting this into Eqs.~\ref{RG1}--\ref{RG4}\ 
and dropping the $O(1/N)$ terms, one finds that the forward and
umklapp scattering vertices are exactly marginal (unrenormalized).
The Cooper channel interactions obey the simplified equations
\begin{eqnarray} 
  \partial_l\hat{c}^{\rho}_{ij}
  =&& -\frac{1}{N} \sum _k (\hat{c}^{\rho}_{ik}\hat{c}^{\rho}_{kj}
  +\frac{3}{16} 
  \hat{c}^{\sigma}_{ik}\hat{c}^{\sigma}_{kj}), 
  \\ 
  \partial_l\hat{c}^{\sigma}_{ij} =&&
  -\frac{1}{N} \sum_k (
  \hat{c}^{\rho}_{ik}\hat{c}^{\sigma}_{kj}+\hat{c}^{\sigma}_{ik}
  \hat{c}^{\rho}_{kj} +  \frac12
  \hat{c}^{\sigma}_{ik}\hat{c}^{\sigma}_{kj} ).   
  \label{2d_cooper}
\end{eqnarray} 

The Hubbard model initial condition that
$4c^{\rho}_{ij}(0) = c^{\sigma}_{ij}(0)$ is preserved by
Eq.~\ref{2d_cooper}, so they may be collapsed into the single
non-trivial RG equation
\begin{equation}
  \frac{d\hat{c}^{\sigma}_{ij}}{dl} = -\frac{1}{N} \sum_k
  \hat{c}^{\sigma}_{ik}\hat{c}^{\sigma}_{kj}. 
  \label{2dRG}
\end{equation} 
Note that the right-hand-side of Eq.~\ref{2dRG}\ has the form of
matrix multiplication.  This implies that, provided the initial
couplings matrix is diagonalizable, each eigenvalue $\lambda_i$ of the
matrix $c^\sigma$ evolves independently according to
\begin{equation}
  \frac{d\lambda_{i}}{dl} = -\frac{1}{N} \lambda_{i}^2.
  \label{ev_flows}
\end{equation}
Focusing for simplicity on PBCs, the initial value of
$\hat{c}^\sigma$ is 
\begin{equation}
  \hat{c}^{\sigma}_{ij}(0) = (\frac{U}{2\pi}) \frac{1}{\sqrt{v_i v_j}}.
\end{equation}
Since this has the form of an outer product, it is proportional to a
projection operator onto the vector $1/\sqrt{v_i}$.  It thus has
$N_f-1$ null vectors and a single non-trivial eigenvector
proportional to $1/\sqrt{v_i}$.\cite{KL:foot,Zanchi96}\  The eigenvalues are 
\begin{equation}
  \lambda_{i}(0) = \big (\frac{U}{2\pi} \sum_{k} \frac{1}{v_k}\big ),0,...,0.
  \label{eigen}
\end{equation}

$ $From Eq.~\ref{ev_flows}, the $N_f$ zero eigenvalues are unchanged
under the RG, while $\lambda_1$ obeys
\begin{equation}
  \lambda_{1}(l) = \frac{N\lambda_1(0)}{N+\lambda_1(0) l},
\end{equation}
where $\lambda_{1}(0)=(U/2\pi)\sum_{k} 1/v_{k}$.  Since $\lambda_1(0) >0$, it
is marginally irrelevant and flows to zero.  This implies that all the
Cooper interactions flow logarithmically to zero.

To connect with previous two-dimensional treatments, we may define a
simple continuum limit:
\begin{eqnarray}
  \hat{c}_{ij}^\sigma & \rightarrow & V(k_{yi},k_{yj}), \\
  {1 \over N}\sum_i & \rightarrow & \int_{-\pi}^\pi \! {{dk_{yi}}
    \over {2\pi}},
\end{eqnarray}
which gives the RG equation
\begin{equation}
  \frac{d}{dl}V(k_y, k'_y) = -\int_{-\pi}^\pi {{dk''_y} \over {2\pi}}
  V(k_y,k''_y)V(k''_y, k'_y),
\end{equation}
as derived previously by Shankar\cite{Shankar94} directly in 2d.

\subsection{Two-dimensional weak-coupling limit}

On reflection, the agreement with approaches directly in two
dimensions is perhaps surprising, since the RG equations used above
are valid only for $U/t \lesssim 1/\ln N$ as $N\rightarrow\infty$.  To
study the true 2d weak-coupling limit (with $U/t \ll 1$ but fixed as
$N\rightarrow\infty$) requires consideration of the additional
shifted interactions (such as ${\cal H}^s$) introduced in section IIC.
Fortunately, one can show that, even upon including these
interactions, the modifications of the RG equations are actually
negligible in weak-coupling.  Rather than belabor this reasoning,
which is essentially discussed already in, e.g. Shankar's review
article,\cite{Shankar94}\
we will only schematically indicate how this comes about.

Once the additional shifted interactions are included in the RG, we
must worry about two questions. How do these new vertices renormalize,
and how do they feed back into the flow equations for the unshifted
couplings?  In answer to the first question, under normal conditions,
the shifted interactions renormalize almost identically to their
unshifted counterparts, at least in the initial stages of the RG.
This is because for each process involving two unshifted vertices
feeding into an unshifted vertex, there is an analogous process
involving the same vertices shifted, usually feeding back into the
analogous shifted vertex.  Next, note that, in weak coupling, the
range of momentum shifts is highly constrained:
\begin{eqnarray}
  |\Delta k_y| = 2\pi|\delta|/N & < & 2\pi\delta_{\rm max}/N \sim \Lambda
  t_\perp/t, \nonumber \\
  \Rightarrow |\Delta k_y| & \lesssim &  |\Delta k_y|_{\rm max} = \pi
  {t_\perp \over t}   e^{-{\rm const. }t/U},
\end{eqnarray}
as can be seen from Eqs.~\ref{shift}\ and \ref{UcondA}.  For any
non-singular interaction, the initial coupling constants are
reasonably smooth in momentum space, so that all the shifted
interactions in the narrow range $|k_{y\delta}| < |\Delta k_y|$ are
essentially equal in magnitude.  Since each shifted vertex then has
the same initial conditions and obeys the same RG equation as an
unshifted coupling, it remains so under the RG, and the original
equations remain sufficient to study their evolution.

\begin{figure}[hbt]
\epsfxsize=3.5in\epsfbox{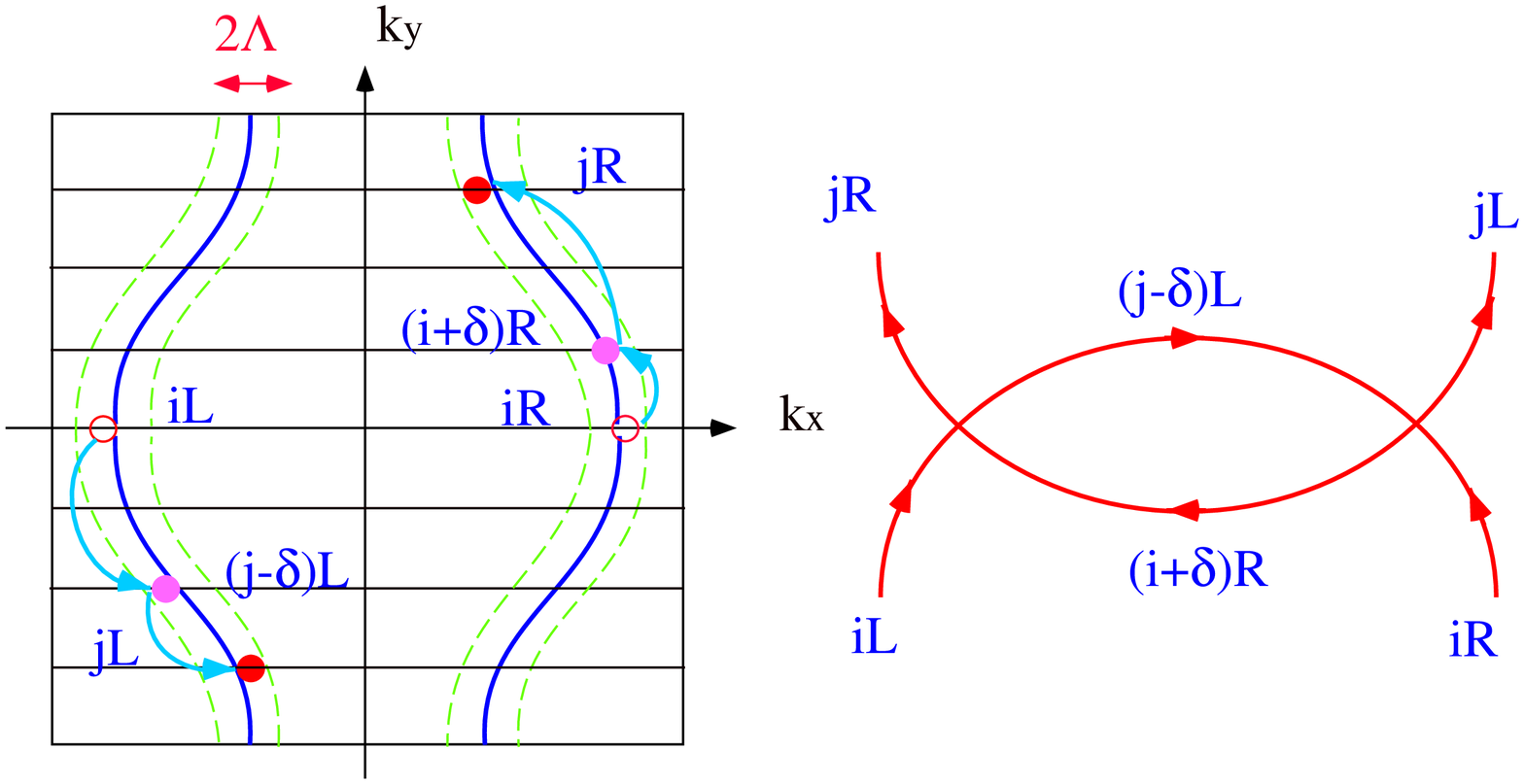}
{\noindent FIG. 11: An example of renormalization of 
$c^{\sigma}_{ij}$ from shifted vertices in Eq.~\ref{shift_loop}.
}
\end{figure}

It remains to answer the second question.  There {\sl are} a few
additional processes involving the shifted interactions, such as the
one shown in Fig.[11], which feed back into the original Cooper
and forward scattering channels.  The feedback into the forward
scattering channel is negligible for the same phase-space reasons that
render them exactly marginal for $N\rightarrow\infty$ above.  More
subtle is the feedback into the unshifted Cooper channel.  Once
shifted vertices are included, an intermediate sum over $\delta$
allows for non-vanishing contributions of the form
\begin{equation}
  \frac{d\hat{c}^{\sigma}_{ij}}{dl} = \cdots + \frac{\rm const.}{N}
  \sum_\delta
  \hat{c}^{\sigma}_{i(j-\delta)}(\delta)\hat{f}^{\sigma}_{i(j-\delta)}
  (\delta) + \cdots,
\label{shift_loop}   
\end{equation}
in which the (shifted) forward scattering vertices feed back into the
Cooper channel even as $N\rightarrow\infty$ (since $\delta_{\rm
  max.}/N$ is finite in this limit). However,
the phase space for these renormalizations is considerably smaller
than the processes already included.  Roughly speaking, these
additional terms are down by a factor of $\Lambda/\pi \sim \exp(-{\rm
  const.}t/U)$ from the others, since the range of angles of the
intermediate momenta are restricted to a width $|\Delta k_y|_{\rm
  max.} \ll 2\pi$.  Furthermore, the number of allowed terms continues
to decrease as the RG proceeds to lower energies, since the band
curvature $t_\perp/t$ effectively grows under rescaling.

\subsection{Instabilities for $1 \ll N \ll \infty$}

In the large $N$ limit, therefore, additional (shifted) interactions
{\sl are} present at weak but finite couplings, but do not modify the
RG equations or their analysis as presented above.  At $N=\infty$,
then, the 2d metal is marginally stable.\cite{KL:foot}\ What occurs
in weak coupling for large but finite $N$?  Our numerical results suggest that
instabilities always persist when feedback of forward-scattering
interactions into the Cooper channel is included.  However, this
cross-coupling is an $O(1/N)$ effect, and sois extremely weak for
large $N$.  Specifically, these
terms can only begin to affect the flows once
the Cooper interactions have themselves renormalized down to order
$1/N$.  Since the $N=\infty$ flows are logarithmic, this occurs only
after a rescaling $b = e^l \sim e^N$, so that the characteristic
energy gaps (and critical temperatures) of any paired states should obey
\begin{equation}
  \Delta_N \lesssim e^{-N},
\end{equation}
with a prefactor which is not determinable by such coarse arguments.

This exponential decrease of the energy scale for pairing is a signature
of the rather robust stability of the generic Fermi liquid.  This
weak-coupling result, however, does not make any statement about
pairing instabilities for {\sl strong} repulsive interactions.
Non-generic situations can, of course, give rise to much larger energy
scales, even at weak coupling.  Of particular importance in highly
anisotropic repulsively interacting systems is the spin-density-wave
instability.  Because this requires nesting in weak coupling, the
associated interaction vertices have been thrown out in our
calculations.

\section{Discussion}

The principal results of this paper are the $N$-chain weak coupling
phase diagrams, described in detail for small and large $N$ in
sections V and VI, respectively.  We now conclude with a discussion of
the {\sl implications} of these results for both ideal finite $N$
systems (accessible via numerical calculations) and for true {\sl
  quasi}-1d systems, where non-zero inter-ladder couplings need to be
taken into account.

\subsection{Numerics}

Numerical calculations have the advantage that direct comparisons
with 1d models can be made.  Recall that each phase is
characterized at the simplest level by its number of gapless charge
and spin modes.  These numbers can be measured numerically in a number
of ways.  Most directly, the lowest lying charge and spin excitation
energies can be determining by comparing ground state energies (in,
e.g. exact diagonalization or density-matrix RG methods) with
particles added or spins flipped.  Such measurements can also be
refined to determine the energies for the lowest-lying excitations
with a definite {\sl parity}, which can be related to the band indices
used here.  The {\sl total number} of gapless modes (both charge and
spin) can in principle be extracted alternatively from the coefficient
of $1/L$ (where $L$ is the chain length) in the finite-size correction
to the ground-state energy density.

The parity of the ground state and low-lying excited states are also 
accessible in weak coupling. Here we focus on the C1S0 phase in a 
2-chain system as an example. The parity operator of a 2-chain system is
\begin{equation}
P = \exp (i \pi N_{2}),
\end{equation} 
where $N_{2}$ is the total number of particles in the anti-bonding band.
The parity operator $P$ commutes with the Hamiltonian, so that
parity is a good quantum number. If the ground state is a linear
superposition of states with odd/even $N_{2}$, it has odd/even parity.
In the C1S0 phase, if total number of particles is even,  $N_{2}$ is 
even because electrons pair up in both bands. Therefore, the ground 
state has even parity. However, if the total number is odd, the 
analysis is complicated and remains an open question for further study.
The parities of excited states is determined by commutation relation
between the corresponding creation operators and parity operator. The
Bosonic field operator $\phi^{\rho+}_{12}(p)$, which is the only 
gapless mode in the C1S0 phase, creates a density excitation with
momenta $p$. If we express $N_{2}$ in terms of the Bosonic fields
\begin{equation}
N_{2}= \sqrt{2\pi} \{ \theta_{2\rho}(\infty) 
-\theta_{2\rho}(-\infty)\},
\end{equation} 
it is easy to show that $\phi^{\rho+}_{12}(p)$ commutes with the parity
operator. This implies that the excited state has the same parity as the 
ground state. In other words, the excitation carries even parity.
Consequently, numerical calculations should find a {\it charge gap} in
the odd parity channel, despite the exsistence of a gapless charge mode
with even parity. A simple test is the correlation function
\begin{equation}
C(x)= \langle \Delta \rho(x) \Delta \rho(0) \rangle,
\end{equation} 
where $\Delta \rho(x) \equiv c^{\dag}_{1\alpha}c_{1\alpha}(x)- 
c^{\dag}_{2\alpha}c_{2\alpha}(x)$. Since $P^{\dag} \Delta \rho P =-\Delta
 \rho$, $C(x)$ has only odd-parity (relative to the ground state)
intermediate states, and should therefore decay {\it exponentially},
$C(x) \sim e^{-\Delta_{\rho -} x/v}$!

It is often more convenient to compute correlation functions rather
than ground state energies (indeed, in a Monte Carlo calculation, this
is essentially the only option).  In this case, the information that
can be most reliably assayed is the presence or absence of charge and
spin gaps.  If there is a true gap in either sector, the corresponding
correlators are expected to decay exponentially in space.  To probe
the charge sector, the correlators of interest are those of the
density $\rho = c^\dagger_\alpha c^{\vphantom\dagger}_\alpha$ and the
pair field $\Delta = c_\uparrow c_\downarrow$.  In the spin sector,
the corresponding operator is simply the spin $\bbox{S} =
c^\dagger_\alpha {{\bbox{\sigma}} \over 2}
c^{\vphantom\dagger}_\beta$.  In principle, a detailed examination of
the Fourier content of the correlations should identify the Fermi
momenta of the gapless spin and charge modes (even more information
than their number), but this is quite difficult in practice due to
finite-size limitations imposed by the numerics.

\begin{figure}[hbt]
\epsfxsize=3.2in
\epsfbox{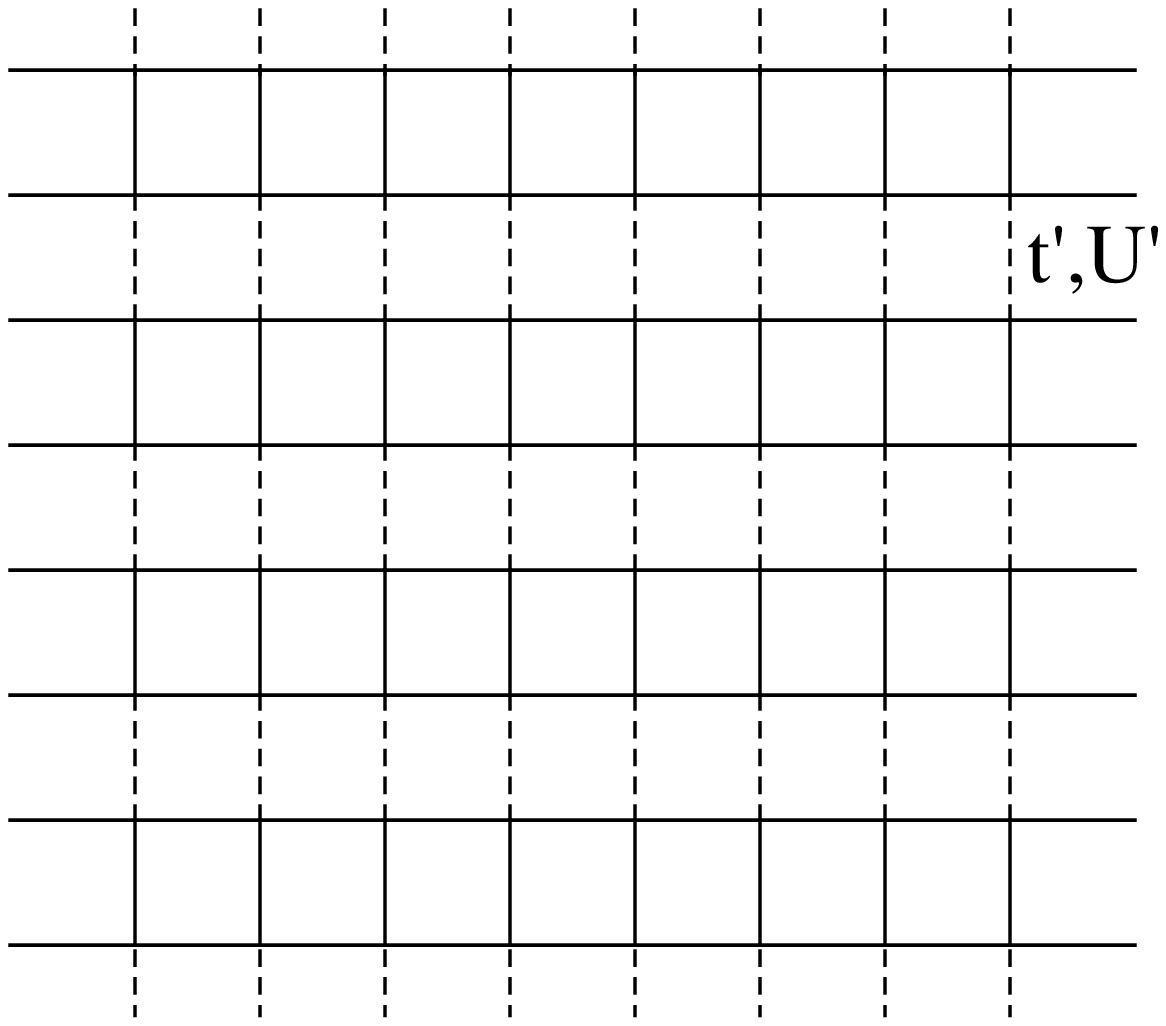}
\vspace{15pt}
{\noindent FIG. 12: A 2d array of Hubbard ladders,
    with weak inter-ladder hopping $t'$ and inter-ladder
    density-density interaction $U'$.}
\label{arrayfig}
\end{figure}

\subsection{Experimental consequences}

Comparison with experiments is more challenging.  In particular, it is 
inevitably the case that in any candidate compound there is at least 
some residual coupling between ladders as shown in Fig.  12.  In this 
sense, all real materials are at best {\sl quasi}-1d.  It is to the 
features of such quasi-1d ladder materials and the regime of validity 
of the previous results in this context to which we now turn.  The 
discussion will be kept at a general level, using only scaling 
considerations.  We take as a model a two- or three-dimensional 
regular array of $N$-chain ladders, the precise geometry of which is 
not crucial, although special cases resulting in Fermi-surface nesting 
will not be addressed.  A microscopic electronic model for such an 
array would involve the hopping amplitudes and interactions both on 
and between the ladders.  We will assume that the former are of the 
Hubbard form studied in the previous sections.  The latter generically 
introduce two new energy scales: an inter-ladder hopping amplitude 
$t'$ and an inter-ladder density-density interaction $U'$.  For now, 
we assume that, {\sl at least} $t'/t, U'/t \ll 1$.

\subsubsection{Weak interactions}

To proceed, let us imagine repeating the weak-coupling RG with these
additional interactions.  Upon first integrating out the large
wavevector modes ($|k_x| > \Lambda$) on each ladder, $t'$ and $U'$
will suffer small renormalizations, and other new interactions will
also be generated.  Two of these are of particular importance:
two-particle hopping processes, in which two Fermions are
simultaneously transferred from one ladder to a neighbor, and
four-particle ``pair-density'' interactions, in which Fermions
interact energetically on neighboring ladders in a manner quartic in
the density, but no charge is transferred .  These will occur with a
pair-hopping amplitude $t''$ and a pair-density
interaction $U''$, which are approximately
\begin{eqnarray}
  t'' & \sim & (t')^2 U^2/t^3, \label{est1}\\
  U'' & \sim & (U')^2 U^2/t^3, \label{est2}
\end{eqnarray}
in the weak-coupling limit (see Fig.[13]).  For
generality, we shall keep $t''$ and $U''$ as independent parameters.
Other interactions are of course also generated, but are either of
similar type but much smaller magnitude than those already considered,
or are higher-order and hence at least perturbatively irrelevant.

\begin{figure}[hbt]
\epsfxsize=3.2in
\epsfbox{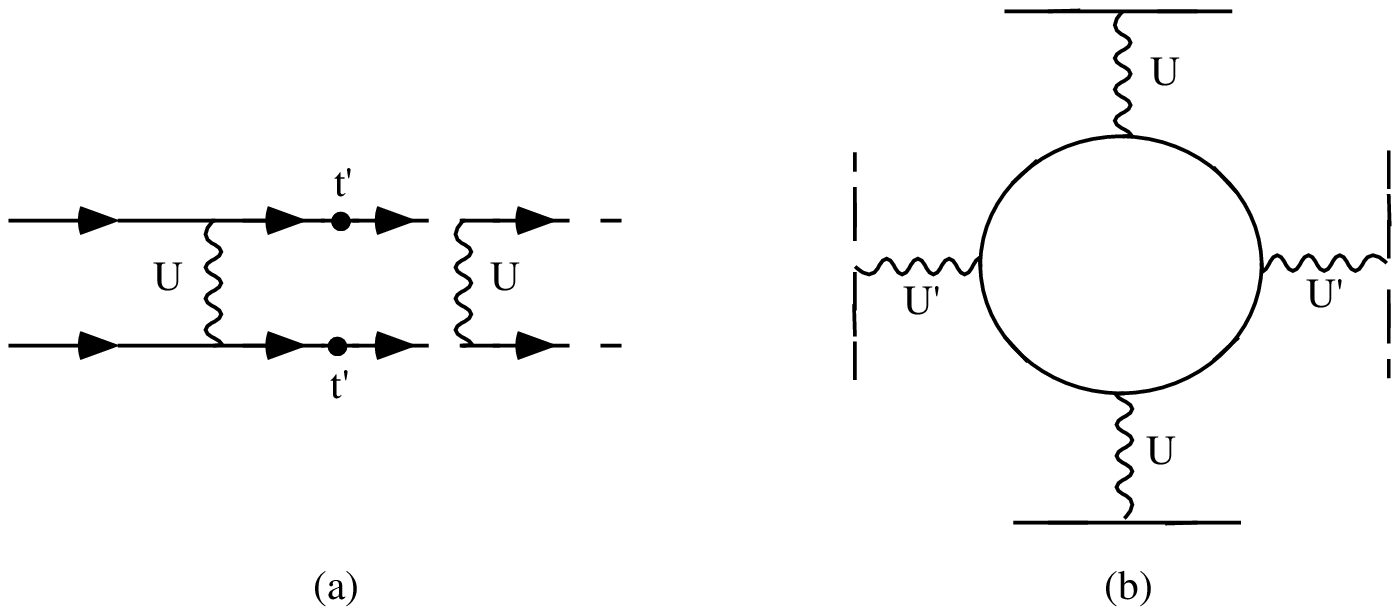}
\vspace{15pt}
{\noindent FIG. 13: Lowest-order diagrams illustrating the generation
  of inter-chain pair-hopping $t''$ and pair-density interactions
  $U''$ from the bare inter-chain hopping $t'$ and density interaction
  $U'$.}
\label{pair_generate_fig}
\end{figure}

At this point, we proceed with the RG as before,
working perturbatively in $t'/t$, $t''/t$, $U'/t$, and $U''/t$.  Like the
original ladder parameters, these will also rescale and non-linearly
renormalize themselves and other couplings.  The corrections to
Eqs.~\ref{RG1}--\ref{RG4}\ will, however, be negligible provided the
values of the {\sl running} couplings $t'/t|_l, t''/t|_l, U'/t|_l, U''/t|_l
\lesssim g_i(l)$.  Initially, this of course requires $t', t'', U' ,
U'' \ll U$, but the constraints become stronger as we iterate
to lower energy scales.  In particular, for the divergences
encountered purely within the ladder RG to be essentially unchanged
requires that the running inter-ladder couplings be negligible
compared to one (since the relevant ladder couplings become of order
one) at the scale $l^* \sim c t/U$, where $c$ is a constant.  Simple
power counting gives
\begin{eqnarray}
{\partial \over {\partial l}} \left({t' \over t}\right) & \approx & {t' \over
  t}, \label{tprimeRG} \\
{\partial \over {\partial l}} \left({t'' \over t}\right) & \approx &
\left({t'' \over  t}\right)^2, \label{tppRG} \\
{\partial \over {\partial l}} \left({U' \over t}\right) & \approx &
\left({U' \over  t}\right)^2, \label{UprimeRG} \\
{\partial \over {\partial l}} \left({U'' \over t}\right) & \approx &
-2{U'' \over  t}, \label{UppRG} 
\end{eqnarray}
since $t'$, $t''$, $U'$ and $U''$ represent two, four, four, and eight-
Fermion operators.
The hopping $t'/t$ thus grows exponentially, the two-particle
processes $t''/t$,$U'/t$ scale only logarithmically (are marginal),
and the pair-density interaction $U''$ is strongly irrelevant.
The 1d RG results are then valid up to the instability scale $l^*$
provided 
\begin{eqnarray}
  t' & \lesssim & t e^{-c t/U} \sim \Delta, \label{req1}\\
  t'' & \lesssim & U, \\
  U' & \lesssim & U, \label{req3} \\
  U'' & \lesssim & t e^{2ct/U} \sim t^3/\Delta^2, \label{req4}
\end{eqnarray}
where $\Delta$ is the energy scale of the gap in the 1d system (note
that factors of $(t/U)$ in the prefactor are not captured within a
one-loop RG treatment, so this is a rough estimate and not a strict
asymptotic statement).  The requirement of $t' \lesssim \Delta$ has a
simple physical interpretation:  for $t' \gtrsim \Delta$, it is
favorable for singlet pairs to break up to reduce their kinetic
energy, destroying the paired state.

Provided Eqs.~\ref{req1}--\ref{req4}\ are satisfied, the
strong-coupling analysis of section IV holds, and the paired bands are
adequately described by the single collective phase mode $\phi^{\rho+}$ (and
its conjugate $\theta^{\rho+}$).  For concreteness, we now specialize to the
two-chain case, where there are no additional bands.  The
single-particle tunneling operator conjugate to $t'$ then involves
exponentials of the dual fields $\phi_{a\sigma}$ (c.f.
Eq.~\ref{massive_hamiltonian}), which fluctuate wildly and are
exponentially suppressed (strongly irrelevant).  Similarly, the
density-density interaction $U'$ is also negligible due to strong
fluctuations of the relative-displacement mode $\theta^{\rho-}$.  The
remaining two couplings ($t''$ and $U''$) survive, and have simple
interpretations.  The pair-hopping, $t''$ simply hops a single Boson
between neighboring ladders, and is hence like a Josephson coupling.
The pair-density coupling $U''$ is effectively a density-density
interaction between Bosons on neighboring ladders, which are created
by the pair fields of section IV.  In terms of the phases, the
effective Hamiltonian is
\end{multicols}
\begin{equation}
   H_{\rm eff.} = \int \! dx \; \bigg\{ \sum_n \left[{{Kv} \over 2}
    (\partial_x \phi_n^{\rho+})^2 + {v \over
      {2 K}}(\partial_x\theta_n^{\rho+})^2\right] 
   -\sum_{\langle n n'\rangle} \left[t''\cos
    \sqrt{\pi}(\phi_n^{\rho+}-\phi^{\rho+}_{n'}) + {{U''} \over \xi^2} \cos
    2\sqrt{\pi} (\theta^{\rho+}_n-\theta^{\rho+}_{n'})\right] \bigg\},
  \label{boson_ham}  
\end{equation}
\begin{multicols}{2}
\noindent where the index $n$ labels the different ladders, and we have taken
the cut-off scale $\Lambda = 1$ for simplicity.  The factor of
$\xi^{-2}$ in the pair-density interaction reflects its irrelevance at
the non-interacting Fermion fixed point, and results directly from
integration of Eq.~\ref{UppRG}.  As discussed in
Ref.~\onlinecite{Balents96}\ the stiffness $K$ is not exactl
determinable within the weak-coupling RG for generic parameters.
However, various arguments suggest $K>1/2$,\cite{Noack96}\ and in
particular $K \approx 1$ close to half-filling in the two-leg ($N=2$)
ladder\cite{Schulz96a} at weak coupling. 

To determine the nature of the ground state of the system, we must now
address the physics at energy scales below $\Delta$.  To do this, we
use a different RG rescaling, using standard sine-Gordon
techniques.\cite{Jose77}\ This gives the linear flow
equations
\begin{eqnarray}
  {\partial \over {\partial l}} t'' & \approx &
  (2-{1 \over {2K}})t'' +O[(t'')^2,(U'')^2], \label{tppRG2} \\
  {\partial \over {\partial l}} U'' & \approx &
  (2-2K)U'' +O[(t'')^2,(U'')^2]. \label{UppRG2} 
\end{eqnarray}
Note that the Josephson coupling $t''$ is relevant for $K>
1/4$, while the pair-density interactions are relevant for $K<1$.
There is therefore no region of stability for truly one-dimensional
behavior, regardless of $K$.  Most probably $1/2 < K < 1$, and {\sl
  both} perturbations are relevant.  If both the dimensionless bare
interactions are weak, $t''/t, U''/(t\xi^2) \ll 1$, then the nature of
the instability is determined by the interaction which renormalizes to
large (order one) values first.  Simple algebra thus predicts that
pair-tunneling dominates for
\begin{equation}
  \left({t'' \over t}\right)^{2-2K} \gtrsim \left({U'' \over
      {t\xi^2}}\right)^{2-1/(2K)}, \label{SC_condition}
\end{equation}
and the ladders phase lock into a bulk superconducting (SC) state.  In
the opposite limit, pair-density interactions dominate and lead to a
paired-insulator or CDW state.  Note that for $K$ close to $1$, the SC
state dominates for all but extremely small $t''$.  Generically,
though, as $t''$ is reduced below the limit of Eq.~\ref{SC_condition},
the system makes a transition to the CDW state.  

Let us now use the estimates in Eqs.~\ref{est1}--\ref{est2}\ to
determine the bulk phase diagram for the two-chain system in the
weak-coupling limit.  As $t'$ is decreased, the pairing instability
occurs first, according to Eq.~\ref{req1}, when $t' \lesssim \Delta$.
Just below this scale, it is straightforward to show that this
instability always leads to a SC rather than CDW, provided $U/t \ll
1$, as supposed.  This is because Eq.~\ref{SC_condition}\ can be
rewritten, using Eqs.~\ref{est1}--\ref{est2}\ and the scaling of the
dimensionless coherence length $\xi \sim t/\Delta$, as
\begin{equation}
  {t' \over t} \gtrsim {t \over U} \left[ {{U' U \Delta} \over
      t^3}\right]^{(1-1/4K)/(1-K)}, 
\end{equation}
and $(1-1/(4K))/(1-K) \geq 1$ for $K> 1/2$.  Only for much smaller
$t'$ does this inequality cease to hold and the system go over into a
CDW state.  A schematic zero-temperature phase diagram for fixed small
$U/t$ with these features is shown in Fig.[14]. 

\begin{figure}[hbt]
\epsfxsize=3.2in
\epsfbox{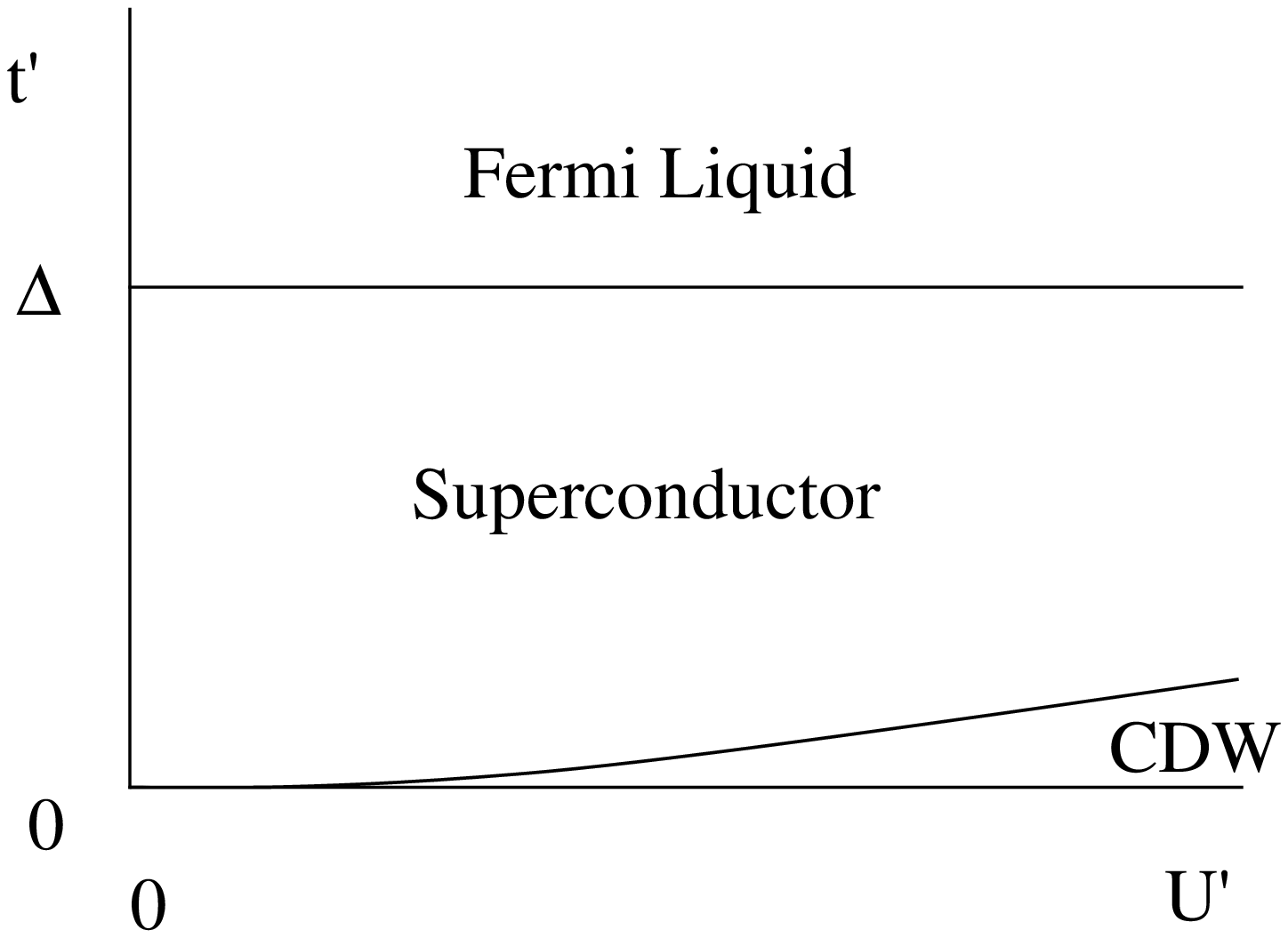}
\vspace{15pt}
{\noindent FIG. 14: Schematic zero temperature phase diagram in the
  $t'$--$U'$ plane for an array of coupled ladders.}
\label{schematic_pd_fig}
\end{figure}

Supposing the system is in the SC phase at zero temperature, what is
the expected phenomenology as the temperature is varied?  In weak
coupling, we expect several large crossover ranges.  For $T \gg T_{\rm
  pair} \sim \Delta$, the system acts approximately in a
non-interacting 1d fashion, with small logarithmic corrections which
are precursors of the instability to be encountered for $T \simeq
T_{\rm pair}$.  Below that temperature scale pairing effectively
occurs, and measurements (e.g.  magnetic susceptibility or electron
tunneling ) probing single-particle and spin excitations should
exhibit activated behavior.  However, although pairs form at $T_{\rm
  pair}$, superconducting coherence sets in only at a lower
temperature, $T_{sc}$.  

To see this, we continue rescaling after reaching $l^*$, i.e. to
energies below the spin gap.  We can now rescale further, as indicated
above, until the rescaled temperature grows to the order of the energy
cut-off, at which point the zero-temperature RG fails.  Having already
rescaled to $l^* \sim \ln (t/\Delta)$, the effective temperature has
already been increased to $(t/\Delta)T$, but for $T \lesssim \Delta$,
this is still small relative to the cut-off $t$, and we can rescale
further by the factor $b \sim \Delta/T$.  At this point the
temperature is on the order of the cut-off, and thermal smearing is
sufficient to remove any remaining quantum coherence at lower
energies.  The corresponding length $L_{qc} \sim v/T$ represents a
quantum to classical crossover scale at temperature $T$.  Fluctuations
of larger size behave essentially classically (as can be shown
explicitly by restricting the rescaled effective action to the zero
Matsubara frequency modes) and can be studied using the rescaled
classical (phase only) model
\begin{eqnarray}
  &\beta& H_{\rm class.}  =  \int \! dx \bigg\{ \sum_n {K \over 2}
  (\partial_x \phi^{\rho+}_n)^2 \nonumber \\
  & & - \sum_{\langle n n'\rangle}{t'' \over t}
  ({\Delta \over T})^{2-1/2K} \cos \sqrt{\pi} (\phi^{\rho+}_n -
  \phi^{\rho+}_{n'}) \bigg\}.
\end{eqnarray}
This classical model has only a single dimensionless coupling
constant, as can be seen by rescaling $x \rightarrow x/K$.  The
superconducting transition must occur when this dimensionless value is
order one, giving the critical temperature
\begin{equation}
  T_{\rm sc} \sim \Delta \left( t'' \over t\right)^{2K/(4K-1)},
\end{equation}
valid within the SC region of the phase diagram away from the zero
temperature quantum SC-CDW transition.  Note that for $t'' \ll t$,
there is a large temperature range $T_{\rm sc} \ll T \ll T_{pair}$
over which the system has a ``pseudo-gap''-like behavior.  The
difference between the exponent $2K/(4K-1)$ and $1/2$ (the classical
result for weakly coupled 1d XY chains) represents a suppression of
the transition temperature due to quantum fluctuations.

\subsubsection{Beyond weak interactions}

Although the analysis of this paper has assumed $U \ll t$, many
two-chain models have now been convincingly demonstrated, numerically
and in some limits analytically, to display spin gaps when weakly
doped even for relatively strong interactions (e.g. $U \gtrsim t$).
In fact, most aspects of the phenomenology in this conclusion are
expected to continue to hold even in this limit, provided that the
{\sl inter}-chain couplings are small, $U',t' \ll t$.  In particular,
on physical grounds, we expect a transition to a Fermi liquid (or at
least two-dimensional behavior) for $t' \gtrsim \Delta$.  For $t'
\lesssim \Delta$, inter-chain couplings essentially never break pairs,
and the physics will still be well-described by Eq.~\ref{boson_ham}\ 
at these and lower energy scales.  Of course, in strong coupling the
parameters $t''$ and $U''$ cannot be estimated using the weak-coupling
diagrams of Fig.[13] leading to Eqs.~\ref{est1}--\ref{est2}.  In
addition, it is difficult in strong coupling to determine $K$: recent
numerical simulations with $U/t =8$ suggest values of $K \approx
1/2$,\cite{Noack96}\ somewhat smaller (and hence less superconducting)
than in weak-coupling.  With these caveats, the remaining
phenomenology should continue to hold, both for the zero-temperature
phase diagram and for the crossovers and transitions at $T>0$.

\acknowledgements We are grateful to Doug Scalapino for illuminating
conversations.  This work has been supported by the National Science
Foundation under grant Nos. PHY94-07194, DMR-9400142 and DMR-9528578.

\appendix
\section{Current Algebra} 

Current algebra methods allow, among other things, an $algebraic$ 
calculation of the one-loop RG equations. 
Here we give a very terse description of this method. All
currents are defined in terms of the fermion fields $\psi_{R/Li\alpha}
\; (i=1,2, ..., N)$, which obey the operator products, 
\begin{eqnarray}
  \psi_{Ri\alpha}(x,\tau)\psi^{\dag}_{Rj\beta}(0,0)  &&\sim
  \frac{\delta_{ij}\delta_{\alpha \beta}}{2\pi z_i} +O(1),
  \nonumber\\
  \psi_{Li\alpha}(x,\tau)\psi^{\dag}_{Lj\beta}(0,0)  &&\sim
  \frac{\delta_{ij}\delta_{\alpha \beta}}{2\pi z^{*}_i} +O(1),
\end{eqnarray} 
where $z_i = v_i\tau -ix$. The operator products should
be understood to hold when two points $(x,\tau)$ and $(0,0)$ are
brought close together. We therefore only need to keep the singular
terms as replacement within correlation functions. As an example,
consider the product $J_{ij} J_{lm}$. Performing all possible
contractions gives
\begin{eqnarray} 
J_{ij}(x,\tau)J_{lm}(0,0) &&\sim
  :\psi^{\dag}_{i\alpha} \psi^{\vphantom\dag}_{j\alpha}: :\psi^{\dag}_{l\beta}
  \psi^{\vphantom\dag}_{m\beta}: 
  \nonumber\\ 
  && \sim \frac{2}{4\pi^2 z_i z_j}
  \delta_{im}\delta_{jl} +\frac{\delta_{im}}{2\pi
    z_i}:\psi_{j\alpha}\psi^{\dag}_{l\alpha}:
  \nonumber\\ 
  &&+\frac{\delta_{jl}}{2\pi z_j}:\psi^{\dag}_{i\alpha}\psi_{m\alpha}:
  +:J_{ij}J_{lm}:
  \nonumber\\ 
  &&\sim (\frac{\delta_{jl}}{2\pi z_j} J_{im}
  -\frac{\delta_{im}}{2\pi z_i} J_{lj})\nonumber\\
  &&+\delta_{im}\delta_{jl} \frac{2}{4\pi^2 z_i z_j}+O(1). 
\end{eqnarray} 
We can
compute the full set of operator products in a similar way. The ones
needed are: 
\begin{eqnarray} 
  J^{a}_{ij} J^{b}_{lm} &&\sim
  \frac{1/2}{4\pi^2 z_i z_j} \delta_{im}\delta_{jl}
  +\frac{\delta_{ab}}{4} ( \frac{\delta_{jl}}{2\pi z_j} J_{im}
  -\frac{\delta_{im}}{2\pi z_i} J_{lj} )\nonumber\\
  &&+\frac{i\epsilon^{abc}}{2}( \frac{\delta_{jl}}{2\pi z_j} J^{c}_{im}
  +\frac{\delta_{im}}{2\pi z_i} J^{c}_{lj} ),
  \\
  J^{a}_{ij} J_{lm} &&\sim
  \frac{\delta_{jl}}{2\pi z_j} J^{a}_{im} -\frac{\delta_{im}}{2\pi z_i}
  J^{a}_{lj}, 
\end{eqnarray} 
where the coordinates of two operators on
each left hand side are consecutively $(x,\tau)$ and $(0,0)$. Similar
forms hold for the left-moving currents, but with $z_i \rightarrow
z^{*}_{i}$.


The RG equations can be obtained very simply from the operator product
expansions.  We use the functional integral formulation which
results in the Euclidean action, $ S_E=
\int dxd\tau {\cal H}$, and the partition function, 
\begin{equation}
  {\cal Z} = \int [d{\bar \psi}][d\psi] e^{-S_E}. 
\end{equation} 
To perform the RG, the exponential is expanded to quadratic order in
${\cal H}$. A typical term takes the form, 
\begin{equation} 
  \frac12 {\tilde f}^{\sigma}_{ij} {\tilde f}^{\sigma}_{lm} \int_{z,w}
  \langle   J^{a}_{Rii}(z)J^{a}_{Ljj}(z)J^{b}_{Rll}(w)J^{b}_{Lmm}(w)
  \rangle, 
  \label{RG_term}
\end{equation} 
where $\int_{z,w}$ denotes a 4-dimensional integral over the two
complex planes $z$ and $w$. As in any RG, we wish to integrate out the
short-scale degrees of freedom to derive the effective theory at long
wavelengths and low energy. Here this is accomplished by considering
the contributions to Eq.~\ref{RG_term}\ when the two points $z$ and
$w$ are close together (near the cut-off scale). We make use of the
operator product expansion to integrate out the short-scale degrees of
freedom, which gives,
\begin{equation} 
  \frac12 ({\tilde
    f}^{\sigma}_{ij})^2 i\epsilon^{abc} i\epsilon^{abd} \int_{z,w}
  \frac{1}{2\pi(z_i-w_i)} \frac{1}{2\pi(z^*_j-w^*_j)}
  J^{c}_{Rii}J^{d}_{Ljj}. 
\end{equation} 
We choose a short distance
cut-off $a=\Lambda^{-1}$ in space, but none in imaginary time. For a rescaling
factor $b$, we must then perform the integral, 
\begin{equation}
  I_{ij}
  = \int_{a<|x|<ba} dx \int^{\infty}_{-\infty} d\tau \frac{1}{(2\pi)^2
    z_i z^*_j}= \frac{\ln b}{\pi(v_i+v_j)}, 
\label{renormalized_integral}
\end{equation}
where $z$ is the
relative coordinates in Eq.~\ref{renormalized_integral}. 
The contribution to the RG equation after integration is, 
\begin{equation} 
  -\frac{({\tilde
      f}^{\sigma}_{ij})^2}{\pi(v_i+v_j)} dl \int_z {\bf J}_{Rii} {\bf
    J}_{Ljj}, 
\end{equation}
where $dl = \ln b$ is the logarithmic length scale. This term, when
re-exponentiated, renormalizes ${\tilde f}^{\sigma}_{ij}$ and gives
the first term in Eq.~\ref{RG2}.  All other terms in the RG equations
can be carried out by similar steps.

\section{RG equations of umklapp interactions}

For PBC, when the number of chains is even, we need to study 
the transverse umklapp interactions in addition to the
forward and Cooper vertices. We then obtain additional terms on
the right hand sides of Eqs.~\ref{RG1}-\ref{RG4}.   Denoting these by 
$\delta f_{ij}, \delta c_{ij}$, one finds
\begin{eqnarray}
  \delta {\dot f}^{\rho}_{ij} = &&-\delta_{i\bar{j}}\sum_{k}\alpha_{ii,k}
  \big\{(u^{1\rho}_{ik})^2 +\frac{3}{16}(u^{1\sigma}_{ik})^2 \big\}
  \nonumber\\
  &&-\big\{(u^{2\rho}_{ij})^2 +\frac{3}{16}(u^{2\sigma}_{ij})^2 \big\}
  \nonumber\\
  &&+\sum_{a} \big\{(u^{a\rho}_{i\bar{j}})^2 
  +\frac{3}{16}(u^{a\sigma}_{i\bar{j}})^2\big\},
\end{eqnarray}
\begin{eqnarray}
  \delta {\dot f}^{\sigma}_{ij} = &&-\delta_{i\bar{j}}\sum_{k}\alpha_{ii,k}
  \big\{2u^{1\rho}_{ik} u^{1\sigma}_{ik} +\frac12 (u^{1\sigma}_{ik})^2 
  \big\}
  \nonumber\\
  &&-\big\{2u^{2\rho}_{ij} u^{2\sigma}_{ij} +\frac12 
  (u^{2\sigma}_{ij})^2 \big\}
  \nonumber\\
  &&+\sum_{a}
  \big\{2u^{a\rho}_{i\bar{j}} u^{a\sigma}_{i\bar{j}}
  -\frac12 (u^{a\sigma}_{i\bar{j}})^2 \big\},
\end{eqnarray}
\begin{eqnarray}
  \delta {\dot c}^{\rho}_{ij}=&&\delta_{ij}
  \big\{ (u^{1\rho}_{i\bar{i}})^2 
  +\frac{3}{16}(u^{1\sigma}_{i\bar{i}})^2\big\}
  \nonumber\\
  &&+2(u^{1\rho}_{i\bar{j}}u^{2\rho}_{i\bar{j}}
  +\frac{3}{16}u^{1\sigma}_{i\bar{j}}u^{2\sigma}_{i\bar{j}}),
\end{eqnarray}
\begin{eqnarray}
  \delta {\dot c}^{\sigma}_{ij}=&&\delta_{ij}
  \big\{2u^{1\rho}_{i\bar{i}} u^{1\sigma}_{i\bar{i}} 
  -\frac12 (u^{1\sigma}_{i\bar{i}})^2 \big\}
  \nonumber\\
  &&+2( u^{2\rho}_{i\bar{j}} u^{1\sigma}_{i\bar{j}} 
  +u^{1\rho}_{i\bar{j}} u^{2\sigma}_{i\bar{j}}
  -\frac12 u^{1\sigma}_{i\bar{j}} u^{2\sigma}_{i\bar{j}}).
\end{eqnarray}

We also need the RG equations for the umklapp couplings themselves,
which are,
\begin{eqnarray}
  {\dot u}^{1\rho}_{ij} =&& (q^{\rho}_{ij} u^{1\rho}_{ij} 
  +\frac{3}{16} q^{\sigma}_{ij} u^{1\sigma}_{ij} )
  \nonumber\\
  &&+2(c^{\rho}_{i\bar{j}} u^{2\rho}_{ij}
  +\frac{3}{16}c^{\sigma}_{i\bar{j}} u^{2\sigma}_{ij}),
\end{eqnarray}
\begin{eqnarray}
  {\dot u}^{1\sigma}_{ij} =&&( q^{\rho}_{ij} u^{1\sigma}_{ij}
  +q^{\sigma}_{ij} u^{1\rho}_{ij}
  -\frac12 q^{\sigma}_{ij} u^{1\sigma}_{ij})
  \nonumber\\
  &&-(f^{\sigma}_{i\bar{i}}+f^{\sigma}_{j\bar{j}}) u^{1\sigma}_{ij}
  \nonumber\\
  &&+2(c^{\rho}_{i\bar{j}} u^{2\sigma}_{ij}
  +c^{\sigma}_{i\bar{j}} u^{2\rho}_{ij}
  -\frac12 c^{\sigma}_{i\bar{j}} u^{2\sigma}_{ij}),
\end{eqnarray}
\begin{eqnarray}
  {\dot u}^{2\rho}_{ij}=&&2(c^{\rho}_{i\bar{j}}u^{1\rho}_{ij}
  +\frac{3}{16} c^{\sigma}_{i\bar{j}}u^{1\sigma}_{ij})
  \nonumber\\
  &&+2(p^{\rho}_{ij}u^{2\rho}_{ij}
  +\frac{3}{16} p^{\sigma}_{ij}u^{2\sigma}_{ij}),
\end{eqnarray}
\begin{eqnarray}
  {\dot u}^{2\sigma}_{ij} = &&
  2(c^{\rho}_{i\bar{j}} u^{1\sigma}_{ij}
  +c^{\sigma}_{i\bar{j}} u^{1\rho}_{ij}
  -\frac12 c^{\sigma}_{i\bar{j}} u^{1\sigma}_{ij})
  \nonumber\\
  &&+2(p^{\rho}_{ij} u^{2\sigma}_{ij}
  +p^{\sigma}_{ij} u^{2\rho}_{ij}
  -\frac12 p^{\sigma}_{ij} u^{2\sigma}_{ij})
  \nonumber\\
  &&-2f^{\sigma}_{ij}u^{2\sigma}_{ij},
\end{eqnarray}
where $p^{\alpha}_{ij} \equiv f^{\alpha}_{i\bar{j}}-f^{\alpha}_{ij}$
and $q^{\alpha}_{ij} \equiv 2\delta_{i\bar{j}} c^{\alpha}_{ii}
+(2f^{\alpha}_{i\bar{j}}-f^{\alpha}_{i\bar{i}}-f^{\alpha}_{j\bar{j}})$,
$\alpha = \rho, \sigma$. 

\section{initial values of the couplings}

Upon changing to the band basis, the on-site Hubbard repulsion
is transformed into a set of interactions
between the different bands. Using Eq.~\ref{trans_matrix}, 
we have
\begin{eqnarray}
  -{\cal H}_{int} &&= -U \sum_{i}: c^{\dag}_{i\uparrow}(x)
  c^{\vphantom\dag}_{i\uparrow}(x) 
  c^{\dag}_{i\downarrow}(x) c^{\vphantom\dag}_{i\downarrow}(x):
  \\
  &&=-U\sum_{ijkl} A_{ijkl} : \psi^{\dag}_{i\uparrow}
  \psi^{\vphantom\dag}_{j\uparrow} 
  \psi^{\dag}_{k\downarrow} \psi^{\vphantom\dag}_{l\downarrow}:,
  \nonumber\\
  \text{where}\:\: 
  &&A_{ijkl} \equiv \sum_{m} S^{*}_{mi}S_{mj}S^{*}_{mk}S_{ml}.
\label{Afunction}
\end{eqnarray}

Consider first OBCs.  After linearizing around the Fermi points, each
operator is split into left and right moving pieces.  In terms of these,
\begin{eqnarray}
  -{\cal H}_{int} =&&-U\sum_{ijkl} \sum_{P_i=\pm} A_{ijkl} : 
  \psi^{\dag}_{P_1i\uparrow} \psi^{\vphantom\dag}_{P_2j\uparrow}
  \psi^{\dag}_{P_3k\downarrow} \psi^{\vphantom\dag}_{P_4l\downarrow}:
  \nonumber\\
  &&e^{i(-P_1k_{F1}+P_2k_{F2}-P_3k_{F3}+P_4k_{F4})}.
\label{int_form}
\end{eqnarray}
Now we are ready to compare the coefficients of interactions in
Eqs.~\ref{int1}, \ref{int_form}.  For example, comparing the
coefficient in front of the term
$\psi^{\dag}_{Ri\uparrow}\psi^{\vphantom\dag}_{Ri\downarrow}\psi^{\dag}_{Li\downarrow}\psi^{\vphantom\dag}_{Li\uparrow}$
yields the relation
\begin{equation}
  \frac12 {\tilde f}^{\sigma}_{ij} = U A_{ijji}=U A_{iijj}.
\end{equation}
Comparing the coefficients of
$\psi^{\dag}_{Ri\uparrow}\psi^{\vphantom\dag}_{Ri\uparrow}
\psi^{\dag}_{Li\downarrow}\psi^{\vphantom\dag}_{Li\downarrow}$ gives
\begin{equation}
  -({\tilde f}^{\rho}_{ij}+\frac14 {\tilde f}^{\sigma}_{ij}) = -U A_{iijj}.
\end{equation}
We can then solve for the initial values of the forward couplings.
\begin{eqnarray}
  {\tilde f}^{\sigma}_{ij} =&& 4 {\tilde f}^{\rho}_{ij} = 2 U B_{ij},
  \nonumber\\
  \text{where} \:\: B_{ij} \equiv&& \sum_{m} |S_{mi}|^2 |S_{mj}|^2.
\end{eqnarray}
A straightforward computation gives
\begin{equation}
  B_{ij} = \frac{1}{N+1} ( 1 +\frac12 \delta_{i+j, N+1} +\frac12 \delta_{i,j}).
\end{equation}
By similar comparisons, we obtain the initial values of the Cooper
couplings,
\begin{equation}
  {\tilde c}^{\sigma}_{ij} = 4 {\tilde c}^{\rho}_{ij} = 2U A_{ijij} = 2U B_{ij},
\end{equation}
where in the last step we use the fact that the transformation matrix
$S_{ij}$ in Eq.~\ref{SOBC} is real for OBCs.

For PBCs, similar results can be obtained by this method.  Taking care
to note the different conventions for left and right movers for PBCs
(see Eq.~\ref{XfourierPBC}), the initial values
of the forward and Cooper couplings are
\begin{eqnarray}
  {\tilde f}^{\sigma}_{ij} =&& 4{\tilde f}^{\rho}_{ij} = 2U
  A_{ii\bar{j}\bar{j}},
  \\
  {\tilde c}^{\sigma}_{ij} =&& 4{\tilde c}^{\rho}_{ij} = 2U
  A_{ij\bar{i}\bar{j}}, 
\end{eqnarray}
where $\bar{i} =-i$.
$ $From Eq.~\ref{SPBC}, one can compute these initial values.  These are
\begin{eqnarray}
  {\tilde f}^{\sigma}_{ij} =&& 4{\tilde f}^{\rho}_{ij} = \frac{2U}{N},
  \\
  {\tilde c}^{\sigma}_{ij} =&& 4{\tilde c}^{\rho}_{ij} =\frac{2U}{N}.
\end{eqnarray}
If the number of chains is even, we also need the initial values of the
transverse umklapp couplings.  In fact, the initial values
are the same as those for the forward and Cooper couplings:
\begin{eqnarray}
  {\tilde u}^{1\sigma}_{ij} =&& 4{\tilde u}^{1\rho}_{ij} = \frac{2U}{N},
  \\
  {\tilde u}^{2\sigma}_{ij} =&& 4{\tilde u}^{2\rho}_{ij} =\frac{2U}{N}.
\end{eqnarray}
Note that, in all case, the initial values of the rescaled couplings
in Eqs.~\ref{RG1}-\ref{RG4}\ (without the tildes) are obtained by
multiplying the factor $1/\pi(v_i+v_j)$.

\section{Representations for the Klein factors}

The Klein factors defined in Eq.~\ref{bosonization_formula}\ satisfy the 
commutation relations,
\begin{equation}
  \{ \eta_{i\alpha}, \eta_{j\beta} \} = 2 \delta_{ij} \delta_{\alpha \beta}.
\end{equation}
In order to bosonize the relevant interactions in 
Eq.~\ref{relevant_int1}, we need to prove that the products of the Klein
factors in different terms commute with each other. Then, they can 
be simultaneously diagonalized with a specific choice of representation.

The products of the Klein factors for the first term in 
Eq.~\ref{relevant_int1} are,
\begin{equation}
  \eta_{d\uparrow}\eta_{d\downarrow}\eta_{d\downarrow}\eta_{d\uparrow}
  =\eta_{d\downarrow}\eta_{d\uparrow}\eta_{d\uparrow}\eta_{d\downarrow}
  =1.
\end{equation}
For the second term, the Klein factors we need are,
\begin{equation}
  \eta_{a\uparrow}\eta_{a\downarrow}
  \eta_{b\downarrow}\eta_{b\uparrow}
  =\eta_{a\downarrow}\eta_{a\uparrow}
  \eta_{b\uparrow}\eta_{b\downarrow}
  \equiv g.
\end{equation}
A simple computation gives $g^{2}=1$. Thus, all the products of
Klein factors in the above equations commute with each other.  It is
therefore consistent to choose the trivial representation $g=1$.
For the cases of interest, then, no special signs or auxiliary Fermion
fields are necessary in the bosonized Hamiltonian.

\end{multicols}


\begin{thebibliography}{10}
\bibitem{KL:foot}
The one-step integration gives $O(U^2)$ corrections to $c_{ij}^\sigma(0)$ which
  could induce negative eigenvalues in some of the $N_f-1$ directions whose
  bare eigenvectors vanish. This mechanism can lead to extremely weak
  superconducting instabilities (the Kohn-Luttinger effect) even in
  weak-coupling, as discussed in this context by Zanchi\cite{Zanchi96}.
\end{thebibliography}
\end{document}